\documentclass[11pt]{article}

\newif\iflabstyle
\labstyletrue

\usepackage[margin=1in]{geometry}
\usepackage{amsmath,amssymb,amsfonts}
\usepackage{graphicx}
\graphicspath{{figures/}}
\usepackage{float}
\usepackage{wrapfig}
\usepackage{array}
\usepackage{booktabs}
\usepackage{longtable}
\usepackage{pdflscape}
\usepackage{ragged2e}
\usepackage{hyperref}
\usepackage{xcolor}
\usepackage{multirow}
\usepackage{subcaption}
\usepackage[super,comma,sort&compress]{natbib}

\usepackage{lineno}
\usepackage{etoc}
\usepackage{tcolorbox}
\tcbuselibrary{breakable}
\usepackage{soul}
\usepackage{microtype}
\usepackage{xspace}

\newsavebox{\fitbox}
\newcommand{\fitwidth}[1]{%
  \sbox{\fitbox}{#1}%
  \ifdim\wd\fitbox>\linewidth
    \resizebox{\linewidth}{!}{\usebox{\fitbox}}%
  \else
    \usebox{\fitbox}%
  \fi}

\newcolumntype{P}[1]{>{\RaggedRight\arraybackslash}p{#1}}
\newcolumntype{C}[1]{>{\Centering\arraybackslash}p{#1}}

\tcbuselibrary{skins}
\definecolor{methodslice}{HTML}{93C0A0}

\iflabstyle
  \usepackage{setspace}
  \usepackage{titlesec}
  \definecolor{accentink}{HTML}{5E8C6E}   %
  \definecolor{ruleink}{HTML}{B9CFC1}     %
  \definecolor{marginink}{HTML}{9A9A9A}   %

  \newlength{\headnumsep}
  \titleformat{\section}
    {\normalfont\sffamily\large\bfseries}
    {\llap{\textcolor{accentink}{\thesection}\hspace{\headnumsep}}}{0pt}{}
  \titleformat{\subsection}
    {\normalfont\sffamily\normalsize\bfseries}
    {\llap{\textcolor{accentink}{\thesubsection}\hspace{\headnumsep}}}{0pt}{}
  \titleformat{\paragraph}[runin]
    {\normalfont\sffamily\bfseries\color{accentink}}{}{0pt}{}
  \titlespacing*{\section}{0pt}{1.5\baselineskip}{0.45\baselineskip}
  \titlespacing*{\subsection}{0pt}{1.15\baselineskip}{0.3\baselineskip}
  \titlespacing*{\paragraph}{0pt}{0.9\baselineskip}{0.6em}

\fi

\tcbset{qualbox/.style={
  enhanced,
  sharp corners=south,
  rounded corners=north,
  arc=3pt,
  boxrule=0.6pt,
  colframe=black!45,
  colback=white,
  left=7pt,
  right=7pt,
  top=5pt,
  bottom=6pt
}}

\usepackage[nolist,nohyperlinks]{acronym}

\usepackage[capitalize]{cleveref}

\usepackage{newfloat}
\DeclareFloatingEnvironment[name={Extended Data Fig.},placement=htbp]{edfigure}
\DeclareFloatingEnvironment[name={Extended Data Table},placement=htbp]{edtable}
\DeclareCaptionSubType{edfigure}
\crefname{edfigure}{Extended Data Fig.}{Extended Data Figs.}
\Crefname{edfigure}{Extended Data Fig.}{Extended Data Figs.}
\crefname{edtable}{Extended Data Table}{Extended Data Tables}
\Crefname{edtable}{Extended Data Table}{Extended Data Tables}
\crefname{subedfigure}{Extended Data Fig.}{Extended Data Figs.}
\Crefname{subedfigure}{Extended Data Fig.}{Extended Data Figs.}
\crefname{appendix}{Supplementary Note}{Supplementary Notes}
\Crefname{appendix}{Supplementary Note}{Supplementary Notes}

\iflabstyle
  \hypersetup{
    colorlinks,
    linkcolor=accentink,
    citecolor=accentink,
    urlcolor=accentink,
  }
\else
  \hypersetup{
    colorlinks,
    linkcolor=blue,
    citecolor=blue,
    urlcolor=magenta,
  }
\fi

\newtcolorbox[auto counter,crefname={Box}{Boxes},Crefname={Box}{Boxes}]{promptbox}[2][]{
  colback=blue!5,
  colframe=blue!30,
  fonttitle=\bfseries,
  boxrule=0.5pt,
  arc=2pt,
  left=6pt, right=6pt, top=4pt, bottom=4pt,
  title={Prompt Box~\thetcbcounter: #2},
  #1
}

\newcommand{\methodfont}[1]{\textsf{#1}}

\DeclareMathOperator*{\argmin}{arg\,min}

\newcommand{\benchmark}{\methodfont{MorphoRecoveryBench}\xspace}
\newcommand{\ourmethod}{\methodfont{SCOPE}\xspace}
\newcommand{\SIMIL}{\methodfont{SI-MIL}\xspace}
\newcommand{\SIMILDense}{\methodfont{tile-dense}\xspace}
\newcommand{\SIMILSparse}{\methodfont{tile-sparse}\xspace}
\newcommand{\tiletopk}{\methodfont{tile-top25}\xspace}
\newcommand{\ourmethodcbm}{\methodfont{CBM-SCOPE}\xspace}
\newcommand{\virchowtwo}{VirchowV2\xspace}
\newcommand{\pathgenclip}{\methodfont{PathGen-CLIP-L}\xspace}
\newcommand{\splice}{\methodfont{SpLiCE}\xspace}

\makeatletter
\providecommand{\todo}[1]{\textcolor{red}{\textbf{TODO:} #1}\@latex@warning{TODO: #1}}
\providecommand{\TODO}[1]{}

\definecolor{julius}{rgb}{0.8, 0.2, 0.2}    %
\definecolor{steph}{rgb}{0, 0.6, 0.6}   %
\definecolor{kenza}{rgb}{0.2, 0.6, 0.2}    %
\definecolor{daniel}{rgb}{0.8, 0.4, 0.0}   %
\definecolor{shruthi}{rgb}{0.4, 0.4, 0.8}    %
\definecolor{tristan}{rgb}{0.7, 0.3, 0.7}  %

\newcommand{\coloredcomment}[2]{{\color{#1}#2}\@latex@warning{#2}}

\newcommand{\wip}[1]{{\color{gray}#1}\@latex@warning{WIP segment}}
\makeatother

\begin{acronym}
    \acro{MIL}{multiple instance learning}
    \acro{WSI}{whole-slide image}
    \acro{LLM}{large language model}
    \acro{VLM}{vision--language model}
\end{acronym}

\newcommand{\RubyAurocSplice}{0.90}

\newcommand{\RubyAurocAbmil}{0.96}

\newcommand{\RubyAurocCbmSlice}{0.93}
\newcommand{\RubyEfficiencyTilesPerSlide}{12,040}
\newcommand{\RubyEfficiencySIMILSparseSeconds}{66.5}
\newcommand{\RubyEfficiencySLiCEMilliseconds}{5.5}

\newcommand{\RubyCbmCalibrationSeedCount}{5}
\newcommand{\RubyCbmCalibrationRtwoMin}{0.72}
\newcommand{\RubyCbmCalibrationRtwoMax}{0.97}
\newcommand{\RubyJfONESplice}{0.47}
\newcommand{\RubyJfONECosine}{0.11}
\newcommand{\RubyJfONECosineTopkTwentyFive}{0.31}
\newcommand{\RubyJfONERandom}{0.09}

\newcommand{\RubyCptacBrcaAuroc}{0.86}
\newcommand{\RubyCptacMsiAuroc}{0.70}
\newcommand{\RubyCptacNsclcAuroc}{0.97}
\newcommand{\RubyDiscoveryAtlasTaskCount}{26}
\newcommand{\RubyDiscoveryAtlasOrganCount}{8}
\newcommand{\RubyDiscoveryAtlasHypothesisCount}{34}

\newcommand{\RubyDiscoveryAtlasAurocMin}{0.611}
\newcommand{\RubyDiscoveryAtlasAurocMax}{0.969}
\newcommand{\RubyDiscoveryAtlasHnscHpvAuroc}{0.835}
\newcommand{\RubyDiscoveryAtlasHnscHpvHypothesis}{Lesion composed of basaloid ``blue'' epithelial cells with scant cytoplasm arranged in solid nests/sheets showing peripheral palisading and retraction clefting, consistent with a basal cell carcinoma--type pattern. A prominent lichenoid/dense lymphoid infiltrate with intraepithelial lymphocytes is present, with occasional surface scale crust/ulceration and focal necrosis.}
\newcommand{\RubyDiscoveryAtlasCandidateTaskCount}{35}
\newcommand{\RubyDiscoveryAtlasExcludedTaskCount}{9}
\newcommand{\RubyDiscoveryAtlasAurocThreshold}{0.6}
\newcommand{\RubyTemplateJudgeNRepeats}{20}
\newcommand{\RubyTemplateJudgeSigmaJfOne}{0.09}
\newcommand{\RubyTemplateJudgeSeedSigmaJfOne}{0.08}

\newcommand{\RubyJudgeOnlyNRepeats}{10}
\newcommand{\RubyJudgeOnlyFrozenRepeat}{4}
\newcommand{\RubyJudgeSigmaJfOne}{0.05}
\newcommand{\RubyTemplaterSigmaJfOne}{0.06}
\newcommand{\RubyJudgeVarSharePct}{39}
\newcommand{\RubyTemplaterVarSharePct}{61}
\newcommand{\RubyPvalSIMILSparseVsSIMILDense}{$p<0.001$}
\newcommand{\RubyPvalSliceVsSIMILSparseJfOne}{$p=0.70$}
\newcommand{\RubyPvalSliceVsSIMILSparseNdcg}{$p=0.12$}
\newcommand{\RubyJfONESlice}{0.50}
\newcommand{\RubyNdcgTenSlice}{0.62}
\newcommand{\RubyNdcgTenSplice}{0.53}

\newcommand{\RubyPvalBackbonePathgenVsMusk}{$p=0.04$}

\newcommand{\RubyBackbonePathgenJfOne}{0.50}
\newcommand{\RubyBackboneMuskJfOne}{0.27}
\newcommand{\RubyBackboneConchJfOne}{0.22}

\newcommand{\RubyPvalSIMILDenseVsRandom}{$p=0.58$}
\newcommand{\RubyPvalSIMILSparseVsRandom}{$p<0.001$}
\newcommand{\RubyPvalSliceVsRandom}{$p<0.001$}

\newcommand{\RubyHumanPearsonFOne}{0.87}
\newcommand{\RubyHumanPearsonPFOne}{$p<0.001$}
\newcommand{\RubyHumanSpearmanFOne}{0.85}
\newcommand{\RubyHumanSpearmanPFOne}{$p<0.001$}
\newcommand{\RubyPvalGtaugJfOne}{$p=0.009$}
\newcommand{\RubyPvalGtaugRecall}{$p=0.006$}
\newcommand{\RubyCorrGtaugMissingR}{0.65}
\newcommand{\RubyCorrGtaugMissingP}{$p=0.03$}
\newcommand{\RubyPvalBankSaturates}{$p=0.35$}
\newcommand{\RubyPvalBankSmall}{$p<0.001$}
\newcommand{\RubySparsityOptimalTargetMin}{2}
\newcommand{\RubySparsityOptimalTargetMax}{100}

\iflabstyle\rightlinenumbers*\fi

\begin{document}

\title{Sparse concept attribution for histomorphological hypothesis generation from whole-slide classifiers}
\iflabstyle
  \makeatletter
  \renewcommand{\@maketitle}{%
    \begin{flushleft}%
      {\LARGE\bfseries\setstretch{1.12}\@title\par}%
      \vspace{1.1em}%
      {\setstretch{1.15}\@author\par}%
    \end{flushleft}%
    \vspace{0.6em}%
  }
  \makeatother
\fi
\author{%
  \iflabstyle
    Tristan Lazard\,$^{1}$ \quad
    Kenza Bouzid\,$^{1}$ \quad
    Julius Hense\,$^{2,3,\ast}$ \quad
    Shruthi Bannur\,$^{1}$ \quad
    Daniel Coelho de Castro\,$^{1}$ \quad
    Daniel Shao\,$^{4}$ \quad
    Rajesh Jena\,$^{5}$ \quad
    Drew Williamson\,$^{6}$ \quad
    Stephanie Hyland\,$^{1,\dagger}$
  \else
    Tristan Lazard\,$^{1}$ \quad
    Kenza Bouzid\,$^{1}$ \quad
    Julius Hense\,$^{2,3,\ast}$ \\[0.3em]
    Shruthi Bannur\,$^{1}$ \quad
    Daniel Coelho de Castro\,$^{1}$ \quad
    Daniel Shao\,$^{4}$ \\[0.3em]
    Rajesh Jena\,$^{5}$ \quad
    Drew Williamson\,$^{6}$ \quad
    Stephanie Hyland\,$^{1,\dagger}$
  \fi
  \\[1em]
  {\normalsize
    \begin{minipage}{\iflabstyle\textwidth\else\dimexpr\textwidth-2\tabcolsep\relax\fi}
      \raggedright\setlength{\parindent}{0pt}%
      $^{1}$Microsoft Research, Cambridge, UK \\
      $^{2}$Berlin Institute for the Foundations of Learning and Data (BIFOLD), Berlin, Germany \\
      $^{3}$Machine Learning Group, Technische Universit\"at Berlin, Berlin, Germany \\
      $^{4}$Harvard--MIT Division of Health Sciences and Technology, Cambridge, MA, USA \\
      $^{5}$Department of Oncology, University of Cambridge, Cambridge CB2 0QQ, UK \\
      $^{6}$Department of Pathology \& Laboratory Medicine, Emory University School of
      Medicine, Atlanta, GA, USA \\[0.35em]
      $^{\ast}$Work done during an internship at Microsoft. \\
      $^{\dagger}$Corresponding author: \href{mailto:stephanie.hyland@microsoft.com}{stephanie.hyland@microsoft.com}
    \end{minipage}}%
}
\date{}
\maketitle

\iflabstyle
  \renewenvironment{abstract}{%
    \vspace{0.4em}%
    {\color{ruleink}\hrule height 0.7pt}%
    \vspace{0.9em}%
    \begin{spacing}{1.15}\noindent\ignorespaces
  }{%
    \end{spacing}%
    \vspace{0.4em}%
    {\color{ruleink}\hrule height 0.7pt}%
    \vspace{0.6em}%
  }
\fi

\begin{abstract}
Histology images contain rich morphological information and can provide insights into pathological processes. However, deriving hypotheses relating morphological phenotypes to clinical attributes is bottlenecked by a manual image interpretation step. Here, we demonstrate that this process can be automated through interpretable deep learning. We present \ourmethod, a method to interpret slide-level classifiers by combining pathology-specific vision--language models with sparse concept attribution onto a generalist histomorphological concept bank. To measure whether such explanations recover known morphology, we introduce \benchmark, a benchmark of seven tasks with pathologist-curated reference descriptions. On this benchmark, dense concept attribution is indistinguishable from a random baseline, whereas sparse attribution recovers substantial known morphology; decomposing the pooled slide embedding reaches similar explanation correctness at a fraction of the computational cost. Post-hoc interpretation of whole-slide classifiers can thus generate morphological hypotheses at scale, for expert validation.

\end{abstract}


\bigskip

\iflabstyle
\fi

\etocsettocdepth.toc{none}
\section{Introduction}
\label{sec:introduction}
\Acp{WSI} capture the cellular and tissue morphology of a macroscopic biological specimen. However, understanding how this morphology relates to other biological or clinical quantities remains a challenging research question. Decades of research have allowed pathologists to draw generalisable links between tissue morphology and clinical outcomes. %
\Ac{MIL}, pathology foundation models, and large-scale multimodal public datasets (such as TCGA and CPTAC) \citep{weinstein2013tcga, clarkCancerImagingArchive2013} have enabled the emergence of powerful slide-level predictive models. It is now possible to accurately predict a wide range of variables, from cancer grades to molecular biomarkers and patient outcomes \citep{katherPancancerImagebasedDetection2020, yanSparseHierarchicalTransformer2023, kather2019predicting, 
Bouzid2024, titan2025, lazardDeepLearningIdentifies2022a, wangPathologyFoundationModel2024}, directly from H\&E slides. The success of these models suggests a learnable link between morphology and the variable of interest, even when such a link has not been characterised by pathologists. 
Recovering the morphological patterns a model has learned to rely on is thus one way to generate hypotheses about the discriminative morphology of a task.
However, to move from prediction to understanding, we must be able to explain these models in terms a human can interpret and verify.

Inspection of attention-based MIL models typically uses slide heatmaps, localizing the tiles with the highest attention scores~\citep{abmil, clam}. These spatial maps can be contrasted with expert expectations, supporting qualitative validation and error analysis \citep{campanellaClinicalgradeComputationalPathology2019b}. They can also be used for discovery by having experts manually identify morphological patterns in highly attended tiles \citep{chenPancancerIntegrativeHistologygenomic2022a, saillardPredictingSurvivalHepatocellular2020}. 
While valuable, such approaches are limited: heatmaps demonstrate which regions were important to the model for a single prediction, but not the specific histological features responsible. Laborious and subjective human interpretation is therefore required to extrapolate to dataset-level morphological hypotheses~\citep{jamshidiidajiAttentionHeatmapsHow2026}. Prototype-based slide representations share this limitation: they summarise a slide into a small set of recurring morphological prototypes~\citep{panther2024}, but these prototypes are unnamed visual clusters that a pathologist must still inspect and label.
Two more recent families share this property. Sparse autoencoders trained on pathology foundation-model embeddings discover concept directions automatically~\citep{picasso2026}, and have been used to explain subtyping~\citep{bisson2026sarcoma} or survival~\citep{redlich2026gbm}, but these directions are unlabelled and require post-hoc annotation by a pathologist. Counterfactual diffusion models likewise show what would have to change for a prediction to flip~\citep{mopadi2026}, leaving the pathologist to name the difference they illustrate.

We propose to automatically express these hypotheses in language, the native medium in which pathologists describe and reason about morphology.
The development of \acp{VLM} for histopathology~\citep{clip, conch, pathgen2024, musk} provides a mechanism to express images in textual terms, through a learned mapping into a shared representation space.
These models already encode a rich morphological vocabulary, achieving strong zero-shot subtyping performance~\citep{conch, musk, pathgen2024}.
However, the geometry of this representation space is sufficiently complex that image--text translation is not trivial: many unrelated concepts share high similarity, so standard similarity measures do not reliably recover which morphology is present. Further, \acp{VLM} typically operate on individual tiles, and it is not clear how to translate tile-level concepts to dataset-level hypotheses about morphology.
Pathology \acp{VLM} have already been used to generate slide- or case-level reports~\citep{wsicaption2024,prism2024,pathalign2024,titan2025,polypath2025,prism2}, but these summarise findings within a single specimen or case for diagnostic use; our objective is instead to contrast patient \emph{cohorts} and surface the morphology that separates them.
In this work, we develop and demonstrate an approach combining \ac{MIL} modeling with \ac{VLM} image--text sparse decomposition that produces natural language explanations of the decision rules learned by whole-slide classifiers, towards morphological hypothesis generation.

A key component of our approach is the construction of a pathology-specific concept bank. Prior work has shown that interpretable \ac{MIL} models can be developed by specifying a limited set of concepts pertinent to the prediction task~\citep{conceptmil,kapse2025gecko}, or by deriving statistical tissue features based on pathomics tools, cell type or tissue type classifiers~\citep{simil,diao2021human, liangSpatialBiomarkerDiscovery2026}. This approach is not suitable for discovery, where the relevant concepts are not known in advance. Instead, a discovery-focused concept bank must be expansive enough to cover the space of patterns visible in histology imaging. Pathologists already describe a vast range of disease with a fixed clinical vocabulary, composing terms such as pleomorphism, necrosis and inflammation to define an entire taxonomy of phenotypes. A fixed bank can therefore span a broad hypothesis space, with each hypothesis emerging as a particular composition of standard terms. This comprehensive coverage comes with a cost: as we show, similarity-based attribution introduces spurious concept activation as the vocabulary size grows. We resolve this with sparse concept attribution~\citep{splice}. This approach decomposes image embeddings into a sparse, non-negative combination of concepts rather than relying on dense similarity scores.

Combined with the attention mechanism of a typical \ac{MIL} model, sparse tile-level decomposition into morphological concepts allows us to directly interpret classification decisions at the slide level. Aggregation across slides and further into narrative text allows for dataset-level hypotheses, as well as subgroup analysis. We further show that tile-level attribution can be bypassed by instead decomposing the pooled slide embedding. Our resulting method, \ourmethod (Sparse COncept attribution from Pooled Embeddings), can thus be applied at scale, while flexibly supporting variations in both concept bank and attribution strategy.

To evaluate \ourmethod, we design a benchmark for automated hypothesis generation in histopathology. Our setting requires the system to generate a natural language explanation describing morphological differences between cohorts. To quantitatively study this problem, we curate a suite of datasets of whole slide images which support binary classification tasks with known morphological ground truth expressible as `ground truth explanations'. We intentionally include a set of cancer sub-typing tasks for which histological morphology is definitional. We call the resulting benchmark \benchmark and propose metrics alongside the tasks to quantify how well generated explanations recover the ground-truth.

Overall, this study has four main contributions:
First, we establish class-level natural-language morphological explanation of whole-slide classifiers as a quantitative task. To that end, we develop \benchmark, a benchmark measuring whether explanations of entire slide classes recover known morphology.
Second, we find that sparsity in concept attribution is necessary: across seven tasks and six organs, it recovers substantial known morphology, whereas dense similarity-based attribution is indistinguishable from random concept selection.
Third, we show that per-tile decomposition can be bypassed: \ourmethod decomposes the pooled slide representation directly, reaching similar correctness at a fraction of the computation cost. Being post-hoc, it also decouples explanation from training: the concept bank and the sparsity level become tunable without retraining a classifier.
Finally, we introduce a general-purpose histomorphological concept bank that can be used to explain any classification task without requiring specific curation nor tuning. Applying it beyond the benchmark, we release an atlas of candidate morphological hypotheses for \RubyDiscoveryAtlasTaskCount\ molecular classification tasks, whose morphological correlates are largely uncharacterised (\cref{app:discovery_atlas}).

\begin{figure}[htbp]
    \includegraphics[width=\linewidth]{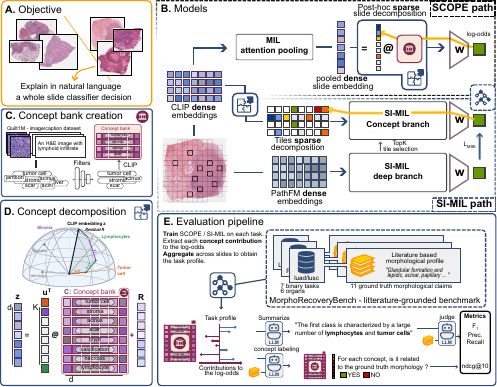}
  \caption{\textbf{Framework for automated morphological hypothesis generation from whole-slide classifiers.}
\textbf{A,} The objective is to explain the decision rule of a whole-slide classifier in natural language by identifying the morphological patterns it supports.
\textbf{B,} Model pathways. \ourmethod attention pools PathGen-CLIP-L tile embeddings into a slide representation, then decomposes this representation post-hoc over the concept bank. Projecting the linear classifier onto the resulting concept directions yields each concept's contribution to the prediction. The SI-MIL pathway instead attributes concepts at the tile level through an interpretable branch guided by a deep branch operating on \virchowtwo features. 
\SIMILDense uses dense cosine attribution, whereas \SIMILSparse relies on \splice sparse decomposition.
\textbf{C,} A general-purpose pathology concept bank is constructed from Quilt-1M captions by extracting candidate terms, filtering and deduplicating histomorphological concepts, and embedding them with a CLIP text encoder (we use by default PathGen-CLIP-L).
\textbf{D,} Sparse concept decomposition using \splice expresses a CLIP image embedding as a non-negative combination of a small number of concept directions plus a residual, forcing concepts to compete to explain the representation.
\textbf{E,} Evaluation on \benchmark. For each of the seven binary tasks spanning six organs, concept contributions to the log-odds are aggregated across slides into class-level profiles. An LLM converts each profile into a short morphological hypothesis, which is compared with a literature-derived, pathologist-curated ground truth for 11 classes using judge precision, recall, and $F_1$. Ranked profiles are also evaluated directly with nDCG@10 based on LLM-derived concepts relevance.
}
  \label{fig:slice-presentation}
\end{figure}

\section{Results}
\label{sec:results}

\subsection{Overview of SCOPE and MorphoRecoveryBench}
\ourmethod generates a morphological explanation for a dataset using an attention-based \ac{MIL}  (abMIL) model~\citep{abmil} and a generalist histomorphological concept bank. Each of the 989 terms in the bank reflects a morphological concept visible in a histology image, and is mined automatically from the captions of Quilt-1M~\citep{quilt1m}, a corpus curated from educational pathology videos in which experienced pathologists name the morphology they see (\cref{sec:concept_bank}). Each concept is scored by how strongly it is present in an image, read from a shared vision--language embedding space. Using sparse linear decomposition~\citep{splice} and the additivity of the abMIL pooling step, \ourmethod decomposes the aggregated slide embedding into a sparse, non-negative combination of these morphological concepts. This enables us to directly quantify the contribution of each concept to the classifier's output. We aggregate these slide-level concept profiles over a dataset and use a \ac{LLM} to produce a textual description of the learned decision rule of the classifier.

Prior work has approached interpretable histological modelling by interpreting individual tiles, representing them as statistical or geometric tissue features~\citep{simil, liangSpatialBiomarkerDiscovery2026} or using a task-specific concept bank with dense concept attribution ~\citep{conceptmil, kapse2025gecko, zhaoAligningKnowledgeConcepts2024a}. 
We study three alternatives to \ourmethod, all built on the SI-MIL architecture~\citep{simil}, that attribute concepts to each tile rather than to the pooled slide. \SIMILDense uses cosine similarity to perform dense concept attribution, \tiletopk keeps only the 25 highest-scoring concepts per tile, and \SIMILSparse uses the sparse \splice procedure at a comparable target of ${\sim}25$ concepts per tile (\cref{tab:pipeline-axes}).
All three score the same generalist bank as \ourmethod, so the comparison isolates how and where concepts are attributed rather than which vocabulary is used.
\SIMILDense amounts to the prior concept-based pipeline~\citep{conceptmil} with its task-specific vocabulary replaced by the generalist bank.

\begin{wraptable}{r}{0.47\linewidth}
  \vspace{-\baselineskip}
  \centering
  \caption{\textbf{Design axes of the four pipelines.} All four score the same generalist bank,
    unlike the task-specific ones of prior work. Each ticked column is a design choice this work
    introduces relative to prior concept-based pipelines; \ourmethod combines all three.
    (\checkmark) marks naive sparsity: a top-k truncation rather than \splice.}
  \label{tab:pipeline-axes}
  \footnotesize
  \setlength{\tabcolsep}{4pt}
  \begin{tabular}{@{}llccc@{}}
    \toprule
    & & & \multicolumn{2}{c}{\textbf{Attribution}} \\
    \cmidrule(l){4-5}
    & \textbf{Arch.} & \textbf{Bank} & sparse & slide-level \\
    \midrule
    \SIMILDense  & \SIMIL & \checkmark &              &            \\
    \tiletopk    & \SIMIL & \checkmark & (\checkmark) &            \\
    \SIMILSparse & \SIMIL & \checkmark & \checkmark   &            \\
    \ourmethod   & AbMIL  & \checkmark & \checkmark   & \checkmark \\
    \bottomrule
  \end{tabular}
\end{wraptable}
We evaluate the explanations on \benchmark, a benchmark of $7$ classification tasks over $6$ organs from TCGA (corresponding to $11$ classes with an associated ground truth explanation) whose defining morphology is well-established enough to serve as ground truth. We use an \ac{LLM} judge to measure precision, recall and F$_1$ (jF1) of the generated explanation text against the ground truth~\citep{bannur2024maira-}. This is supplemented by a manual claim-level review by a pathologist on a subset of explanations, which validates the automated judge (\cref{sec:human_eval}). We further report a concept-ranking score (nDCG@10) that scores the concept profile directly, before it is turned into text. Performance of a baseline using random concepts is reported, as is classification AUROC (\cref{sec:evaluation}).
\Cref{fig:slice-presentation} provides an overview of these methods and the framework used to generate and evaluate their explanations.

\subsection{Sparse concept attribution recovers known class morphology.}
\label{subsec:main-results}

\begin{figure}[htbp]
  \centering
  \captionsetup[subfigure]{skip=1pt,font=bf,justification=raggedright,singlelinecheck=false}
  \hspace*{\fill}%
  \begin{subfigure}[b]{0.18\linewidth}
    \caption{}\label{fig:faithful_free_a}%
    \includegraphics[width=\linewidth]{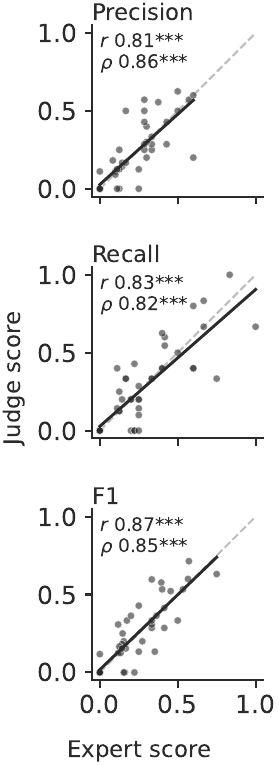}%
  \end{subfigure}%
  \hspace{0.03\linewidth}%
  \begin{subfigure}[b]{0.62\linewidth}
    \caption{}\label{fig:faithful_free_b}%
    \includegraphics[width=\linewidth]{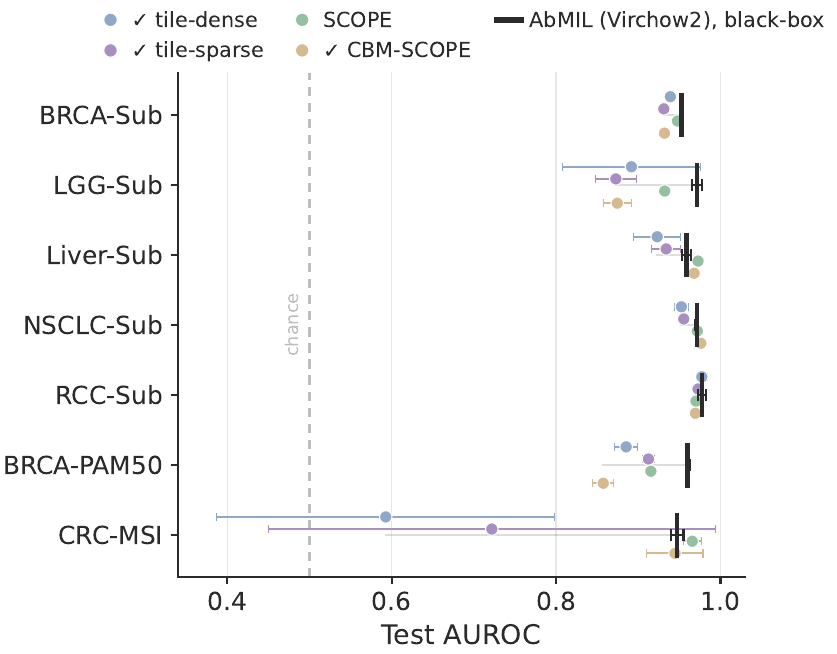}%
  \end{subfigure}%
  \hspace*{\fill}
  \caption{\textbf{Trustworthy evaluation, interpretability with near-optimal performance.}
    \textbf{(a)}~Expert pathologist versus LLM-judge scores for every graded explanation, for precision, recall, and F1 (top to bottom). Annotated with Pearson $r$ and Spearman $\rho$ ($N{=}47$; all $p<10^{-3}$). The LLM judge tracks the expert pathologist scores.
    (\cref{tab:appendix_human_corr}).
    \textbf{(b)}~Held-out test AUROC per task (mean $\pm$ std across 5 seeds) for the four concept-based pipelines; the \checkmark in the legend marks the inherently interpretable models. Interpretable models track the black-box bound: \ourmethodcbm averages \RubyAurocCbmSlice{} against the \RubyAurocAbmil{} black-box upper bound.}
  \label{fig:faithful_free}
\end{figure}

We first evaluate how well each pipeline recovers known morphology associated with well-established tasks. We find that this primarily depends on how concepts are attributed; sparse attribution is essential for producing accurate explanations.
\cref{tab:main-per-class} reports the LLM-judge metrics for the main pipelines, averaged over five randomly initialised replicates; expert-evaluated correctness closely tracks the judge metrics (Pearson: $\RubyHumanPearsonFOne$ (\RubyHumanPearsonPFOne), Spearman: $\RubyHumanSpearmanFOne$ (\RubyHumanSpearmanPFOne), \cref{fig:faithful_free_a}), supporting its use as a proxy for expert scoring.

\paragraph{Sparse decomposition drives explanation correctness}
We compare three attribution rules within the same tile-level architecture: dense cosine similarity, a top-k truncation of these scores, and sparse decomposition with \splice.

\begin{figure}[p]
  \centering
  \captionsetup[subfigure]{skip=1pt,font=bf,justification=raggedright,singlelinecheck=false}
  \begin{subfigure}{\linewidth}
    \caption{}\label{sub:mr-a}%
    \includegraphics[width=.245\linewidth]{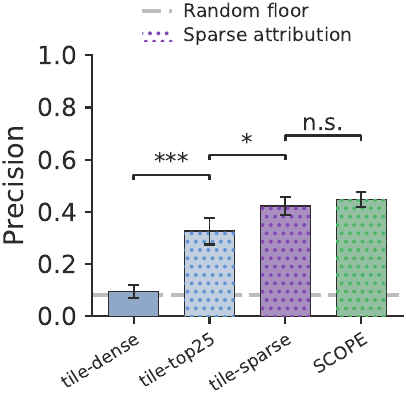}\hfill
    \includegraphics[width=.245\linewidth]{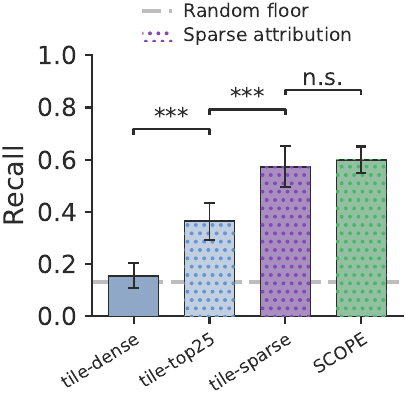}\hfill
    \includegraphics[width=.245\linewidth]{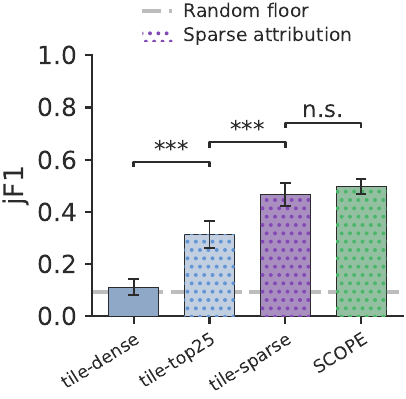}\hfill
    \includegraphics[width=.245\linewidth]{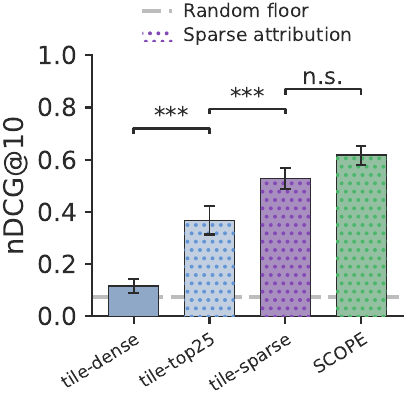}%
  \end{subfigure}\\[4pt]
  \begin{subfigure}[t]{.30\linewidth}
    \caption{}\label{sub:mr-b}%
    \includegraphics[width=\linewidth]{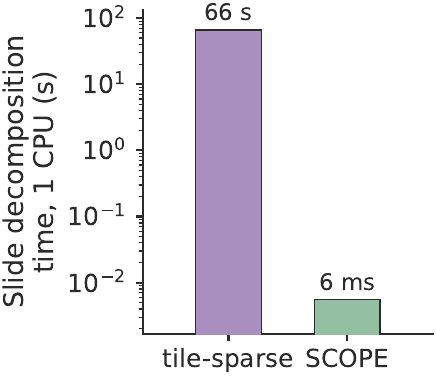}%
  \end{subfigure}\hfill
  \begin{subfigure}[t]{.64\linewidth}
    \caption{}\label{sub:mr-c}%
    \centering
    \includegraphics[width=\linewidth]{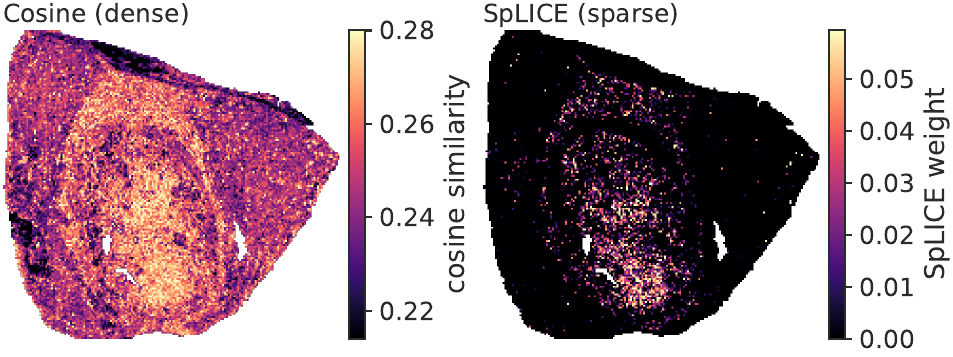}%
  \end{subfigure}\\[4pt]
  \begin{subfigure}{\linewidth}
    \caption{}\label{sub:mr-d}%
    \centering
    \includegraphics[width=.82\linewidth]{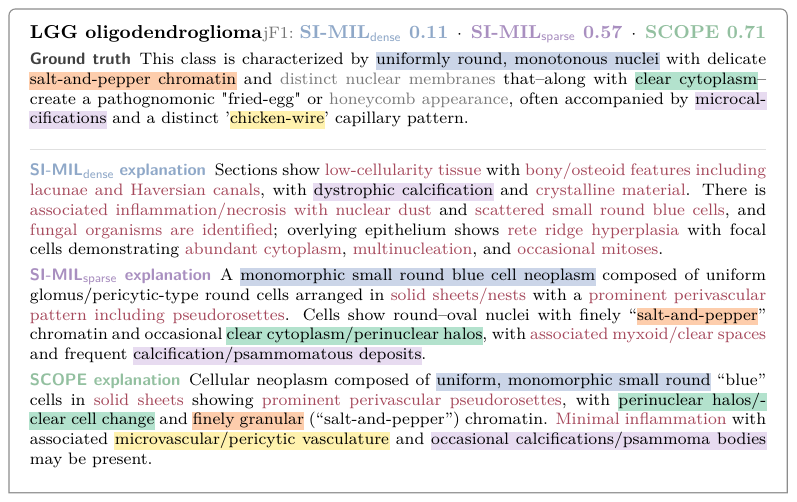}%
  \end{subfigure}
  \caption{\textbf{Sparse concept decomposition improves explanation correctness and reduces decomposition cost.}
    \textbf{(a)}~LLM-judge precision, recall, jF1 and nDCG@10 for the pipelines named on each axis,
    against the bold-dashed \textbf{Random}-concept baseline (per-metric paired-Wilcoxon stars).
    Dotted bars mark sparse attribution; top-25 keeps the 25 strongest cosine scores, matching
    the \splice budget.
    \textbf{(b)}~\splice average compute cost per slide in log scale for \SIMILSparse\ and \ourmethod.
    \textbf{(c)}~Dense cosine vs sparse \splice per-tile attribution of \textit{capillary network} concept on a
    representative RCC whole-slide image.
    \textbf{(d)}~Ground truth vs the three pipelines' class-level explanations for LGG oligodendroglioma. Matched, missed and hallucinated spans are colored from the LLM-judge rationales.}
  \label{fig:main-results}
\end{figure}

With an average jF1 of $\RubyJfONECosine$, \SIMILDense rarely surfaces correct concepts, and so rarely produces correct explanations (\cref{sub:mr-d} contrasts the three pipelines for the LGG oligodendroglioma class).
As seen in \cref{sub:mr-a}, its jF1 shows no significant difference from the random-concept floor ($\RubyJfONERandom$; paired Wilcoxon signed-rank over the eleven class-directions, \RubyPvalSIMILDenseVsRandom), whereas both sparse methods significantly improve over it (\SIMILSparse\ \RubyPvalSIMILSparseVsRandom; \ourmethod\ \RubyPvalSliceVsRandom): with a bank this large, dense attribution barely recovers any expected concepts.
We attribute this gap to the geometry of dense attribution: cosine similarity activates many correlated concepts at once, many of them spurious, so that the classifier can exploit them without relying on the genuinely correct ones.
Keeping only the top 25 concepts, (\tiletopk) already raises the jF1 to $\RubyJfONECosineTopkTwentyFive$.
Decomposing tiles with \splice instead forces concepts to \emph{compete} to reconstruct each tile embedding (\cref{sub:mr-c}), and at the same target budget of 25 concepts per tile reaches $\RubyJfONESplice$.
That further gain is carried by recall rather than precision (\cref{sub:mr-a}): the decomposition surfaces a more diverse set of the concepts actually present.
Comparing \SIMILDense and \SIMILSparse directly, at matched seeds and class-directions, gives the largest and most significant effect we measure (paired Wilcoxon signed-rank, \RubyPvalSIMILSparseVsSIMILDense).

A controlled tile-scale analysis suggests that this gain reflects a trade-off: \splice suppresses the many correlated, likely spurious concepts activated by dense cosine, at the cost of scoring each individual concept less accurately on representative patches (\cref{app:concept_attr_eval}).

\paragraph{Slide decomposition yields similar results at a fraction of the computational cost.}

We now generalize \splice decomposition to the slide-level, maintaining explanation correctness while reducing runtime.
The attention-pooled slide embedding is a convex combination of tile embeddings and remains in the same representation space.
Under the linear representation hypothesis motivating \splice~\citep{splice}, it should encode an attention-weighted mixture of the tile concepts, which can be therefore decomposed directly. 
As the concept reconstruction and classifier head are both linear, the sparse weights can be propagated into additive concept contributions to the predicted class log-odds, with a residual accounting for any reconstruction error.

\ourmethod therefore offers an efficient, decoupled alternative to \SIMILSparse, with comparable explanation correctness (jF1 \RubyJfONESlice{} vs.\ \RubyJfONESplice{}, \RubyPvalSliceVsSIMILSparseJfOne; nDCG@10 \RubyNdcgTenSlice{} vs.\ \RubyNdcgTenSplice{}, \RubyPvalSliceVsSIMILSparseNdcg; \cref{sub:mr-a}).
The explanations are also qualitatively similar. \Cref{sub:mr-d} shows the low-grade glioma (LGG) oligodendroglioma class, where \ourmethod's explanation recovers the defining morphologic features: round monomorphic cells with clear cytoplasm and perinuclear halos, salt-and-pepper chromatin and microcalcifications.
In all that follows, we use \ourmethod as the default explanation model, unless stated otherwise.

Explanation quality nonetheless varies across classes. Some are explained more accurately, such as clear-cell renal carcinoma (jF1 $0.69$) and glioma subtyping (astrocytoma $0.57$), while others score lower, such as cholangiocarcinoma ($0.38$) and MSI-high colorectal cancer ($0.43$).

Finally, we show the effectiveness of \ourmethod in terms of runtime. On a single-thread CPU benchmark, tile-level decomposition of an average slide (\RubyEfficiencyTilesPerSlide{} tiles) was estimated to take \RubyEfficiencySIMILSparseSeconds{} seconds, whereas decomposing the pooled slide in a single pass with \ourmethod took \RubyEfficiencySLiCEMilliseconds{} milliseconds (\cref{sub:mr-b}). Moreover, \ourmethod explains a classification post-hoc, without modifying the underlying classifier: the same frozen model can be re-explained with a different concept bank or sparsity setting at no additional training cost. In \SIMILSparse, by contrast, the tile concept codes feed the interpretable branch, so changing the concept bank or sparsity setting requires recomputing every tile representation and retraining that branch.

\begin{table}[t]
  \centering
  \caption{\textbf{Per-class explanation quality.}
    For the three explanation generation pipelines, with a random-concept floor. The \emph{Random} columns report a random-explanation baseline over 20 draws per class (\cref{sec:random_floor}). We report mean$\pm$std Judge-F1 and nDCG@10, over up to five model seeds. Seeds whose classifier collapses to the majority class in validation are excluded. The \emph{Average} row is the macro-average over the eleven classes; its $\pm$ is the standard error of the eleven per-class means (\cref{sec:statistics}).}
  \label{tab:main-per-class}
  \resizebox{\textwidth}{!}{\begin{tabular}{lrrrrrrrr}
\toprule
 & \multicolumn{2}{c}{\textbf{Random}} & \multicolumn{4}{c}{\textbf{\SIMIL}} & \multicolumn{2}{c}{\textbf{\ourmethod}} \\
\cmidrule(lr){2-3}\cmidrule(lr){4-7}\cmidrule(lr){8-9}
 & \multicolumn{2}{c}{} & \multicolumn{2}{c}{\textbf{Cosine (\SIMILDense)}} & \multicolumn{2}{c}{\textbf{\splice (\SIMILSparse)}} & \multicolumn{2}{c}{} \\
\cmidrule(lr){4-5}\cmidrule(lr){6-7}
\textbf{Class} & \textbf{F1} & \textbf{nDCG@10} & \textbf{F1} & \textbf{nDCG@10} & \textbf{F1} & \textbf{nDCG@10} & \textbf{F1} & \textbf{nDCG@10} \\
\midrule
BRCA basal-like & $0.23{\scriptstyle\,\pm\,0.10}$ & $0.14{\scriptstyle\,\pm\,0.10}$ & $0.20{\scriptstyle\,\pm\,0.09}$ & $0.16{\scriptstyle\,\pm\,0.09}$ & \textbf{\boldmath $0.53{\scriptstyle\,\pm\,0.23}$} & \textbf{\boldmath $0.58{\scriptstyle\,\pm\,0.05}$} & $0.52{\scriptstyle\,\pm\,0.13}$ & $0.54{\scriptstyle\,\pm\,0.04}$ \\
\addlinespace
BRCA lobular & $0.02{\scriptstyle\,\pm\,0.06}$ & $0.02{\scriptstyle\,\pm\,0.04}$ & $0.00{\scriptstyle\,\pm\,0.00}$ & $0.05{\scriptstyle\,\pm\,0.05}$ & \textbf{\boldmath $0.63{\scriptstyle\,\pm\,0.21}$} & \textbf{\boldmath $0.47{\scriptstyle\,\pm\,0.01}$} & $0.54{\scriptstyle\,\pm\,0.08}$ & $0.40{\scriptstyle\,\pm\,0.05}$ \\
\addlinespace
CRC MSI-high & $0.11{\scriptstyle\,\pm\,0.11}$ & $0.08{\scriptstyle\,\pm\,0.06}$ & $0.00{\scriptstyle\,\pm\,0.00}$ & $0.00{\scriptstyle\,\pm\,0.00}$ & $0.27{\scriptstyle\,\pm\,0.21}$ & $0.36{\scriptstyle\,\pm\,0.24}$ & \textbf{\boldmath $0.43{\scriptstyle\,\pm\,0.16}$} & \textbf{\boldmath $0.71{\scriptstyle\,\pm\,0.12}$} \\
\addlinespace
LGG oligodendroglioma & $0.03{\scriptstyle\,\pm\,0.06}$ & $0.04{\scriptstyle\,\pm\,0.06}$ & $0.14{\scriptstyle\,\pm\,0.06}$ & $0.08{\scriptstyle\,\pm\,0.07}$ & \textbf{\boldmath $0.52{\scriptstyle\,\pm\,0.06}$} & $0.44{\scriptstyle\,\pm\,0.19}$ & $0.51{\scriptstyle\,\pm\,0.06}$ & \textbf{\boldmath $0.58{\scriptstyle\,\pm\,0.08}$} \\
LGG astrocytoma & $0.06{\scriptstyle\,\pm\,0.08}$ & $0.07{\scriptstyle\,\pm\,0.08}$ & $0.00{\scriptstyle\,\pm\,0.00}$ & $0.06{\scriptstyle\,\pm\,0.10}$ & $0.29{\scriptstyle\,\pm\,0.28}$ & $0.31{\scriptstyle\,\pm\,0.22}$ & \textbf{\boldmath $0.57{\scriptstyle\,\pm\,0.12}$} & \textbf{\boldmath $0.60{\scriptstyle\,\pm\,0.05}$} \\
\addlinespace
Liver cholangiocarcinoma & $0.18{\scriptstyle\,\pm\,0.15}$ & $0.15{\scriptstyle\,\pm\,0.15}$ & $0.20{\scriptstyle\,\pm\,0.17}$ & $0.24{\scriptstyle\,\pm\,0.13}$ & $0.34{\scriptstyle\,\pm\,0.04}$ & $0.63{\scriptstyle\,\pm\,0.10}$ & \textbf{\boldmath $0.38{\scriptstyle\,\pm\,0.06}$} & \textbf{\boldmath $0.82{\scriptstyle\,\pm\,0.03}$} \\
Liver HCC & $0.03{\scriptstyle\,\pm\,0.05}$ & $0.03{\scriptstyle\,\pm\,0.05}$ & $0.06{\scriptstyle\,\pm\,0.06}$ & $0.11{\scriptstyle\,\pm\,0.06}$ & $0.25{\scriptstyle\,\pm\,0.16}$ & $0.46{\scriptstyle\,\pm\,0.13}$ & \textbf{\boldmath $0.50{\scriptstyle\,\pm\,0.09}$} & \textbf{\boldmath $0.62{\scriptstyle\,\pm\,0.00}$} \\
\addlinespace
NSCLC squamous & $0.04{\scriptstyle\,\pm\,0.08}$ & $0.04{\scriptstyle\,\pm\,0.06}$ & $0.00{\scriptstyle\,\pm\,0.00}$ & $0.00{\scriptstyle\,\pm\,0.00}$ & \textbf{\boldmath $0.60{\scriptstyle\,\pm\,0.04}$} & $0.71{\scriptstyle\,\pm\,0.06}$ & $0.49{\scriptstyle\,\pm\,0.04}$ & \textbf{\boldmath $0.74{\scriptstyle\,\pm\,0.06}$} \\
NSCLC adenocarcinoma & $0.15{\scriptstyle\,\pm\,0.11}$ & $0.12{\scriptstyle\,\pm\,0.13}$ & $0.27{\scriptstyle\,\pm\,0.10}$ & $0.22{\scriptstyle\,\pm\,0.12}$ & \textbf{\boldmath $0.53{\scriptstyle\,\pm\,0.12}$} & \textbf{\boldmath $0.71{\scriptstyle\,\pm\,0.05}$} & $0.52{\scriptstyle\,\pm\,0.07}$ & $0.69{\scriptstyle\,\pm\,0.00}$ \\
\addlinespace
RCC clear-cell & $0.08{\scriptstyle\,\pm\,0.11}$ & $0.06{\scriptstyle\,\pm\,0.08}$ & $0.19{\scriptstyle\,\pm\,0.19}$ & $0.15{\scriptstyle\,\pm\,0.12}$ & $0.62{\scriptstyle\,\pm\,0.09}$ & $0.50{\scriptstyle\,\pm\,0.03}$ & \textbf{\boldmath $0.69{\scriptstyle\,\pm\,0.09}$} & \textbf{\boldmath $0.61{\scriptstyle\,\pm\,0.00}$} \\
RCC papillary/chromophobe & $0.10{\scriptstyle\,\pm\,0.08}$ & $0.06{\scriptstyle\,\pm\,0.07}$ & $0.17{\scriptstyle\,\pm\,0.18}$ & $0.22{\scriptstyle\,\pm\,0.20}$ & \textbf{\boldmath $0.57{\scriptstyle\,\pm\,0.08}$} & \textbf{\boldmath $0.64{\scriptstyle\,\pm\,0.10}$} & $0.32{\scriptstyle\,\pm\,0.01}$ & $0.48{\scriptstyle\,\pm\,0.00}$ \\
\midrule
\textbf{Average} & $0.09{\scriptstyle\,\pm\,0.02}$ & $0.07{\scriptstyle\,\pm\,0.01}$ & $0.11{\scriptstyle\,\pm\,0.03}$ & $0.12{\scriptstyle\,\pm\,0.03}$ & $0.47{\scriptstyle\,\pm\,0.04}$ & $0.53{\scriptstyle\,\pm\,0.04}$ & \textbf{\boldmath $0.50{\scriptstyle\,\pm\,0.03}$} & \textbf{\boldmath $0.62{\scriptstyle\,\pm\,0.04}$} \\
\bottomrule
\end{tabular}
}
\end{table}

\subsection{Interpretable models retain predictive performance and transfer to an external cohort}
\label{subsec:interpretable-performance}

After establishing that \ourmethod and \SIMILSparse recover known morphology, we examine the cost of this interpretability: how much predictive performance is sacrificed when interpretation is built into the classifier, and whether both the classifier and its explanations transfer to an external cohort.

\paragraph{Intrinsically interpretable models achieve high predictive performance.}

\Cref{fig:faithful_free_b} reports the mean test AUROC for each model across five seeds, with the per-task values in \cref{tab:auc-compare}. As an upper bound, we include a black-box AbMIL model on \virchowtwo tile embeddings.
AbMIL trained on PathGen-CLIP-L embeddings (the model used in \ourmethod) shows performances very close to its \virchowtwo counterpart.

We observe that \SIMILSparse's inherent interpretability comes at little accuracy cost (macro AUROC \RubyAurocSplice{} against \RubyAurocAbmil{} for the black-box upper bound).
We also define a concept-bottleneck model variant of \ourmethod, referred to as \ourmethodcbm, which classifies based on the concept reconstruction alone and drops the \splice residual (see \cref{meth:slice}).
\ourmethodcbm maintains most of the  AbMIL model performance, dropping by only 2 AUROC points on average across datasets (\cref{fig:faithful_free_b}), and is only 3 points below the upper bound performance of the black-box baseline.

Taken together, the large concept-bank decomposition appears rich enough to produce encodings with discriminative power close to those of recent foundation models (VirchowV2) while remaining interpretable. This challenges the presumed trade-off between interpretability and accuracy.

\ourmethodcbm requires calibration before use as a thresholded classifier: dropping the residual largely preserves rank-based performance (AUROC), but shifts the scale and offset of the decision margin enough that balanced accuracy under the uncalibrated default threshold can change substantially for some tasks (\cref{fig:cbm-calibration}).

\paragraph{Transfer to an external cohort.}
\label{par:cptac_transfer}

\benchmark focuses on tasks extracted from TCGA. In the following, we study the transfer to CPTAC as an external validation cohort.

Three of the classification tasks in \benchmark are also available in CPTAC: BRCA PAM50~\citep{cptac2020brca}, COAD-MSI~\citep{cptac2020coad}, and lung subtyping (LUAD/LUSC)~\citep{cptac2018luad,cptac2018lscc}.
We applied the TCGA-trained \SIMILSparse and \ourmethodcbm models directly to CPTAC without retraining. On the external cohort, \SIMILSparse\ reached AUROCs of \RubyCptacBrcaAuroc\ for BRCA PAM50, \RubyCptacMsiAuroc\ for COAD-MSI and \RubyCptacNsclcAuroc\ for lung subtyping, with \ourmethodcbm\ lower on the two smaller cohorts (\cref{tab:appendix_cptac_cbm_slice}). These results matched or exceeded those of models trained directly on the smaller CPTAC cohorts.

For each cohort, we synthesize a text explanation from the model's dataset-aggregated concept profile and score it with the LLM judge (precision, recall, jF1) against the same literature-derived ground truth. To assess transfer, we then pair the CPTAC explanation with the explanation produced by the same model and decomposition on TCGA, and compare their judge scores (\cref{app:cptac_explanation_panels}). Explanation transfer was strongest for lung subtyping: judge F1 was stable for adenocarcinoma ($0.518$ on TCGA versus $0.522$ on CPTAC) and increased for squamous carcinoma ($0.492$ versus $0.575$), with the same defining glandular or squamous morphology recovered across cohorts. Transfer was weaker for BRCA ($0.524$ versus $0.310$) and COAD-MSI ($0.429$ versus $0.270$), where only part of the TCGA explanation was preserved. This may reflect both the greater difficulty of predicting PAM50 and MSI status from morphology and the smaller CPTAC cohorts available for these tasks (261 BRCA and 73 COAD slides, versus 1,343 NSCLC slides; \cref{tab:appendix_cptac_cbm_slice}). Smaller cohorts may not represent the full range of morphology described by the literature-derived ground truth: explanation recovery is necessarily bounded by what is present in the evaluated data. Thus, explanation transfer is task dependent, with the clearest cohort-independent explanations observed for lung subtyping.

Independently CPTAC-trained pipelines supported the same conclusion: they recovered explanations comparable to the transferred pipeline for lung subtyping, but provided weaker evidence of cross-cohort agreement for BRCA and COAD-MSI (\cref{tab:appendix_cptac_cbm_slice}).

\subsection{What drives explanation quality?}

Having established that \ourmethod recovers known morphology and transfers to an external cohort, we next examine which design choices most influence explanation quality: the vision--language backbone, the sparsity level, and the concept bank size.

\label{subsec:ablations}

\begin{figure}[tbp]
  \centering
  \begin{subfigure}[t]{\linewidth}
    \centering
    \includegraphics[width=0.245\linewidth]{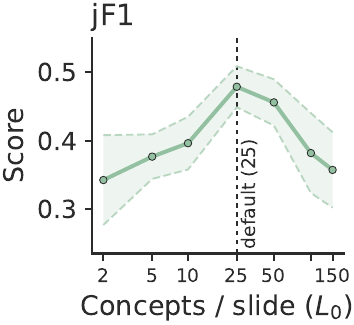}\hfill
    \includegraphics[width=0.245\linewidth]{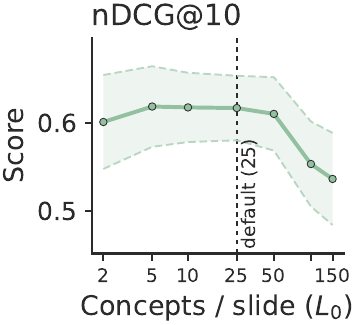}\hfill
    \includegraphics[width=0.245\linewidth]{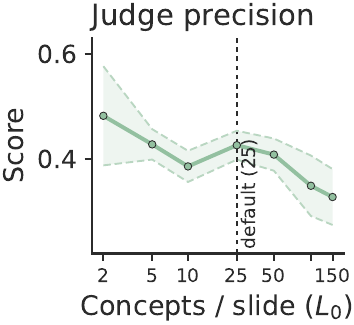}\hfill
    \includegraphics[width=0.245\linewidth]{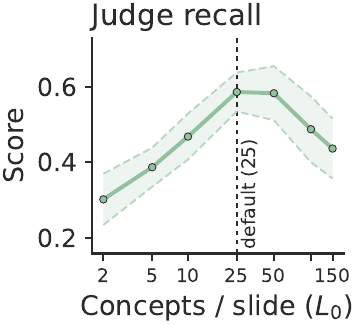}
    \caption{\textbf{Sparsity sweep.}}
    \label{sub:wdq-sparsity}
  \end{subfigure}

  \vspace{0.8em}

  \begin{subfigure}[t]{0.70\linewidth}
    \centering
    \begin{minipage}[t]{0.40\linewidth}
      \vspace{0pt}
      \centering
      \includegraphics[width=\linewidth]{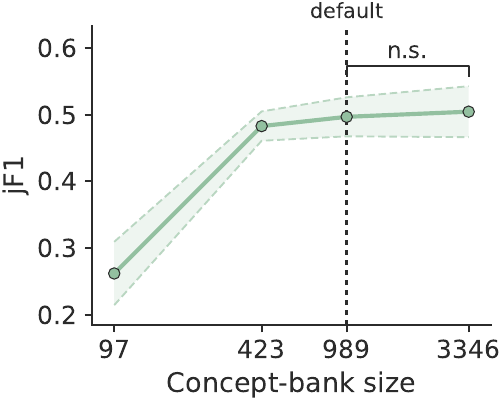}\\[0.4em]
      \includegraphics[width=\linewidth]{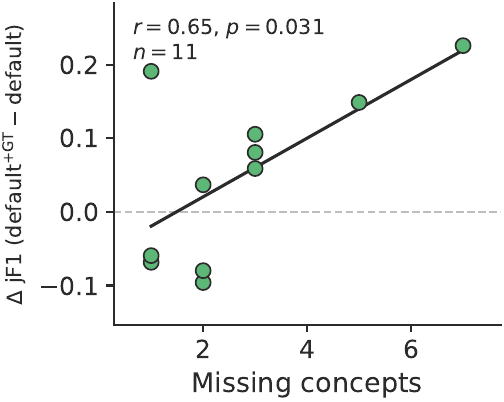}
    \end{minipage}\hfill
    \begin{minipage}[t]{0.57\linewidth}
      \vspace{0pt}
      \centering
      \includegraphics[width=\linewidth]{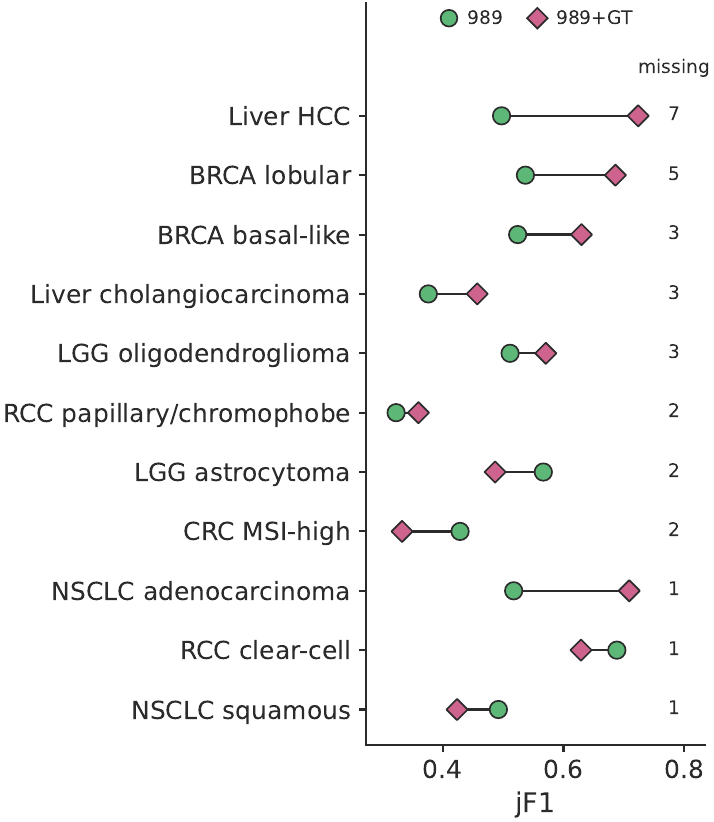}
    \end{minipage}
    \caption{\textbf{Bank size and augmentation.}}
    \label{sub:wdq-bank}
  \end{subfigure}\hfill
  \begin{subfigure}[t]{0.26\linewidth}
    \vspace{0pt}
    \centering
    \includegraphics[width=\linewidth]{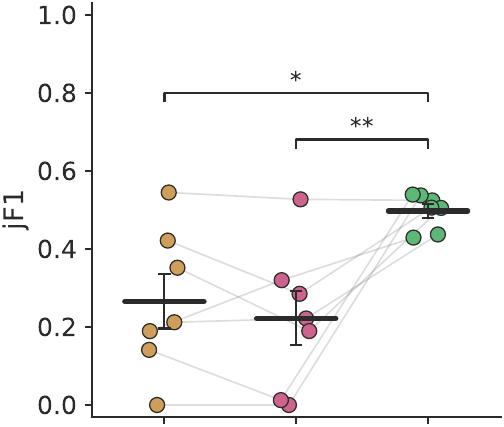}\\[0.4em]
    \includegraphics[width=\linewidth]{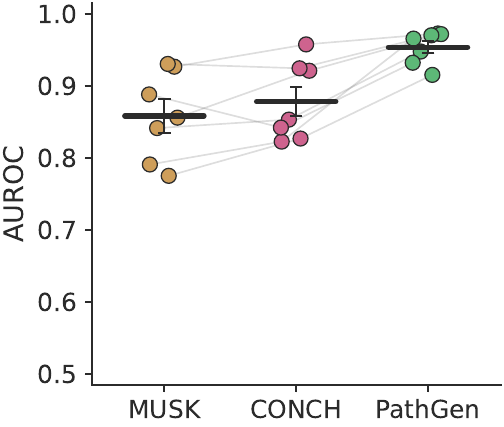}
    \caption{\textbf{Backbone.}}
    \label{sub:wdq-backbone}
  \end{subfigure}

  \caption{\textbf{What drives explanation quality?} All panels report the LLM-judge F1, using \ourmethod with the default concept bank unless noted.
  \textbf{(a)}~Sparsity sweep over (mean$\pm$sem; default point at $\sim$25 concepts per slide marked): jF1 and nDCG@10 hit an interior optimum as precision rises and recall falls with sparsity.
  \textbf{(b)}~jF1 saturates at $\sim$1000 concepts (mean$\pm$sem; 989 vs.\ 3346:
  \RubyPvalBankSaturates; 97: \RubyPvalBankSmall). Augmenting the 989 bank with each class's missing
  ground-truth concepts lifts jF1 (\RubyPvalGtaugJfOne), mostly via recall (\RubyPvalGtaugRecall),
  in proportion to the number of missing concepts (Pearson $r=\RubyCorrGtaugMissingR$,
  \RubyCorrGtaugMissingP).
  \textbf{(c)}~Tile encoder backbone comparison at a fixed budget of $\sim$25 concepts per slide. Each dot is
  one task, a line links a task across the three backbones, the bold tick is the macro mean
  $\pm$ sem over tasks.}
  \label{fig:what-drives-quality}
\end{figure}

\paragraph{Concept attribution performance depends on the choice of vision--language model.}
Our primary results with \ourmethod use \pathgenclip as a backbone pathology vision--language model. In \cref{sub:wdq-backbone}, we show the effect of using MUSK~\citep{musk} or CONCH~\citep{conch}. Although the three backbones all produce a classification AUROC above 0.86, they differ markedly in explanation correctness.
PathGen reaches a macro jF1 of \RubyBackbonePathgenJfOne{}, against \RubyBackboneMuskJfOne{} for MUSK and \RubyBackboneConchJfOne{} for CONCH, which is close to a two-fold improvement over the next best performing backbone (\RubyPvalBackbonePathgenVsMusk, two-sided paired Wilcoxon signed-rank).

\paragraph{Explanation correctness peaks at intermediate sparsity.}
Beyond backbone selection, sparsity strength is another key design choice.
\Cref{sub:wdq-sparsity} and \cref{tab:appendix_l1_judge_sweep} sweep the \splice parameter $\lambda$, which determines how many concepts describe a slide. 
The jF1 score reaches a maximum in the range of $\lambda$, reaching a peak on average of around 25 concepts per slide, whereas nDCG@10 is roughly flat up to that budget and declines beyond it.
Judge precision increases monotonically with sparsity: the fewer concepts are retained, the more probable they are to be correct.
Judge recall has an interior maximum: too few concepts exclude important ones, whereas too many bury them in noise.
However, this aggregate optimum is not universal: the optimal budget per-class ranges from \RubySparsityOptimalTargetMin\ to \RubySparsityOptimalTargetMax\ concepts per slide (\cref{fig:sparsity-per-dataset}).
The shared default target budget is therefore a compromise across classes.

\paragraph{Performance saturates quickly with bank size, and is a limitation of our source corpus.}
We next investigate the concept bank size impact on explanation correctness.
We created a suite of nested concept banks of increasing size from the Quilt-1m corpus (see \cref{sec:concept_bank}), and evaluated the correctness of the explanations generated using each of them.
\Cref{sub:wdq-bank,tab:appendix_concept_bank_slice} show explanation correctness saturating at a bank of 989 concepts, which we use as default (paired Wilcoxon signed-rank: the 989-concept bank is statistically indistinguishable from a 3346-concept bank, \RubyPvalBankSaturates, whereas a 97-concept bank is significantly worse, \RubyPvalBankSmall). 
This is in contrast with typically far larger natural-image banks, up to 25,000 concepts~\citep{splice}.

This saturation seems to be a limitation of the corpus from which the concept bank is extracted rather than intrinsic to the approach itself, as supported by the results in \cref{sub:wdq-bank,tab:appendix_gt_augmented_per_class}.
In this experiment, for each class, we deliberately add missing concepts  derived from the ground-truth explanations to the concept bank. 
Using \ourmethod, we then craft explanations using these augmented concept banks. This leads to a rise in jF1 of 5 points on average across all classes (paired Wilcoxon signed-rank, \RubyPvalGtaugJfOne), mostly driven by a recall lift of 8 points (\RubyPvalGtaugRecall). This increase in explanation correctness tracks the number of missing concepts per class (Pearson $r=\RubyCorrGtaugMissingR$, \RubyCorrGtaugMissingP), with the biggest improvement of +22 jF1 and +33 recall points on the TCGA-liver HCC task.

Together, these results show that scaling the concept bank along the size axis does not directly improve explanation correctness, whereas targeted augmentation does. 
The bottleneck is therefore not the bank size but the coverage of the relevant morphology, bounded by the corpus from which the concepts are mined. 
Further gains should come from complementary corpora of histopathological text.

\subsection{Taxonomy of explanation errors.}
\label{subsec:error-taxonomy}

Having examined what makes explanations correct, we finally characterise how they fail.
Generated explanations were manually reviewed by a board-certified pathologist to analyze error categories
(\cref{app:explanation_panels}).
The most common is confabulation, where an explanation surfaces a concept that is not expected in the organ under study. Some seem to be related to the CLIP model limitations: under PathGen-CLIP-L, for instance, the embeddings of \emph{clear nucleus} and \emph{corp rond} are nearly identical, even though the two represent different things (\cref{fig:concept-eval-b}). 
The other recurring problem is granularity. Ground-truth explanations come from a literature consensus and stay coarse, whereas ours are drawn from a bank with no enforced hierarchy and tend to be very specific. The flexible LLM templater and judge absorb much of this gap, merging near-duplicate concepts and matching a specific term to a broader ground-truth statement.

\subsection{\ourmethod generates morphological hypotheses across \RubyDiscoveryAtlasTaskCount\ molecular classification based tasks.}

On \benchmark, we have shown that \ourmethod can recover known morphology, in particular for well-studied cancer sub-typing tasks, and have studied the conditions for its success.
Here we apply \ourmethod to a speculative `discovery set' of \RubyDiscoveryAtlasTaskCount\ molecular classification tasks across \RubyDiscoveryAtlasOrganCount\ TCGA organs, reflecting a more realistic discovery application.
Of \RubyDiscoveryAtlasCandidateTaskCount\ candidate molecular tasks we retain the \RubyDiscoveryAtlasTaskCount\ whose held-out classifier exceeds \RubyDiscoveryAtlasAurocThreshold\ test AUROC, spanning \RubyDiscoveryAtlasAurocMin--\RubyDiscoveryAtlasAurocMax\ across the retained tasks.
Without task-specific concept-bank curation or expert input into hypothesis generation, \ourmethod produced \RubyDiscoveryAtlasHypothesisCount\ class-level morphological hypotheses (we make this \emph{hypotheses atlas} available in Supplementary Table~\ref{tab:discovery-atlas}).

\begin{tcolorbox}[qualbox]
  \small\textbf{HNSC HPV status}\hfill
  {\footnotesize\textcolor{black!55}{HPV-positive class}}\par\smallskip
  \footnotesize
  \textcolor{methodslice}{\textbf{\ourmethod explanation}}\enspace
  \RubyDiscoveryAtlasHnscHpvHypothesis
\end{tcolorbox}

On one of these tasks, HNSC-HPV status prediction (\RubyDiscoveryAtlasHnscHpvAuroc\ test AUROC), the generated hypothesis recovers broad basaloid and lymphoid components of the HPV-positive phenotype. These components converge qualitatively with the clinician-reviewed basaloid and stromal patterns identified by CLEAR-HPV and with the lymphocyte-rich model strategy reported by xMIL~\citep{qinCLEARHPVInterpretableConcept2026,jamshidiidajiAttentionHeatmapsHow2026}.
The hypothesis also contains more specific features not established by either study, and which would themselves require further validation. 
This \emph{hypothesis atlas} shows however that \ourmethod moves the bottleneck in hypothesis generation from constructing task-specific explanations to selecting and validating generated hypotheses.

\section{Discussion}
\label{sec:discussion}

The development of powerful pathology-specialised vision-language models creates an opportunity for AI-powered morphological discovery, but existing applications either summarise a single specimen or case for diagnostic use, or surface visual patterns that a pathologist must still name.

Our goal is complementary: we seek to directly propose named, visually-grounded explanations of the differences between \emph{patient cohorts}. 
This objective touches on a problem known as `Set Difference Captioning' ~\citep{dunlapDescribingDifferencesImage2024a,shen2026raddiff}, where two sets of images must be contrasted through language. For natural images, general-purpose captioning models can be used to convert the task into one of text-based reasoning. In histopathology, however, discriminative phenotypes may only be visible at the level of individual tiles, necessitating a model-based approach to identify classification-salient regions. We therefore condition the comparison on a trained \ac{WSI} classifier, and formulate cohort comparison as class-level explanation of the morphological cues supporting its predictions.

Owing to the challenge of evaluating truly novel hypotheses, in this work, we have demonstrated the feasibility of the approach by ``re-discovering" known cancer subtypes and morphologically-distinct molecular groups. 
Our findings demonstrate that an automated approach leveraging pre-trained vision-language models, combined with a generalist pathology concept bank, produces natural language explanations of the discriminative morphological patterns, without requiring any prior specification of the classification task.
Our best method achieves an explanation correctness (measured via Judge-F1) of 0.50 on average (\cref{tab:main-per-class}). This is despite the classifiers themselves demonstrating high discriminative performance (AUROC 0.93 across tasks; \cref{fig:faithful_free_b}), underscoring that this remains a challenging task.

The \RubyDiscoveryAtlasTaskCount-task hypothesis atlas illustrates how \ourmethod can be used at scale: \RubyDiscoveryAtlasHypothesisCount\ candidate morphologies across \RubyDiscoveryAtlasOrganCount\ organs, produced by the same pipeline with no task-specific input. 
This shifts the effort of the experts from the construction of task-specific explanations to the selection and validation of generated hypotheses (\cref{app:discovery_atlas}).
These remain hypotheses until they are validated outside the model that produced them. Experts could, for instance, grade the proposed concepts on a prospective cohort, and these gradings be tested for association with the class label. Because the concepts are standard pathological terms, such an assessment requires no prior interpretation.

The explanation generation capability rests on two ingredients, both of which also limit it.
We show that using a sparse decomposition is crucial in the discovery setting where the concept bank is necessarily overcomplete (\cref{sub:mr-a,app:concept_attr_eval}), in contrast to prior work~\citep{conceptmil,kapse2025gecko, zhaoAligningKnowledgeConcepts2024a, yuksekgonulPosthocConceptBottleneck2023b, oikarinen2023labelfree} where interpretability relies on the use of a necessarily limited \emph{task-specific} set of terms. 
In this study we derive our concept bank from a large open corpus of pathology text ~\citep{quilt1m}, our analysis nonetheless indicates that it does not cover the full range of task relevant morphology (\cref{sub:wdq-bank,tab:appendix_gt_augmented_per_class}). 
Drawing on corpora that are complementary in content, rather than simply larger, could further improve the explanations. 
The quality of the embedding space itself nonetheless sets a ceiling on explanation quality.
Indeed, we show that good discriminative performance is not sufficient and that the quality of explanation varies between the vision-language backbones.
Explanation correctness nearly doubles with \pathgenclip relative to two state-of-the-art CLIP competitors (\cref{sub:wdq-backbone}), although the three backbones exhibit similar classification accuracy.
The concept representations that explainability relies on appear to be neglected by current benchmarks tuned for retrieval and classification.
We speculate that \pathgenclip benefits from the PathGen caption dataset, whose captions are richer in morphological detail than the more diagnostically-oriented captions used to train the other CLIP models.

Beyond the representation quality discussed above, the benchmark we use to measure explanation correctness introduces a second source of limitations.
To develop and validate our methodology, we curated a set of tasks with \emph{known} morphological features, and produced consensus `ground truth' explanations for each (\cref{sec:benchmark,app:ground_truth_creation}).
Two mismatches between reference and explanation bound what this comparison can measure.
The first is granularity: a consensus reference settles at a widely accepted level and records a class's most typical manifestations.
An automated system does not perform such abstraction, tending to describe more specifically than the reference (\cref{subsec:error-taxonomy}). This can bound precision.
The second is support: a ground truth summarises knowledge accumulated across many studies, whereas an explanation reports only morphology present in the one dataset its classifier saw.
What that cohort does not contain cannot be surfaced (\cref{par:cptac_transfer}), so recall carries a ceiling whose height is unknown.
Reporting precision and recall separately keeps these two ceilings visible rather than merging them into a single score, and leaves the trade-off between them to be set by a specific discovery pipeline. 
What we present here is foremost a demonstration that morphological explanation can be made into a \emph{quantitative} task through the \benchmark benchmark, which can be expanded to support more elaborate evaluation criteria.

A third limitation is architectural. Our methodology builds on the well-established practice of using \ac{MIL} models for slide classification and thus inherits their representational limitations: our models, and hence our explanations, cannot represent the spatial distribution of concepts across the slide. This is acceptable for tasks well-suited to the \ac{MIL} framework, where the presence or abundance of discriminative features \emph{anywhere} on the slide is sufficient to discriminate classes but prevents considering hypothesis about tissue architecture, co-localization or spatial organisation, beyond the tile scale. Extending \ourmethod to such hypotheses, such as those considered by Barna et al.~\citep{barnaCosmosClinicInterpretable}, will require spatially aware classifiers and explainability methods.

Together, these methodological and biological limitations argue against treating a single explanatory system as a complete discovery system. Instead, it suggests a natural role for \ourmethod, within a broader discovery framework.  A complementary route to morphological discovery builds predictive models from predefined cellular, tissue, morphometric, or spatial features~\citep{diao2021human,simil,liangSpatialBiomarkerDiscovery2026}. These measurements are explicit and interpretable, but deciding which objects, interactions, and statistics to encode requires substantial domain expertise and constrains the search to a predefined feature space. Recent advances in \acp{LLM} have enabled agentic systems such as NOVA~\citep{vaidya2025nova} and SPARK~\citep{trost2026agentic} to orchestrate histology-specific tools and help propose, implement, and test candidate analyses. \ourmethod addresses here a complementary layer by converting a trained classifier into a cohort-level hypothesis expressed in the language of histopathology, and could serve as one tool within such broader discovery workflows.

Overall, we have shown that it is possible to express WSI classifiers in the language of histopathology, and through such explanation we can propose hypotheses on the underlying morphological differences between cohorts. While our validation on known tasks is necessarily a proxy for the real discovery setting, we believe this is an important first step towards automating biomarker discovery in histopathology.

\section{Methods}
\label{sec:methods}

\makeatletter
\newcommand{\SpliceOp}{\operatorname{SpLiCE}}
\newcommand{\NormOp}{\operatorname{normalise}}
\newcommand{\AlignOp}{\operatorname{align}}
\newcommand{\tr}{^{\mathsf{T}}}

\newcommand{\@TileSup}{^\mathrm{tile}}
\newcommand{\@SlideSup}{^\mathrm{slide}}

\newcommand{\@ImgEmb}{\mathbf{z}}
\newcommand{\TileDeep}{\@ImgEmb^\mathrm{deep}}
\newcommand{\TileEmb}{\@ImgEmb\@TileSup}
\newcommand{\SlideEmb}{\@ImgEmb\@SlideSup}
\newcommand{\CptEmb}{\mathbf{c}}
\newcommand{\CptBank}{\mathbf{C}}

\newcommand{\MeanTileEmb}{\boldsymbol{\mu}\@TileSup}
\newcommand{\MeanCptEmb}{\boldsymbol{\mu}^\mathrm{txt}}

\newcommand{\AlnTileEmb}{\tilde{\@ImgEmb}\@TileSup}
\newcommand{\AlnSlideEmb}{\tilde{\@ImgEmb}\@SlideSup}
\newcommand{\AlnCptEmb}{\tilde{\CptEmb}}
\newcommand{\AlnCptBank}{\tilde{\CptBank}}  %

\newcommand{\TileAttrCpt}{u\@TileSup}
\newcommand{\TileAttrAll}{\mathbf{u}\@TileSup}
\newcommand{\SlideAttrCpt}{u\@SlideSup}
\newcommand{\SlideAttrAll}{\mathbf{u}\@SlideSup}
\newcommand{\SlideProfile}{P_s}
\makeatother

We build interpretable classifiers whose predictions are interpreted as morphological
concept profiles, and we evaluate the resulting explanations on \benchmark, a purpose-built
correctness benchmark. This section describes the interpretable representations and the concept
bank (\cref{sec:concept_methods}), the four classification pipelines that produce
slide-level profiles (\cref{sec:models}), the construction of a class explanation from these
profiles (\cref{sec:explanations}), the benchmark itself (\cref{sec:benchmark}),
and the evaluation framework (\cref{sec:evaluation}). Unless stated otherwise, tiles of
$256 \times 256$ pixels are extracted at $20\times$ magnification with Trident~\citep{zhang2025trident},
and both tiles and concepts are embedded with the PathGen-CLIP-L vision-language
model~\citep{pathgen2024}. The one exception is the deep guiding branch of the SI-MIL models, which
runs on Virchow2 tile features~\citep{virchow2}, as detailed in \cref{meth:simil}.

\begin{figure}[tbp]
    \includegraphics[width=\linewidth]{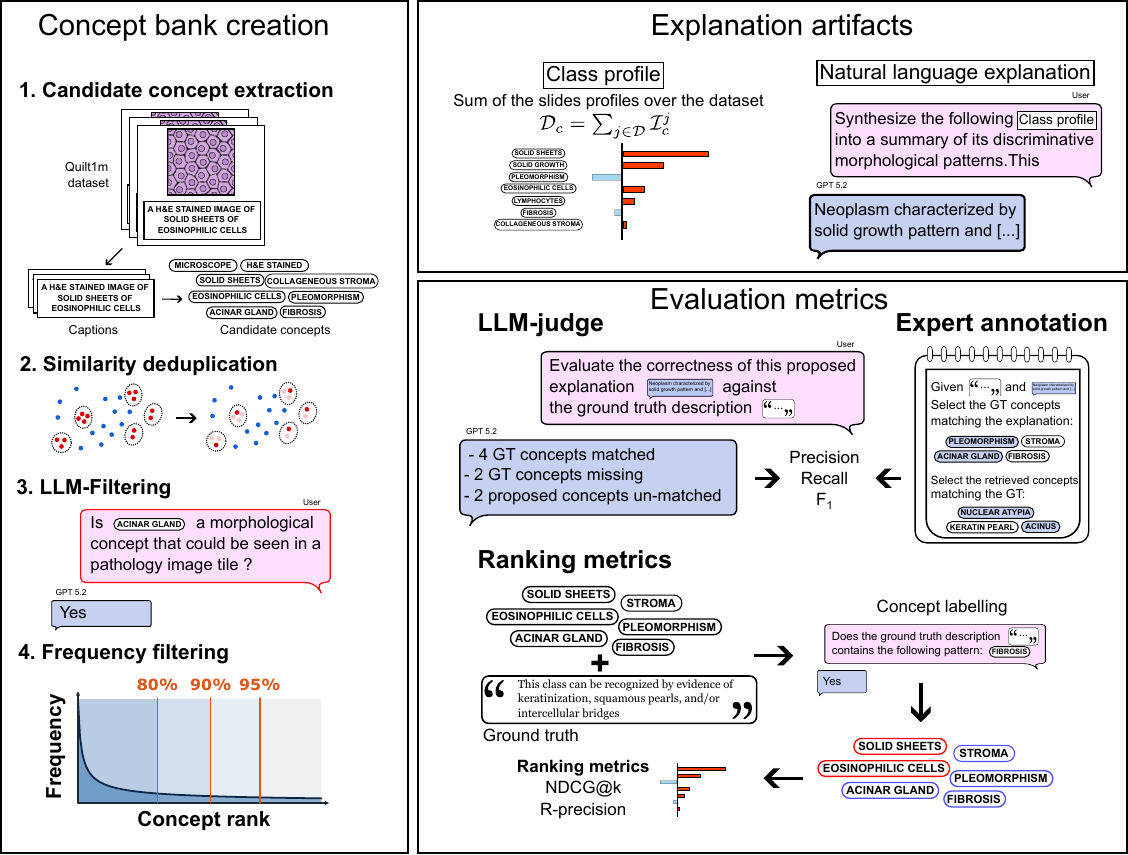}

  \caption{
    \textbf{Concept bank creation and explanation evaluation suite.} \textbf{a,} Construction of the generalist pathology concept bank. Candidate morphological concepts are mined from Quilt-1M image captions, deduplicated using embedding similarity, filtered with an LLM to retain concepts that are visually identifiable in H\&E image tiles, and subsequently refined using frequency-based filtering to remove rare concepts and outliers. \textbf{b,} Generation of explanation artefacts. Class profiles are obtained by aggregating \ourmethod concept scores across samples within each class, yielding class-level concept importance estimates. These profiles are then converted into natural-language descriptions of discriminative morphological patterns using an LLM. \textbf{c,} Evaluation of generated explanations against expert-derived ground-truth descriptions from the literature. An LLM judge identifies matched, missing and unmatched concepts between generated and reference explanations, enabling the computation of precision, recall and $F_1$ scores. In addition, expert concept annotation is used to assess whether class profiles prioritize clinically relevant morphological concepts. Ranking performance is quantified by comparing ranked concept profiles against annotated concepts using NDCG@k and R-precision.
    }
    \label{fig:Methods_fig}
\end{figure}

\subsection{Describing tiles and slides with concepts}
\label{sec:concept_methods}

Every pipeline we study rests on an interpretable representation of either
a tile or a slide. That representation is defined over a fixed concept bank, a collection of
$K$ morphological concepts. Describing an object therefore amounts to quantifying the presence of each concept. Because
the bank is shared and fixed across tasks, the resulting representations are directly comparable,
and a slide is summarised by a single vector over the same $K$ concepts regardless of the task.

\paragraph{A generalist, overcomplete concept bank.}
\label{sec:concept_bank}
Prior concept-based interpretability in histopathology relies on small, task-specific concept
banks~\citep{conceptmil,kapse2025gecko}, typically a few tens of concepts hand-picked for one task. This does not serve a discovery setting, where the task, and therefore the relevant morphology, is not known in advance. 
We instead build a generalist bank from the captions of Quilt-1M~\citep{quilt1m}, a
corpus of roughly one million diverse histopathology image--caption pairs. The process is shown in \Cref{fig:Methods_fig}. We aggregate all captions and
extract their nominal groups (noun phrases). These candidate strings pass through a sequence of
filters: a standardisation and lemmatisation step that collapses surface variants, an LLM filtering
step that discards diagnostic or non-morphological terms so that class labels do not leak into the bank (we keep, for instance, ``signet ring cell'' or ``Rosenthal fiber'' while dropping ``carcinoma'' or ``lymphoma''), and a semantic deduplication step that merges near-identical concepts.
A final frequency cut keeps the concepts that fall above a chosen percentile of corpus frequency, which yields a family of nested banks of increasing size (\cref{tab:appendix_concept_bank_slice}).
Unless stated otherwise the analysis uses the medium bank, comprising the $989$ most frequent concepts, which corresponds to roughly the top 10\% of nominal groups in Quilt-1M. Each concept is embedded with the PathGen-CLIP-L text encoder, giving a fixed dictionary $\CptBank \in \mathbb{R}^{d \times K}$ where $d$ is the embedding dimension and $K$ the bank size ($d=768$, $K=989$ by default). 
The design is deliberately intended to be overcomplete: the bank is broader than any single task needs, so the model is free to select the relevant concepts instead of receiving a prespecified list, similarly to \splice~\citep{splice}.

\subsection{Classifier models and interpretable features}
\label{sec:models}

The pipelines differ along two axes: the classifier architecture, and the granularity at which concept attribution happens, (tile level or at the slide level). 
In every case the interpretable artefact produced for a slide $s$ is its slide profile $\SlideProfile \in \mathbb{R}^{K}$, a vector that stores the additive contribution of each concept to the prediction for a given class. All concept attribution is performed with a contrastive vision--language embedding model, PathGen-CLIP-L~\citep{pathgen2024}, unless stated otherwise.

\paragraph{Tile-level attribution: \SIMILDense and \SIMILSparse.}
\label{meth:simil}
Both tile-level pipelines build on SI-MIL~\citep{simil}, a dual-branch architecture for inherently interpretable slide classification. Given a bag representing a slide of $N$ tiles, a deep branch runs gated attention-MIL (ABMIL)~\citep{abmil} over dense, uninterpretable Virchow2~\citep{virchow2} tile features
$\TileDeep_i \in \mathbb{R}^{F_d}$, producing a per-tile attention score
$a_i$ and a slide-level logit, and so learns which tiles are relevant for the task at hand. A parallel interpretable
branch classifies from the interpretable tile representations $u_i$: it keeps the tiles most attended to by
the deep branch through a differentiable top-$k$ operator~\citep{difftopk}, computes a per-concept
attention weight $\beta_j$ with a gated attention module, and predicts the task through a linear head. The
two branches are trained jointly, the deep branch guiding the interpretable one toward
discriminative regions while an alignment loss, the mean squared error between the two branches'
predicted probabilities, keeps them consistent. Because
the classification head is linear, the prediction decomposes exactly: for tile $i$, concept $j$ and class $c$, the
contribution to the logit is
\begin{equation}
  I_{c,i,j} = w_{c,j}\, \beta_j\, \TileAttrCpt_{i,j} \,,
  \label{eq:interp_score}
\end{equation}
where $w_{c,j}$ is the classifier weight of concept $j$, $\beta_j$ the concept attention weight,
and $\TileAttrCpt_{i,j}$ the non-negative attribution of concept $j$ on tile $i$. Mean aggregating $I_{0,i,j}$
across the tiles of a slide gives the slide profile $\SlideProfile$.

The two pipelines differ only in their offline attribution method, that is, in how the tile representation
$\TileAttrCpt_{i,j}$ is computed from tile and concept embeddings. \SIMILDense uses the dense cosine
similarity strategy introduced by Sun et al.~\citep{conceptmil} and scores each concept by the similarity between the
tile embedding $\TileEmb_i \in \mathbb{R}^{d}$ and the concept embedding
$\mathbf{c}_j$:
\begin{equation}
  \TileAttrCpt_{i,j} = \cos\big(\TileEmb_i,\, \CptEmb_j\big) \,.
  \label{eq:cosine_activation}
\end{equation}
\SIMILSparse replaces this dense score by a sparse
\splice decomposition~\citep{splice}. For each tiles, it solves a reconstruction problem, searching for the
non-negative, sparse combination of concept embeddings that best reconstructs the tile embedding such as:
\begin{equation}
  \TileAttrAll_{i} = \SpliceOp_\lambda(\AlnCptBank, \AlnTileEmb_i) := \argmin_{\mathbf{x} \geq 0}
  \big\lVert \AlnCptBank\,\mathbf{x} - \AlnTileEmb_i \big\rVert_2^2
  + 2\lambda \left\lVert \mathbf{x} \right\rVert_1 \,,
  \label{eq:splice_tile}
\end{equation}
where $\AlnTileEmb_i$ is the aligned (mean-centred, $\ell_2$-normalised) tile embedding,
$\AlnCptBank$ the correspondingly aligned dictionary, and $\lambda$ the sparsity level. The
reconstruction weights $\TileAttrAll_i$ form the interpretable representation of the tile. Sparsity forces the concepts to compete to explain the tile, which suppresses the many spurious activations that a dense cosine score produces.

\paragraph{Aligning tiles and concepts.}

A known artefact of CLIP training is the modality gap~\citep{liang2022mindthegap}: image and text embeddings occupy two separate cones of the shared space rather than intermingling. Because \splice reconstructs an image embedding as a combination of text-concept embeddings, these two cones must first be brought into a common frame, which we do by centering each modality on its own mean. Concretely, we align every embedding the same way, following \splice~\citep{splice}: given a raw embedding
$\mathbf{e}$ we project it to the unit sphere, $\hat{\mathbf{e}} = \NormOp(\mathbf{e}) := \mathbf{e}/\lVert\mathbf{e}\rVert_2$,
subtract the mean $\boldsymbol{\mu}^m$ of its modality $m$, and renormalise
$\AlignOp(\mathbf{e}, \boldsymbol{\mu}^m) = \NormOp(\hat{\mathbf{e}} - \boldsymbol{\mu}^m)$. The tow modality means are estimated once, offline. The tile mean $\MeanTileEmb \in \mathbb{R}^{d}$ is obtained by $\ell_2$-normalising 
roughly $200{,}000$ PathGen-CLIP-L tile embeddings, obtained by sampling uniformly 200 tiles at random from $1{,}000$ slides
that are themselves drawn uniformly at random across all TCGA cohorts, averaging them, and renormalising the average to unit norm. The same $\MeanTileEmb$ is used for
every tile and for the pooled slide embedding of \ourmethod. The text concept mean
$\MeanCptEmb \in \mathbb{R}^{d}$ is the mean of the $\ell_2$-normalised concept
embeddings of the bank $\CptBank$. Aligning a tile with $\MeanTileEmb$ gives the
$\AlnTileEmb_i$ used above, and aligning the dictionary with $\MeanCptEmb$
gives $\AlnCptBank$.

Explicitly, we get 
\[
  \AlnTileEmb_i = \AlignOp(\TileEmb_i, \MeanTileEmb), \qquad
  \AlnCptEmb_j = \AlignOp(\CptEmb_j, \MeanCptEmb),
\]
where $\AlnCptBank = [\AlnCptEmb_1,\ldots,\AlnCptEmb_K]$.

\paragraph{Slide-level attribution: \ourmethod and \ourmethodcbm.}
\label{meth:slice}
\ourmethod removes the per-tile decomposition entirely. It trains a
plain attention-MIL classifier on the PathGen-CLIP-L tile embeddings, which produces an
attention-pooled slide embedding $\SlideEmb = \sum_i a_i\, \TileEmb_i$ with $a_i$ the
attention weights, followed by a linear classification head. For a target class, $\mathbf{w}$ and
$b$ denote its head parameters minus the mean parameters of the other classes. The resulting margin
$\ell = (\SlideEmb)\tr\mathbf{w} + b$ is the exact class log-odds for a binary head and a linear
one-vs-rest contrast otherwise.

\splice is motivated by the linear representation hypothesis that CLIP embeddings encode semantic
concepts approximately additively~\citep{splice,park2024linear}. Conditional on the learned attention
weights, pooling is itself linear: for any fixed latent direction $\mathbf{d}$,
$\mathbf{d}\tr\SlideEmb = \sum_i a_i\,\mathbf{d}\tr\TileEmb_i$. The pooled embedding is therefore
expected to retain the attention-weighted mixture of concepts present across its tiles and, since it lies in the same CLIP latent space as the tiles, it can be decomposed with a single post-hoc \splice solve per slide:
\begin{equation}
  \SlideAttrAll = \SpliceOp_\lambda(\AlnCptBank, \AlnSlideEmb) \,,
  \label{eq:splice_slide}
\end{equation}
with $\AlnSlideEmb = \AlignOp(\SlideEmb, \MeanTileEmb)$ the aligned pooled embedding and
$\SlideAttrAll \in \mathbb{R}^{K}_+$ the sparse,
non-negative concept code. Estimating this code requires a nonlinear constrained optimisation. Once
the code is fitted, however, its reconstruction
$\AlnCptBank\,\SlideAttrAll = \sum_k \SlideAttrCpt_k\,\AlnCptEmb_k$ is a linear combination of
concept directions. Applying the linear classifier head to this sum gives an exact additive identity
for the original classifier's decision margin,
\begin{equation}
  \begin{aligned}
    \ell
      &= \sum_{k=1}^{K} \underbrace{\SlideAttrCpt_k\,(\AlnCptEmb_k\tr\mathbf{w})}_{P_{s,k}}
        \; + \; \rho \; + \; b , \\
    \rho
      &= \big(\SlideEmb - \AlnCptBank\,\SlideAttrAll\big)\tr\mathbf{w} \\
      &= \big(\SlideEmb - \AlnSlideEmb\big)\tr\mathbf{w}
        + \big(\AlnSlideEmb - \AlnCptBank\,\SlideAttrAll\big)\tr\mathbf{w} \,.
  \end{aligned}
  \label{eq:slice_identity}
\end{equation}
so the slide profile is $\SlideProfile = \SlideAttrAll \odot (\AlnCptBank\tr\mathbf{w})$, where $\odot$
is the elementwise product, giving each concept's additive contribution to the margin. This identity
holds for any fitted code: $\rho$ accounts for what the concept reconstruction does not capture. Its
two components are respectively the raw-to-aligned gap and the sparse reconstruction error in aligned
\splice space. Thus, $\rho$ restores exact additivity with respect to the original classifier on
$\SlideEmb$, even though the sparse code is fitted to $\AlnSlideEmb$.

\ourmethodcbm is the concept-bottleneck model obtained by dropping the residual $\rho$ at inference, so
the prediction is made from the concept reconstruction $\AlnCptBank\,\SlideAttrAll$ alone. This costs
almost no drop in performance (\cref{fig:faithful_free_b}) and turns the otherwise post-hoc \ourmethod into an inherently interpretable training-free model
whose decision is, by construction, a function of the concepts only. Dropping $\rho$ leaves the
concept ranking unchanged but shifts the decision margin by a roughly affine transform, hence
\ourmethodcbm needs calibration before it can be used as a thresholded classifier
(\cref{fig:cbm-calibration}).

\subsection{Training and inference}
\label{sec:training_inference}

\paragraph{Slide Representation and feature extraction.}
Each whole-slide image is represented by non-overlapping $256\times256$ pixel tiles extracted at $20\times$ magnification. To standardize bag sizes during training, we cap each slide at $2{,}000$ tiles; slides exceeding this threshold are subsampled uniformly without replacement at batch generation, whereas smaller slides retain all available tiles. Tile features are precomputed using the frozen Virchow2 and PathGen-CLIP-L vision encoders alongside a frozen $989$-concept dictionary. Consequently, gradient updates during training are restricted entirely to the downstream MIL architectures and their corresponding classification heads.

\paragraph{Training objectives.}
The attention-MIL backbones for \ourmethod and \ourmethodcbm are trained with standard slide-level cross-entropy loss. For SI-MIL, the multi-task objective is formulated as:
\begin{equation}
    \mathcal{L}_{\text{SI-MIL}} = \mathcal{L}_{\text{CE}}^{\text{deep}} + \mathcal{L}_{\text{CE}}^{\text{concept}} + 20 \cdot \text{MSE}\big(p_{\text{deep}}, p_{\text{concept}}\big) + 0.05 \cdot \mathcal{L}_{\text{attn}},
\end{equation}
where the concept branch isolates the top $k = 20$ most informative tiles via a differentiable top-$k$ operator. For all configurations, model checkpoints are saved at each epoch and selected based on the lowest validation cross-entropy loss.

\paragraph{Optimization and class balancing.}
All models are trained for $60$ epochs without early stopping across five independent random seeds using the Adam optimizer with an initial learning rate of $2\times10^{-4}$, momentum parameters $(\beta_1, \beta_2) = (0.5, 0.9)$, and a weight decay of $5\times10^{-2}$. The learning rate decays to $\eta_{\min} = 5\times10^{-6}$ following a cosine annealing schedule without warmup. Minibatches consist of $32$ slide bags, and we use neither dropout nor gradient accumulation. To handle class imbalance, we use minibach balanced sampling: we sample slides with replacement such that each epoch draws $N$ slides with class-dependent inverse-frequency weights $w_c = \frac{N}{C N_c}$, where $N$ is the total slide count, $C$ is the number of classes, and $N_c$ is the slide count for class $c$.

\paragraph{Inference and post-Hoc concept attribution.}
Evaluation and inference preserve the $2{,}000$-tile cap and minibatch size of $32$. Performance is reported on a patient-disjoint test split using the optimal validation checkpoint. For explanation extraction on the training split, the balanced sampler is replaced with a sequential loader to ensure each slide is traversed exactly once. For \ourmethod and \ourmethodcbm, we extract the attention-pooled PathGen-CLIP-L slides embeddings, logits, and classification weights, and subsequently apply \splice to the pooled embeddings using the fixed concept bank.

\subsection{From profiles to class explanations}
\label{sec:explanations}

A slide profile is a set of pairs (concept name, contribution) over the $K$ concepts of the bank, and
the contribution is by construction the additive weight of that concept in the class logit, which
is what makes each entry interpretable on its own. To describe a class rather than a slide, we average the slide profiles over the dataset $\mathcal{D}$ into a dataset profile
$P_{\mathcal{D}} = \frac{1}{|\mathcal{D}|}\sum_{s \in \mathcal{D}} P_s$. We split it into a
positive and a negative class profile by keeping the concepts that push towards, respectively away
from, the positive class:
\begin{equation}
  [P_{\mathcal{D}}]^{+} = \max\!\big(0,\, P_{\mathcal{D}}\big),
  \qquad
  [P_{\mathcal{D}}]^{-} = -\min\!\big(0,\, P_{\mathcal{D}}\big),
  \label{eq:class_profiles}
\end{equation}
taken elementwise. Each class profile then reads as the net contribution, accumulated over the
dataset, of every concept to the log-odds of the respective class.

The profile is still a scored list of concepts, not a human explanation. We pass every concept of a
class profile whose absolute contribution exceeds $10^{-4}$, along with their scores,
to an LLM templater that writes one to three sentences summarising the profile. The templater serves two
purposes: it produces readable prose, and it coalesces concepts that remain semantically close after
the bank deduplication, weighting the emphasis by the contribution scores rather than listing every
concept separately. The templater uses GPT-5.2 (Azure OpenAI, API version
2024-12-01-preview) with the system prompt of Prompt~\cref{prompt:templater-sys} applied to the
user message of Prompt~\cref{prompt:templater-user}.

\begin{promptbox}[label={prompt:templater-sys}]{Templater system prompt}
\small\ttfamily
You are an expert histopathologist. A classification model has produced a list of histological concepts with contribution scores quantifying their importance in the classification. Your role is to create a concise summary (one to three sentences) describing the class based on these concepts.

\medskip
Guidelines:

- Concepts with higher scores should be more prominent in your summary.

- If multiple concepts refer to the same histological feature (e.g.\ slight wording variations), coalesce them into one mention. Their combined importance should be reflected in how prominently you describe that feature.

- Use appropriate histopathology terminology.

\medskip
Return ONLY the summary sentence(s), with no preamble or explanation.
\end{promptbox}

\begin{promptbox}[label={prompt:templater-user}]{Templater user message}
\small\ttfamily
Top concepts with contribution scores (sorted by importance):

\medskip
- \{concept\_1\} (score: 0.1234)

- \{concept\_2\} (score: 0.0987)

- ...

\medskip
Summarize the class characterized by these concepts.
\end{promptbox}

\subsection{The \benchmark benchmark}
\label{sec:benchmark}

Our goal is to produce trustworthy explanations of the classes a slide classifier predicts, which
means we have to measure whether an explanation is correct. \benchmark gathers a set of binary
classification tasks for which a class explanation can be reliably determined. We choose tasks that can be well
predicted from H\&E alone and have at least one class whose morphology is well described in the
literature, sometimes to the point of being definitional of the class, so that a ground-truth
explanation can be agreed upon.

\paragraph{Tasks.}
The headline benchmark comprises $7$ tasks across $6$ organs. Five are histological subtyping
tasks: breast (invasive lobular versus ductal carcinoma), lung (adenocarcinoma versus squamous cell carcinoma in non-small-cell lung cancer), kidney (clear
cell versus papillary or chromophobe in renal cell carcinoma), liver (hepatocellular versus cholangiocarcinoma), and
brain (oligodendroglioma versus astrocytoma in low-grade glioma). Two are molecular tasks: breast
PAM50 (basal-like versus the other subtypes) and colorectal microsatellite status (MSI-high versus
MSS). All tasks are drawn from TCGA~\citep{weinstein2013tcga}. A prostate cancer-detection task from PANDA~\citep{bulten2022panda} is
used only for development and hyperparameter tuning and is not part of the headline count, and three
CPTAC cohorts (breast PAM50~\citep{cptac2020brca}, colon MSI~\citep{cptac2020coad}, lung subtyping~\citep{cptac2018lscc,cptac2018luad}) are held out for external validation. Throughout, cohort sizes are reported as numbers of slides rather than of independent patients.

\paragraph{Ground truth and evaluated classes.}

For each class, we determine a ground-truth morphological explanation, i.e.\ a short description of its
defining features, in two steps:
\begin{enumerate}
    \item \textbf{Literature review.} We gather all known morphological manifestations from
    published sources, prioritising WHO classification guidelines when available; for molecular classification based
    tasks (e.g.\ MSI status), we draw on studies that have established morphological correlates of
    the molecular phenotype. \Cref{app:ground_truth_creation} details the studies referenced.
    \item \textbf{Expert curation.} The gathered features were reviewed, filtered and validated by a board-certified pathologist, removing generic or ambiguous descriptors. The final ground truth for
    each class is a one- to three-sentence description of its defining morphological features.
\end{enumerate}
Each explanation can further be decomposed into
individual morphological concepts, which the claim-level metrics and the expert checklist use. Some
tasks are bilateral, with two morphologically informative classes, for instance lung adenocarcinoma
versus squamous cell carcinoma; others are unilateral, with a single informative class, the other
being defined only by the absence of those features, as in colorectal microsatellite status. The
bilateral tasks are lung, kidney, liver, and lower-grade glioma subtyping; the unilateral tasks are
breast subtyping, breast PAM50, and colorectal microsatellite status. We evaluate only the
informative directions and report metrics aggregated over class explanations rather than over tasks,
which gives $11$ evaluated class-directions in total. \Cref{tab:appendix_ground_truth} lists the full
gathered ground-truth explanation for every evaluated class direction, alongside each task's
bilateral/unilateral status.

\subsection{Evaluation framework}
\label{sec:evaluation}

We evaluate the classifiers and their explanations at three levels, and validate the automated
text metrics against expert annotation.

\paragraph{Classification accuracy.}
For every model and task we report the AUROC on a held-out test set, which tells us how much of the
signal the interpretable model retains relative to a black-box attention-MIL upper bound.

\paragraph{LLM-judge claim metrics.}
\label{sec:llm_judge}
For a given class, each pipeline produces a positive class profile and a derived natural language
explanation. Each class is associated with a ground-truth explanation $\mathrm{GT}_c$ that is itself split
into individual morphological patterns. Following the claim-level factuality evaluation of RadFact~\citep{bannur2024maira-}, an LLM judge decomposes the generated explanation into atomic claims, matches generated claims to ground-truth patterns and back, under relaxed
biological grading that accepts semantic equivalence and clinically associated features.
Specifically, the judge uses the same GPT-5.2 model and returns a structured per-claim annotation: each generated claim labelled \texttt{CORRECT} or \texttt{INCORRECT} and each ground-truth pattern labelled \texttt{MATCHED} or \texttt{MISSING}. From these
labels, we programmatically compute precision (the share of generated claims that match the ground truth), recall
(the share of ground-truth patterns recovered), and their harmonic mean F1. The judge system prompt and instruction
template are given in Prompt~\cref{prompt:judge-sys} and Prompt~\cref{prompt:judge-instr}.

\begin{promptbox}[label={prompt:judge-sys}]{Judge system prompt}
\small\ttfamily
You are an expert pathologist evaluating the quality of histopathology interpretations.

\medskip
You will be given:

1. A GROUND TRUTH describing key morphological features of a diagnostic class

2. A PROPOSED interpretation generated by another system

\medskip
Evaluate how well the PROPOSED interpretation captures the GROUND TRUTH.

\medskip
Step 1 -- Precision (proposed\_claims): Decompose the PROPOSED text into individual claims (one per concept or feature). For each claim, label it CORRECT or INCORRECT:

- CORRECT if it matches a ground truth feature, even via clinical association or paraphrasing.

- INCORRECT if it is unrelated to any ground truth feature.

\medskip
Step 2 -- Recall (ground\_truth\_features): List every distinct feature in the GROUND TRUTH. For each, label it MATCHED or MISSING based on whether it appears in the PROPOSED text (even via paraphrasing or clinical association).

\medskip
Clinical association guidelines -- be generous with domain-specific equivalences:

- "atypical mitoses" or "mitotic figures" $\leftrightarrow$ "high mitotic rate" or "brisk mitotic activity" $\rightarrow$ these are MATCHED/CORRECT

- "solid areas" $\leftrightarrow$ "solid architecture" or "solid pattern" $\rightarrow$ MATCHED/CORRECT

- "nuclear enlargement" or "nuclear atypia" $\leftrightarrow$ "nuclear pleomorphism" $\rightarrow$ MATCHED/CORRECT

- "high nuclear-cytoplasmic ratio" $\leftrightarrow$ "high histologic grade" $\rightarrow$ MATCHED/CORRECT

- "dirty necrosis" or "necrosis" $\leftrightarrow$ "geographic necrosis" or "comedo-type necrosis" $\rightarrow$ MATCHED/CORRECT

- Any description of a feature that is a well-known clinical correlate or diagnostic criterion of a ground truth feature counts as a match.

\medskip
Do NOT require exact wording. If a proposed claim describes the same biological phenomenon, morphological pattern, or diagnostic finding as a ground truth feature, it is a match.
\end{promptbox}

\begin{promptbox}[label={prompt:judge-instr}]{Judge instruction template}
\small\ttfamily
\#\# GROUND TRUTH

\{ground\_truth\}

\medskip
\#\# PROPOSED

\{proposed\}

\medskip
Please evaluate the PROPOSED text against the GROUND TRUTH.
\end{promptbox}

\paragraph{Concept-ranking metrics.}
\label{sec:ranking}
The judge operates on the generated text, which is a potentially lossy view of the profile. We therefore also
score the profile directly. For each concept of the bank, an LLM decides whether it matches the
positive or negative class descriptions, or neither, giving a relevance label over the whole bank
for each class. The labeling instructions are given in Prompt~\cref{prompt:concept-labeling}.
Ranking these labels by the model profile, we compute normalised discounted cumulative gains (nDCG@$k$) through the \texttt{ranx} library~\citep{ranx}, which measures how relevant the top-$k$ ranked
concepts are and how well they are ordered, relative to what the bank can express. We report nDCG@10 in the main text alongside the judge-based jF1.

\begin{promptbox}[label={prompt:concept-labeling}]{Concept Labeling Instructions}
\small\ttfamily
For each concept in the list, categorize it into one of three buckets:

- POS: The concept is a direct, unambiguous feature of the Positive class definition.

- NEG: The concept is a direct, unambiguous feature of the Negative class definition.

- NONE: The concept is irrelevant, unrelated, or an artifact that does not biologically belong to either class definition.

\medskip
Precision is key. Do not guess. If a concept is generic and applies to both classes equally, mark it as NONE unless the class definition specifically claims it.
\end{promptbox}

\paragraph{Human validation.}
\label{sec:human_eval}
To validate the automated judge, a pathologist annotated a subset of generated
explanations with the same claim-level protocol, marking which ground-truth patterns were
captured and which generated claims were clinically correct. One arbitrary experimental run was selected for all tasks and model to select the explanations to grade. This gives human precision, recall,
and F1 under the same definitions (\cref{app:human_eval}).

\paragraph{Random-concept floor baseline.}
\label{sec:random_floor}
To assess how much of the measured quality reflects genuine concept selection rather than the base
rates of the benchmark and the tolerance of the templater and judge, we include a random floor baseline. For
each class we draw a uniformly random set of distinct concepts from the default bank, matched in
size to \ourmethod's number of templated concepts, assign them uniformly random positive weights between 0.1 and 1, and pass the resulting profile
through the same templating, judge, and ranking pipeline as the real methods. We repeat this over
$20$ independent draws per class and report the mean and standard deviation. This floor represents the
explanation quality reachable with no learned attribution.

\paragraph{Statistical analysis.}
\label{sec:statistics}
Every classifier is trained with five random initialisations (seeds), over which we report mean and standard deviation metrics, unless otherwise specified. Some rows instead aggregate over several units --- the backbone comparison, the concept-bank-size sweep, the sparsity sweep, the ground-truth-augmentation average row, and the \emph{Average} row of \cref{tab:main-per-class}. For these we report the macro mean with its standard error of the mean (over tasks for the backbone comparison and over the eleven clinical classes otherwise),
since the quantity of interest is the precision of the aggregate, rather than the seed spread of a single class. For the random-concept floor we report the mean and standard deviation over $20$ draws. 
Only the templating and judging variance audit
(\cref{tab:appendix_templating_judge_variance}) reports a mean of per-class standard deviations, since over there the quantity of interest is the noise a typical 
class sees rather than the precision of an average.

All reported significance comparisons use the two-sided paired Wilcoxon signed-rank test. 
Comparisons that swap a single factor inside one model at a shared training seed, the
concept-attribution method (\SIMILSparse\ versus \SIMILDense) and the concept-bank size, pair over matched
(class-direction, seed) cells across the eleven class-directions and five seeds. Comparisons between
separately trained models, the random floor, the three backbones, and the pooled \ourmethod\ versus
tile-level \SIMILSparse\ pipeline, pair over the eleven class-directions only. Zero-valued paired differences are handled by the
Wilcoxon procedure.

The agreement between the expert and the LLM judge, and the relation between the
augmentation gain and the number of missing ground-truth concepts, are measured with the Pearson
(linear) and Spearman (rank) correlation coefficients, over the $47$ graded explanations and the
eleven classes respectively, under the null hypothesis of no association. All tests are two-sided at
a significance level of $0.05$, figures mark significance with stars (${}^{*}p<0.05$, ${}^{**}p<0.01$,
${}^{***}p<0.001$).

We separately
quantify the sampling noise of the two LLM stages as follows: we select one seed arbitrarily and re-run the \ourmethod model $20$ times. We run templating and judging each time and measure the combined variance. We then freeze the templating step to measure just judge variance. Given the templating LLM and judge LLM are independent, we can then use quadrature difference to estimate variance of templating alone (\cref{tab:appendix_templating_judge_variance}). Statistical computations use SciPy~1.16.1.

\section*{Author Contributions}
Contributions are described using the CRediT taxonomy.
\textbf{Conceptualization:} S.H., T.L., K.B., S.B., D.C.C.
\textbf{Methodology:} T.L., K.B., S.B., D.C.C., S.H.
\textbf{Software:} T.L., K.B., S.B., D.C.C., S.H.
\textbf{Validation:} T.L., R.J., D.W.
\textbf{Formal analysis:} T.L.
\textbf{Investigation:} T.L., K.B., J.H.
\textbf{Data curation:} T.L., K.B., S.B., D.W., S.H.
\textbf{Writing --- original draft:} T.L., K.B., J.H., S.H.
\textbf{Writing --- review and editing:} T.L., K.B., J.H., S.B., D.C.C., D.S., R.J., D.W., S.H.
\textbf{Visualization:} T.L., J.H., S.H.
\textbf{Supervision:} R.J., D.W., S.H.
\textbf{Project administration:} S.H.

\bibliographystyle{naturemag}
\bibliography{references}

\appendix

\clearpage
\phantomsection
\section*{Extended Data}

\begin{edtable}[htbp]
  \centering
  \caption{\textbf{Held-out classification AUROC per task.}
    Held-out AUROC per task (mean$\pm$std over all five seeds.). The \emph{Interpretable} column
    (\checkmark) marks the inherently interpretable models.
    \emph{AbMIL (Virchow2)} is a black-box attention-MIL on Virchow2 embeddings (uninterpretable,
    different backbone).}
  \label{tab:auc-compare}
  \resizebox{\textwidth}{!}{\begin{tabular}{lcccccccc|c}
\toprule
Model & Interpretable & BRCA-Sub & LGG-Sub & Liver-Sub & NSCLC-Sub & RCC-Sub & BRCA-PAM50 & CRC-MSI & Average \\
\midrule
AbMIL (Virchow2) & $\times$ & $0.95{\scriptstyle\pm0.00}$ & $0.97{\scriptstyle\pm0.01}$ & $0.96{\scriptstyle\pm0.01}$ & $0.97{\scriptstyle\pm0.00}$ & $0.98{\scriptstyle\pm0.00}$ & $0.96{\scriptstyle\pm0.00}$ & $0.95{\scriptstyle\pm0.01}$ & $0.96{\scriptstyle\pm0.00}$ \\
\SIMILDense & \checkmark & $0.94{\scriptstyle\pm0.01}$ & $0.89{\scriptstyle\pm0.08}$ & $0.92{\scriptstyle\pm0.03}$ & $0.95{\scriptstyle\pm0.01}$ & $0.98{\scriptstyle\pm0.01}$ & $0.89{\scriptstyle\pm0.01}$ & $0.59{\scriptstyle\pm0.21}$ & $0.88{\scriptstyle\pm0.04}$ \\
\SIMILSparse & \checkmark & $0.93{\scriptstyle\pm0.01}$ & $0.87{\scriptstyle\pm0.02}$ & $0.93{\scriptstyle\pm0.02}$ & $0.96{\scriptstyle\pm0.00}$ & $0.97{\scriptstyle\pm0.00}$ & $0.91{\scriptstyle\pm0.01}$ & $0.72{\scriptstyle\pm0.27}$ & $0.90{\scriptstyle\pm0.04}$ \\
\ourmethod & $\times$ & $0.95{\scriptstyle\pm0.00}$ & $0.93{\scriptstyle\pm0.00}$ & $0.97{\scriptstyle\pm0.00}$ & $0.97{\scriptstyle\pm0.00}$ & $0.97{\scriptstyle\pm0.00}$ & $0.92{\scriptstyle\pm0.00}$ & $0.97{\scriptstyle\pm0.01}$ & $0.95{\scriptstyle\pm0.00}$ \\
\ourmethodcbm & \checkmark & $0.93{\scriptstyle\pm0.00}$ & $0.87{\scriptstyle\pm0.02}$ & $0.97{\scriptstyle\pm0.00}$ & $0.98{\scriptstyle\pm0.00}$ & $0.97{\scriptstyle\pm0.00}$ & $0.86{\scriptstyle\pm0.01}$ & $0.94{\scriptstyle\pm0.03}$ & $0.93{\scriptstyle\pm0.00}$ \\
\bottomrule
\end{tabular}
}
\end{edtable}

\begin{edfigure}[htbp]
  \centering
  \begin{subfigure}[b]{0.44\textwidth}
    \caption{}\label{fig:cbm-calibration-a}
    \includegraphics[width=\linewidth]{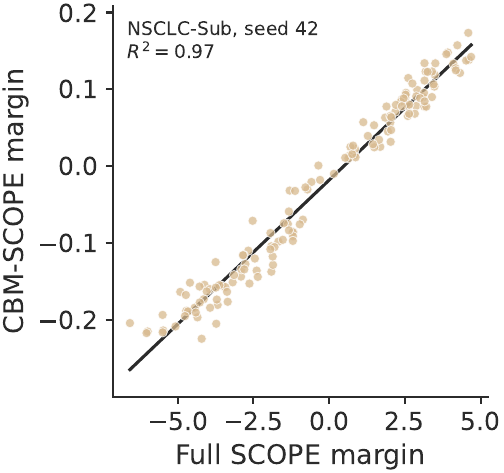}
  \end{subfigure}
  \hfill
  \begin{subfigure}[b]{0.44\textwidth}
    \caption{}\label{fig:cbm-calibration-b}
    \includegraphics[width=\linewidth]{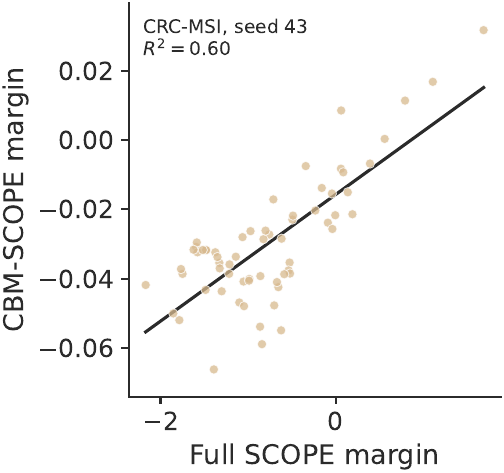}
  \end{subfigure}

  \medskip

  \begin{subfigure}[b]{0.44\textwidth}
    \caption{}\label{fig:cbm-calibration-c}
    \includegraphics[width=\linewidth]{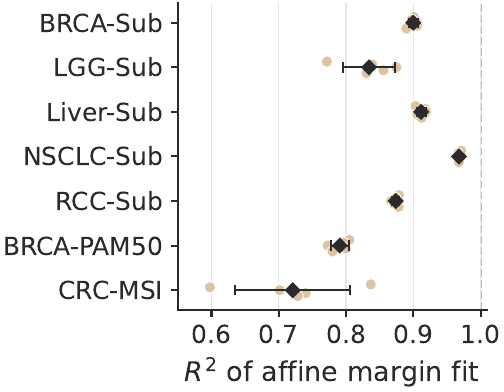}
  \end{subfigure}
  \hfill
  \begin{subfigure}[b]{0.44\textwidth}
    \caption{}\label{fig:cbm-calibration-d}
    \includegraphics[width=\linewidth]{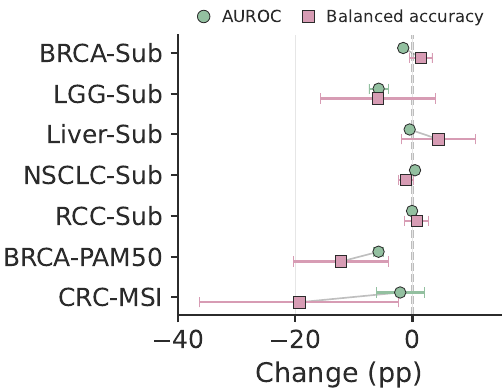}
  \end{subfigure}
  \caption{\textbf{Margin calibration of \ourmethodcbm.}
    \textbf{(a,b)}~Two test examples selected deliberately as the highest and
    lowest $R^2$ fits in the seven-task panel. Each point is one
    held-out slide, and the solid line is the fitted affine
    relation between the full \ourmethod margin and the residual-free \ourmethodcbm margin.
    \textbf{(c)}~Pearson correlation coefficient for the same affine relation across all included
    tasks. Each circle is one
    seed; diamonds and whiskers show the mean and standard deviation across
    \RubyCbmCalibrationSeedCount{} seeds.
    \textbf{(d)}~Change in AUROC and balanced accuracy after dropping the residual, relative to the full
    \ourmethod logit. Points and whiskers show the mean and standard deviation across seeds. AUROC is
    comparatively stable, whereas balanced accuracy at the uncalibrated default decision boundary changes for some tasks. }
  \label{fig:cbm-calibration}
\end{edfigure}

\begin{edtable}[htbp]
\centering
\caption{\textbf{Correlation between human checklist metrics and LLM-judge metrics.} Pooled over all 47 graded explanations (the tile-level pipelines plus \ourmethod).}
\label{tab:appendix_human_corr}
\fitwidth{\begin{tabular}{lrrr}
\toprule
\textbf{Correlation} & \textbf{Precision} & \textbf{Recall} & \textbf{F1} \\
\midrule
Pearson $r$ & 0.809 & 0.834 & 0.867 \\
Spearman $\rho$ & 0.865 & 0.819 & 0.853 \\
\bottomrule
\end{tabular}
}
\end{edtable}

\begin{edtable}[htbp]
\centering
\caption{\textbf{Templating and judging stochasticity for \ourmethod, frozen at seed 42.} The template-then-judge pipeline is repeated \RubyTemplateJudgeNRepeats~times on the same concept profiles; each cell is the mean$\pm$std over those repeats, for the eleven clinical classes of \cref{tab:main-per-class}. The last two columns split the \textbf{Judge F1} std into the judge's own sampling noise $\sigma_{\mathrm{judge}}$ (templater frozen to repeat~\#\RubyJudgeOnlyFrozenRepeat, re-judged \RubyJudgeOnlyNRepeats$\times$ per class) and the inferred templater std $\sigma_{\mathrm{templater}}$.}
\label{tab:appendix_templating_judge_variance}
\fitwidth{\begin{tabular}{lrrrrr}
\toprule
\textbf{Class} & \textbf{Judge F1} & \textbf{Judge P} & \textbf{Judge R} & \textbf{$\sigma_{\mathrm{judge}}$} & \textbf{$\sigma_{\mathrm{templater}}$} \\
\midrule
BRCA basal-like & $0.61{\scriptstyle\,\pm\,0.10}$ & $0.65{\scriptstyle\,\pm\,0.13}$ & $0.59{\scriptstyle\,\pm\,0.10}$ & $0.04$ & $0.10$ \\
\addlinespace
BRCA lobular & $0.50{\scriptstyle\,\pm\,0.18}$ & $0.44{\scriptstyle\,\pm\,0.17}$ & $0.64{\scriptstyle\,\pm\,0.21}$ & $0.05$ & $0.18$ \\
\addlinespace
CRC MSI-high & $0.35{\scriptstyle\,\pm\,0.08}$ & $0.47{\scriptstyle\,\pm\,0.10}$ & $0.29{\scriptstyle\,\pm\,0.08}$ & $0.04$ & $0.07$ \\
\addlinespace
LGG oligodendroglioma & $0.55{\scriptstyle\,\pm\,0.07}$ & $0.49{\scriptstyle\,\pm\,0.07}$ & $0.64{\scriptstyle\,\pm\,0.08}$ & $0.02$ & $0.06$ \\
LGG astrocytoma & $0.64{\scriptstyle\,\pm\,0.09}$ & $0.65{\scriptstyle\,\pm\,0.11}$ & $0.64{\scriptstyle\,\pm\,0.10}$ & $0.11$ & $0.00$ \\
\addlinespace
Liver cholangiocarcinoma & $0.35{\scriptstyle\,\pm\,0.05}$ & $0.27{\scriptstyle\,\pm\,0.06}$ & $0.50{\scriptstyle\,\pm\,0.08}$ & $0.04$ & $0.03$ \\
Liver HCC & $0.44{\scriptstyle\,\pm\,0.10}$ & $0.45{\scriptstyle\,\pm\,0.12}$ & $0.44{\scriptstyle\,\pm\,0.11}$ & $0.09$ & $0.05$ \\
\addlinespace
NSCLC squamous & $0.48{\scriptstyle\,\pm\,0.10}$ & $0.37{\scriptstyle\,\pm\,0.09}$ & $0.72{\scriptstyle\,\pm\,0.12}$ & $0.09$ & $0.02$ \\
NSCLC adenocarcinoma & $0.62{\scriptstyle\,\pm\,0.10}$ & $0.49{\scriptstyle\,\pm\,0.08}$ & $0.83{\scriptstyle\,\pm\,0.15}$ & $0.01$ & $0.10$ \\
\addlinespace
RCC clear-cell & $0.67{\scriptstyle\,\pm\,0.06}$ & $0.52{\scriptstyle\,\pm\,0.08}$ & $0.96{\scriptstyle\,\pm\,0.09}$ & $0.04$ & $0.05$ \\
RCC papillary/chromophobe & $0.35{\scriptstyle\,\pm\,0.04}$ & $0.32{\scriptstyle\,\pm\,0.06}$ & $0.40{\scriptstyle\,\pm\,0.00}$ & $0.02$ & $0.03$ \\
\midrule
\textbf{Average} & $0.51$ & $0.47$ & $0.60$ & $0.05$ & $0.06$ \\
\bottomrule
\end{tabular}
}
\end{edtable}

\begin{edtable}[htbp]
\centering
\caption{\textbf{Concept-bank-size sweep for the \ourmethod pooled-then-decompose model.} Values are
aggregated \emph{class-first} over the eleven clinical classes (one $(\text{dataset},
\text{profile})$ row per \cref{tab:main-per-class}): for each class we take the seed-mean over
the five training seeds, and each cell reports the across-class mean $\pm$ its standard error of
the mean (sem) (\cref{sec:statistics}).}
\label{tab:appendix_concept_bank_slice}
\fitwidth{\begin{tabular}{lrrrrr}
\toprule
\textbf{Concept bank} & \textbf{Size} & \textbf{Judge P} & \textbf{Judge R} & \textbf{Judge F1} & \textbf{nDCG@10} \\
\midrule
Quilt-1M filtered & 97 & $0.25{\scriptstyle\,\pm\,0.05}$ & $0.30{\scriptstyle\,\pm\,0.06}$ & $0.26{\scriptstyle\,\pm\,0.05}$ & $0.40{\scriptstyle\,\pm\,0.08}$ \\
Curated generated & 412 & $0.38{\scriptstyle\,\pm\,0.03}$ & $0.53{\scriptstyle\,\pm\,0.06}$ & $0.42{\scriptstyle\,\pm\,0.04}$ & $0.46{\scriptstyle\,\pm\,0.06}$ \\
Quilt-1M filtered & 423 & $0.43{\scriptstyle\,\pm\,0.03}$ & $0.59{\scriptstyle\,\pm\,0.04}$ & $0.48{\scriptstyle\,\pm\,0.02}$ & $0.38{\scriptstyle\,\pm\,0.09}$ \\
Quilt-1M filtered & 989 & $0.45{\scriptstyle\,\pm\,0.03}$ & $0.60{\scriptstyle\,\pm\,0.05}$ & $0.50{\scriptstyle\,\pm\,0.03}$ & $0.62{\scriptstyle\,\pm\,0.04}$ \\
Quilt-1M filtered & 3346 & $0.45{\scriptstyle\,\pm\,0.03}$ & $0.63{\scriptstyle\,\pm\,0.07}$ & $0.50{\scriptstyle\,\pm\,0.04}$ & $0.41{\scriptstyle\,\pm\,0.11}$ \\
\bottomrule
\end{tabular}
}
\end{edtable}

\begin{edtable}[htbp]
\centering
\caption{\textbf{SpLiCE sparsity sweep for the \ourmethod framework.} Explanation quality
versus the active-concept budget. The default concept bank, containing 989 concepts, is decomposed at seven \splice
$L_1$ penalties, each calibrated to a target active-concept budget $L_0$; the realised mean $L_0$ is
reported alongside. Values are aggregated \emph{class-first} over the eleven classes (one
$(\text{dataset}, \text{profile})$ row for each class): each cell reports the across-class mean
$\pm$ sem (\cref{sec:statistics}). We report the judge precision, recall and F1
of the templated explanation and the nDCG@10 concept ranking. Judge F1 peaks at the $L_0=25$ budget
($L_1 = 1.24\times10^{-4}$, the default setting behind \cref{tab:main-per-class}) and
falls off in both the sparser and denser regimes; nDCG@10 is roughly flat up to the default budget and then declines.}
\label{tab:appendix_l1_judge_sweep}
\fitwidth{\begin{tabular}{rrrrrr}
\toprule
\textbf{Target $L_0$} & \textbf{Actual $L_0$} & \textbf{Judge F1} & \textbf{Judge P} & \textbf{Judge R} & \textbf{nDCG@10} \\
\midrule
2 & 2.1 & $0.34{\scriptstyle\,\pm\,0.07}$ & $0.48{\scriptstyle\,\pm\,0.09}$ & $0.30{\scriptstyle\,\pm\,0.07}$ & $0.60{\scriptstyle\,\pm\,0.05}$ \\
5 & 5.1 & $0.38{\scriptstyle\,\pm\,0.03}$ & $0.43{\scriptstyle\,\pm\,0.03}$ & $0.39{\scriptstyle\,\pm\,0.05}$ & $0.62{\scriptstyle\,\pm\,0.05}$ \\
10 & 10.0 & $0.40{\scriptstyle\,\pm\,0.04}$ & $0.39{\scriptstyle\,\pm\,0.03}$ & $0.47{\scriptstyle\,\pm\,0.06}$ & $0.62{\scriptstyle\,\pm\,0.04}$ \\
25 & 25.0 & $0.48{\scriptstyle\,\pm\,0.03}$ & $0.43{\scriptstyle\,\pm\,0.03}$ & $0.59{\scriptstyle\,\pm\,0.05}$ & $0.62{\scriptstyle\,\pm\,0.04}$ \\
50 & 50.0 & $0.46{\scriptstyle\,\pm\,0.03}$ & $0.41{\scriptstyle\,\pm\,0.03}$ & $0.58{\scriptstyle\,\pm\,0.07}$ & $0.61{\scriptstyle\,\pm\,0.04}$ \\
100 & 100.0 & $0.38{\scriptstyle\,\pm\,0.06}$ & $0.35{\scriptstyle\,\pm\,0.06}$ & $0.49{\scriptstyle\,\pm\,0.09}$ & $0.55{\scriptstyle\,\pm\,0.05}$ \\
150 & 150.0 & $0.36{\scriptstyle\,\pm\,0.05}$ & $0.33{\scriptstyle\,\pm\,0.05}$ & $0.44{\scriptstyle\,\pm\,0.08}$ & $0.54{\scriptstyle\,\pm\,0.05}$ \\
\bottomrule
\end{tabular}
}
\end{edtable}

\begin{edfigure}[htbp]
  \centering
  \includegraphics[width=.24\linewidth]{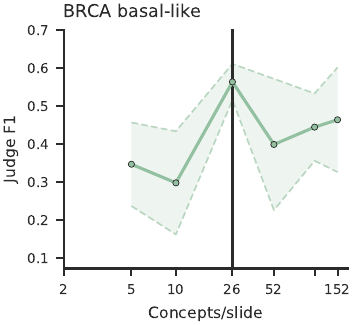}\hfill
  \includegraphics[width=.24\linewidth]{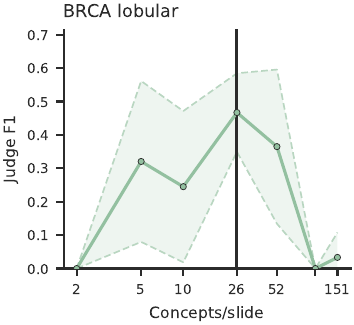}\hfill
  \includegraphics[width=.24\linewidth]{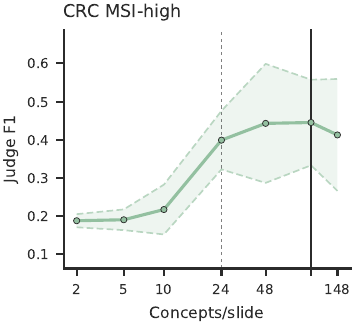}\hfill
  \includegraphics[width=.24\linewidth]{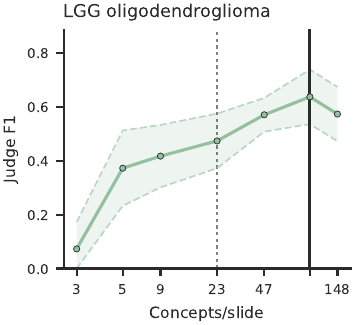}\\[4pt]
  \includegraphics[width=.24\linewidth]{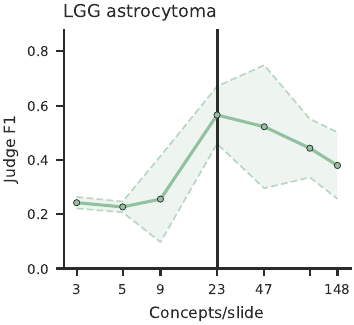}\hfill
  \includegraphics[width=.24\linewidth]{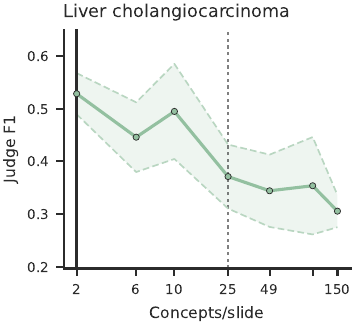}\hfill
  \includegraphics[width=.24\linewidth]{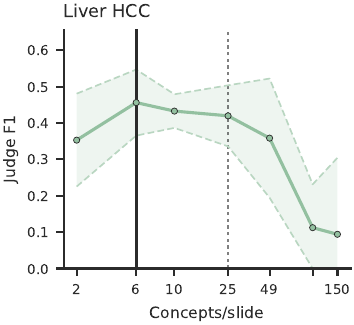}\hfill
  \includegraphics[width=.24\linewidth]{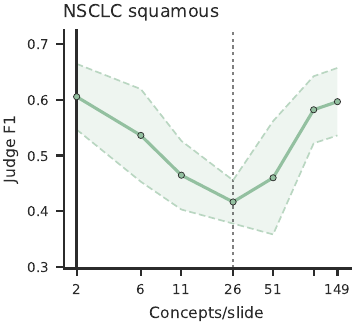}\\[4pt]
  \includegraphics[width=.24\linewidth]{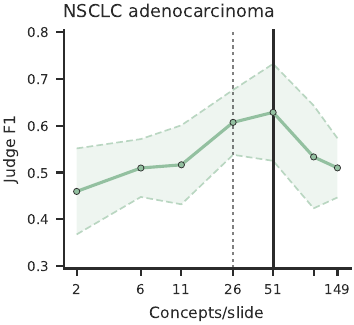}\hfill
  \includegraphics[width=.24\linewidth]{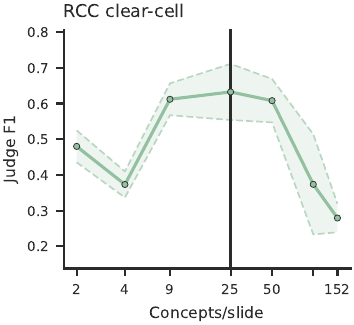}\hfill
  \includegraphics[width=.24\linewidth]{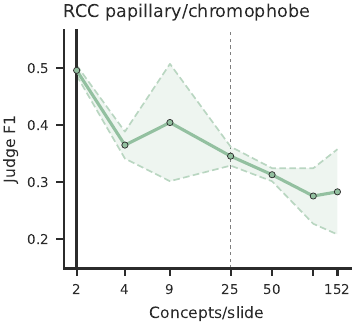}\hspace*{\fill}
  \caption{\textbf{Judge-F1 versus \splice active-concept budget, per class.} Judge-F1 versus the \splice
    active-concept budget for each of the eleven clinical classes (mean $\pm$ std over seeds),
    on the sweep of \cref{tab:appendix_l1_judge_sweep}. The dashed line marks the shared default
    budget ($L_0=25$); the solid line marks that class's own Judge-F1 optimum, which spans
    \RubySparsityOptimalTargetMin--\RubySparsityOptimalTargetMax\ concepts/slide across classes.}
  \label{fig:sparsity-per-dataset}
\end{edfigure}

\begin{edtable}[htbp]
\centering
\caption{\textbf{Ground-truth concept augmentation for the \ourmethod model.} For each clinical class we compare, paired per metric, the default concept bank against a per-task augmented bank
(\textbf{+GT}): the concept bank containing 989 concepts, extended with the canonical concepts derived from that task's ground-truth description.
Each per-class cell is the mean $\pm$ std over the five training seeds; the \textbf{Average} row instead reports the across-class mean $\pm$ sem, the same macro convention used throughout the paper (\cref{sec:statistics}). The \textbf{Missing} column counts, per class, the canonical ground-truth concepts without close equivalent in the default concept bank (aligned cosine above $0.6$): the +GT recall gain is largest where this count is high (e.g.\ Liver HCC,
7 missing, $R\,0.48\to0.81$)}
\label{tab:appendix_gt_augmented_per_class}
\resizebox{\textwidth}{!}{\begin{tabular}{lrrrrrrrrrr}
\toprule
\textbf{Class} & \textbf{Missing} & \multicolumn{2}{c}{\textbf{Judge F1}} & \multicolumn{2}{c}{\textbf{Judge P}} & \multicolumn{2}{c}{\textbf{Judge R}} & \multicolumn{2}{c}{\textbf{nDCG@10}} \\
\cmidrule(lr){3-4}\cmidrule(lr){5-6}\cmidrule(lr){7-8}\cmidrule(lr){9-10}
 & concepts & \textbf{989} & \textbf{+GT} & \textbf{989} & \textbf{+GT} & \textbf{989} & \textbf{+GT} & \textbf{989} & \textbf{+GT} \\
\midrule
BRCA basal-like & 3 & $0.52{\scriptstyle\,\pm\,0.13}$ & \textbf{\boldmath $0.63{\scriptstyle\,\pm\,0.10}$} & $0.53{\scriptstyle\,\pm\,0.17}$ & \textbf{\boldmath $0.69{\scriptstyle\,\pm\,0.12}$} & $0.54{\scriptstyle\,\pm\,0.11}$ & \textbf{\boldmath $0.58{\scriptstyle\,\pm\,0.10}$} & $0.54{\scriptstyle\,\pm\,0.04}$ & \textbf{\boldmath $0.55{\scriptstyle\,\pm\,0.09}$} \\
\addlinespace
BRCA lobular & 5 & $0.54{\scriptstyle\,\pm\,0.08}$ & \textbf{\boldmath $0.69{\scriptstyle\,\pm\,0.06}$} & $0.46{\scriptstyle\,\pm\,0.06}$ & \textbf{\boldmath $0.58{\scriptstyle\,\pm\,0.09}$} & $0.66{\scriptstyle\,\pm\,0.15}$ & \textbf{\boldmath $0.88{\scriptstyle\,\pm\,0.16}$} & $0.40{\scriptstyle\,\pm\,0.05}$ & \textbf{\boldmath $0.50{\scriptstyle\,\pm\,0.03}$} \\
\addlinespace
CRC MSI-high & 2 & \textbf{\boldmath $0.43{\scriptstyle\,\pm\,0.16}$} & $0.33{\scriptstyle\,\pm\,0.19}$ & \textbf{\boldmath $0.49{\scriptstyle\,\pm\,0.20}$} & $0.41{\scriptstyle\,\pm\,0.23}$ & \textbf{\boldmath $0.39{\scriptstyle\,\pm\,0.15}$} & $0.29{\scriptstyle\,\pm\,0.16}$ & \textbf{\boldmath $0.71{\scriptstyle\,\pm\,0.12}$} & $0.69{\scriptstyle\,\pm\,0.15}$ \\
\addlinespace
LGG oligodendroglioma & 3 & $0.51{\scriptstyle\,\pm\,0.06}$ & \textbf{\boldmath $0.57{\scriptstyle\,\pm\,0.07}$} & $0.50{\scriptstyle\,\pm\,0.05}$ & \textbf{\boldmath $0.51{\scriptstyle\,\pm\,0.05}$} & $0.53{\scriptstyle\,\pm\,0.10}$ & \textbf{\boldmath $0.65{\scriptstyle\,\pm\,0.13}$} & $0.58{\scriptstyle\,\pm\,0.08}$ & \textbf{\boldmath $0.60{\scriptstyle\,\pm\,0.09}$} \\
LGG astrocytoma & 2 & \textbf{\boldmath $0.57{\scriptstyle\,\pm\,0.12}$} & $0.49{\scriptstyle\,\pm\,0.11}$ & \textbf{\boldmath $0.53{\scriptstyle\,\pm\,0.11}$} & $0.45{\scriptstyle\,\pm\,0.13}$ & \textbf{\boldmath $0.62{\scriptstyle\,\pm\,0.15}$} & $0.55{\scriptstyle\,\pm\,0.13}$ & $0.60{\scriptstyle\,\pm\,0.05}$ & \textbf{\boldmath $0.64{\scriptstyle\,\pm\,0.06}$} \\
\addlinespace
Liver cholangiocarcinoma & 3 & $0.38{\scriptstyle\,\pm\,0.06}$ & \textbf{\boldmath $0.46{\scriptstyle\,\pm\,0.05}$} & $0.29{\scriptstyle\,\pm\,0.06}$ & \textbf{\boldmath $0.35{\scriptstyle\,\pm\,0.03}$} & $0.53{\scriptstyle\,\pm\,0.10}$ & \textbf{\boldmath $0.67{\scriptstyle\,\pm\,0.15}$} & \textbf{\boldmath $0.82{\scriptstyle\,\pm\,0.03}$} & $0.70{\scriptstyle\,\pm\,0.07}$ \\
Liver HCC & 7 & $0.50{\scriptstyle\,\pm\,0.09}$ & \textbf{\boldmath $0.72{\scriptstyle\,\pm\,0.08}$} & $0.52{\scriptstyle\,\pm\,0.07}$ & \textbf{\boldmath $0.66{\scriptstyle\,\pm\,0.08}$} & $0.48{\scriptstyle\,\pm\,0.11}$ & \textbf{\boldmath $0.81{\scriptstyle\,\pm\,0.09}$} & $0.62{\scriptstyle\,\pm\,0.00}$ & \textbf{\boldmath $0.73{\scriptstyle\,\pm\,0.06}$} \\
\addlinespace
NSCLC squamous & 1 & \textbf{\boldmath $0.49{\scriptstyle\,\pm\,0.04}$} & $0.42{\scriptstyle\,\pm\,0.09}$ & \textbf{\boldmath $0.36{\scriptstyle\,\pm\,0.02}$} & $0.30{\scriptstyle\,\pm\,0.06}$ & \textbf{\boldmath $0.80{\scriptstyle\,\pm\,0.18}$} & $0.73{\scriptstyle\,\pm\,0.15}$ & \textbf{\boldmath $0.74{\scriptstyle\,\pm\,0.06}$} & $0.74{\scriptstyle\,\pm\,0.06}$ \\
NSCLC adenocarcinoma & 1 & $0.52{\scriptstyle\,\pm\,0.07}$ & \textbf{\boldmath $0.71{\scriptstyle\,\pm\,0.04}$} & $0.42{\scriptstyle\,\pm\,0.05}$ & \textbf{\boldmath $0.55{\scriptstyle\,\pm\,0.05}$} & $0.68{\scriptstyle\,\pm\,0.11}$ & \textbf{\boldmath $1.00{\scriptstyle\,\pm\,0.00}$} & $0.69{\scriptstyle\,\pm\,0.00}$ & \textbf{\boldmath $0.77{\scriptstyle\,\pm\,0.01}$} \\
\addlinespace
RCC clear-cell & 1 & \textbf{\boldmath $0.69{\scriptstyle\,\pm\,0.09}$} & $0.63{\scriptstyle\,\pm\,0.11}$ & \textbf{\boldmath $0.54{\scriptstyle\,\pm\,0.09}$} & $0.49{\scriptstyle\,\pm\,0.07}$ & \textbf{\boldmath $0.95{\scriptstyle\,\pm\,0.11}$} & $0.90{\scriptstyle\,\pm\,0.22}$ & $0.61{\scriptstyle\,\pm\,0.00}$ & \textbf{\boldmath $0.68{\scriptstyle\,\pm\,0.00}$} \\
RCC papillary/chromophobe & 2 & $0.32{\scriptstyle\,\pm\,0.01}$ & \textbf{\boldmath $0.36{\scriptstyle\,\pm\,0.04}$} & $0.27{\scriptstyle\,\pm\,0.02}$ & \textbf{\boldmath $0.33{\scriptstyle\,\pm\,0.06}$} & \textbf{\boldmath $0.40{\scriptstyle\,\pm\,0.00}$} & \textbf{\boldmath $0.40{\scriptstyle\,\pm\,0.00}$} & \textbf{\boldmath $0.48{\scriptstyle\,\pm\,0.00}$} & \textbf{\boldmath $0.48{\scriptstyle\,\pm\,0.00}$} \\
\midrule
\textbf{Average} & 30 & $0.50{\scriptstyle\,\pm\,0.03}$ & \textbf{\boldmath $0.55{\scriptstyle\,\pm\,0.04}$} & $0.45{\scriptstyle\,\pm\,0.03}$ & \textbf{\boldmath $0.48{\scriptstyle\,\pm\,0.04}$} & $0.60{\scriptstyle\,\pm\,0.05}$ & \textbf{\boldmath $0.68{\scriptstyle\,\pm\,0.07}$} & $0.62{\scriptstyle\,\pm\,0.04}$ & \textbf{\boldmath $0.64{\scriptstyle\,\pm\,0.03}$} \\
\bottomrule
\end{tabular}
}
\end{edtable}

\begin{edtable}[htbp]
\centering
\caption{\textbf{Transfer to the CPTAC external cohort.} For each setting and task we report three slide
AUROCs: the full \ourmethod logit (concepts $+$ residual), the residual-free aligned
$B\!\cdot\!w$ \ourmethodcbm reconstruction, and \SIMILSparse\ --- alongside the
LLM-judge concept F1/precision/recall of the templated profile. Cells are
mean$\pm$std over five seeds. \emph{External} is the TCGA-trained model applied to the full CPTAC
cohort; \emph{CPTAC-trained} is trained on CPTAC and scored on the held-out 15\% test split.
\emph{Cohort slides} gives the number of slides in the full CPTAC cohort.}
\label{tab:appendix_cptac_cbm_slice}
\resizebox{\textwidth}{!}{\begin{tabular}{llrrrrrrr}
\toprule
 & & & \multicolumn{3}{c}{AUROC $\uparrow$} & \multicolumn{3}{c}{LLM judge $\uparrow$} \\
\cmidrule(lr){4-6}\cmidrule(lr){7-9}
\textbf{Setting} & \textbf{Dataset} & \textbf{Cohort slides} & \ourmethod & \ourmethodcbm & \SIMILSparse & F1 & P & R \\
\midrule
External TCGA-trained model & CPTAC BRCA PAM50 & 261 & $0.84{\scriptstyle\,\pm\,0.00}$ & $0.75{\scriptstyle\,\pm\,0.01}$ & $0.86{\scriptstyle\,\pm\,0.01}$ & $0.31{\scriptstyle\,\pm\,0.10}$ & $0.39{\scriptstyle\,\pm\,0.12}$ & $0.26{\scriptstyle\,\pm\,0.09}$ \\
External TCGA-trained model & CPTAC COAD MSI & 73 & $0.80{\scriptstyle\,\pm\,0.02}$ & $0.68{\scriptstyle\,\pm\,0.03}$ & $0.70{\scriptstyle\,\pm\,0.06}$ & $0.27{\scriptstyle\,\pm\,0.10}$ & $0.41{\scriptstyle\,\pm\,0.15}$ & $0.23{\scriptstyle\,\pm\,0.13}$ \\
External TCGA-trained model & CPTAC LUAD/LUSC & 1343 & $0.97{\scriptstyle\,\pm\,0.00}$ & $0.97{\scriptstyle\,\pm\,0.00}$ & $0.97{\scriptstyle\,\pm\,0.00}$ & $0.55{\scriptstyle\,\pm\,0.04}$ & $0.42{\scriptstyle\,\pm\,0.03}$ & $0.84{\scriptstyle\,\pm\,0.05}$ \\
CPTAC-trained model & CPTAC BRCA PAM50 & 261 & $0.78{\scriptstyle\,\pm\,0.02}$ & $0.79{\scriptstyle\,\pm\,0.04}$ & $0.72{\scriptstyle\,\pm\,0.15}$ & $0.12{\scriptstyle\,\pm\,0.13}$ & $0.17{\scriptstyle\,\pm\,0.19}$ & $0.10{\scriptstyle\,\pm\,0.10}$ \\
CPTAC-trained model & CPTAC COAD MSI & 73 & $0.70{\scriptstyle\,\pm\,0.22}$ & $0.57{\scriptstyle\,\pm\,0.21}$ & $0.62{\scriptstyle\,\pm\,0.16}$ & $0.27{\scriptstyle\,\pm\,0.12}$ & $0.21{\scriptstyle\,\pm\,0.08}$ & $0.38{\scriptstyle\,\pm\,0.18}$ \\
CPTAC-trained model & CPTAC LUAD/LUSC & 1343 & $0.98{\scriptstyle\,\pm\,0.00}$ & $0.96{\scriptstyle\,\pm\,0.00}$ & $0.98{\scriptstyle\,\pm\,0.00}$ & $0.54{\scriptstyle\,\pm\,0.06}$ & $0.39{\scriptstyle\,\pm\,0.06}$ & $0.90{\scriptstyle\,\pm\,0.00}$ \\
\bottomrule
\end{tabular}
}
\end{edtable}

\clearpage
\phantomsection
\section*{Supplementary Information}
\setcounter{section}{0}
\renewcommand{\thesection}{S\arabic{section}}
\crefalias{section}{appendix}
\setcounter{figure}{0}\renewcommand{\thefigure}{S\arabic{figure}}
\setcounter{table}{0}\renewcommand{\thetable}{S\arabic{table}}
\etocsettocdepth.toc{subsection}
\etocsettocstyle{\subsection*{Contents}}{}
\etocsetnexttocdepth{subsection}
\tableofcontents

\section{\ourmethodcbm margin calibration}
\label{app:per_task}

\ourmethodcbm retains most of the black-box model's classification accuracy (\cref{tab:auc-compare}), but one caveat applies. Dropping the residual largely preserves rank-based performance but
changes the scale and offset of the decision margin. Across tasks, the mean $R^2$ of the affine fit
between the full and concept-only margins ranged from $\RubyCbmCalibrationRtwoMin$ to
$\RubyCbmCalibrationRtwoMax$ across seeds (\cref{fig:cbm-calibration-c}). High- and low-$R^2$
examples illustrate the difference between tightly and weakly affine margin relations
(\cref{fig:cbm-calibration-a,fig:cbm-calibration-b}). Accordingly, AUROC changed
comparatively little, whereas balanced accuracy under the uncalibrated default decision rule shifted
substantially for some tasks (\cref{fig:cbm-calibration-d}). \ourmethodcbm therefore requires calibration before it can be used as a thresholded classifier.

\section{Benchmark creation details}
\label{app:ground_truth_creation}

\paragraph{Building the consensus ground truth.}
Each ground truth was built in two stages: we gathered candidate morphological features for the class from the literature (references below), then a board-certified pathologist discarded the generic or ambiguous ones and condensed the rest into a one- to three-sentence description. These sentences are the reference text throughout: the judge grades generated explanations against them, and the expert checklist decomposes them into the features it marks as recovered or missed.

\paragraph{References consulted.}
For each of the tasks in the \benchmark benchmark, we list here the references used to inform the generation of the consensus ground truth. Where a sufficiently clear morphological description was available from a book in the WHO Classification of Tumours series, we preferred this.

Subtyping tasks:
\begin{itemize}
    \item Breast (invasive lobular versus ductal carcinoma): \citep{WHOBreastTumours2021,cserni2020histological,barroso2016differences}.
    \item Lung (adenocarcinoma versus squamous cell carcinoma): \citep{WHOThoracicTumours2021,sharma2025pulmonary,travis2013diagnosis,solis2012histologic}.
    \item Kidney (clear cell versus papillary or chromophobe renal cell carcinoma): \citep{Fletcher2020DiagnosticHistopathology,WHOUrinaryTumours2021}.
    \item Liver (hepatocellular versus cholangiocarcinoma): \citep{WHODigestiveTumours2021,guest2025morphomolecular}.
    \item Brain (oligodendroglioma versus astrocytoma): \citep{WHOCNSTumours2021,AhrendsenAlexandrescuOligodendroglioma,TorkAtkinson2020Oligodendroglioma,wesseling2011pathological,ReesOligodendrogliomas}.
\end{itemize}

Molecular tasks:
\begin{itemize}
    \item Breast (PAM50 basal-like versus not): \citep{schnitt2010classification,rakha2009basal,cakir2012comprehensive}.
    \item Colorectal cancer (microsatellite-instability high versus low): \citep{gustav2025assessing,wagner2023transformer,greenson2009pathologic,faa2024artificial}.
\end{itemize}

\Cref{tab:appendix_ground_truth} lists the full gathered ground-truth explanation for every evaluated class direction, together with whether its task is bilateral or unilateral (\cref{sec:benchmark}).

\begingroup
\small
\setlength{\tabcolsep}{2pt}
\renewcommand{\arraystretch}{1.15}
\begin{longtable}{@{}P{0.14\linewidth}P{0.20\linewidth}C{0.13\linewidth}P{0.45\linewidth}@{}}
\caption{\textbf{Ground-truth morphological explanation for every evaluated class direction.} Each row also gives whether its task is bilateral or unilateral (\cref{sec:benchmark}). Unilateral tasks' uninformative negative class is not evaluated and has no gathered ground-truth explanation.}\label{tab:appendix_ground_truth}\\
\toprule
\textbf{Task} & \textbf{Class} & \textbf{Task type} & \textbf{Ground-truth explanation} \\
\midrule
\endfirsthead
\multicolumn{4}{c}{\tablename\ \thetable\ continued}\\
\toprule
\textbf{Task} & \textbf{Class} & \textbf{Task type} & \textbf{Ground-truth explanation} \\
\midrule
\endhead
\midrule
\multicolumn{4}{r}{Continued on next page}\\
\endfoot
\bottomrule
\endlastfoot
BRCA-Sub & BRCA lobular & Unilateral & This class is characterized by small, non-cohesive cells individually dispersed throughout fibrous connective tissue or arranged in single-file linear cords that invade the stroma (Indian file pattern). The classical form is characterized by a proliferation of small cells that lack cohesion. \\
\addlinespace[0.35em]
BRCA-Sub & BRCA ductal/NST & Unilateral & \textit{Defined by the opposition of the positive class; no ground-truth explanation.} \\
\addlinespace[0.35em]
NSCLC-Sub & NSCLC squamous & Bilateral & This class can be recognized by evidence of keratinization, squamous pearls, and/or intercellular bridges \\
\addlinespace[0.35em]
NSCLC-Sub & NSCLC adenocarcinoma & Bilateral & This class can be recognized by glandular formation and the presence of lepidic, acinar, papillary or complex gland patterns. \\
\addlinespace[0.35em]
RCC-Sub & RCC clear-cell & Bilateral & This class is characterized by sheets and nests of tumor cells with moderate to voluminous clear cytoplasm with interspersed networks of small blood vessels. \\
\addlinespace[0.35em]
RCC-Sub & RCC papillary/chromophobe & Bilateral & This class will lack the network of small blood vessels and may have a papillary architecture, cells with eosinophillic cytoplasm, cell with a thick cytoplasmic membrane, and/or foamy macrophages. \\
\addlinespace[0.35em]
Liver-Sub & Liver cholangiocarcinoma & Bilateral & This class is characterized by biliary differentiation, showing ductal/tubular or cord-like patterns \{sometimes micropapillary\}, abundant desmoplastic fibrous stroma, and small- to medium-sized cuboidal or columnar cells with pale to eosinophilic or vacuolated cytoplasm. \\
\addlinespace[0.35em]
Liver-Sub & Liver HCC & Bilateral & This class is characterized by a well-vascularized tumor made up of malignant hepatocytes with loss of normal hepatic architecture \{portal tracts and reticulin\}, increased arterialization, and characteristic trabecular, solid/compact, pseudoglandular or macrotrabecular growth patterns. \\
\addlinespace[0.35em]
LGG-Sub & LGG oligodendroglioma & Bilateral & This class is characterized by uniformly round, monotonous nuclei with delicate salt-and-pepper chromatin and distinct nuclear membranes that---along with clear cytoplasm---create a pathognomonic "fried-egg" or honeycomb appearance, often accompanied by microcalcifications and a distinct 'chicken-wire' capillary pattern. \\
\addlinespace[0.35em]
LGG-Sub & LGG astrocytoma & Bilateral & This class is characterized by well-differentiated fibrillary glial cells often set in a loosely structured, microcystic matrix, featuring enlarged, hyperchromatic nuclei with irregular contours and uneven chromatin patterns that display variable cellular processes. \\
\addlinespace[0.35em]
BRCA-PAM50 & BRCA basal-like & Unilateral & This class exhibits high histologic grade with solid architecture, absence of tubule formation, marked nuclear pleomorphism and high mitotic rate, frequent geographic or comedo-type necrosis, a continuous pushing margin of invasion, prominent stromal lymphocytic infiltration, and may show metaplastic elements such as squamous or spindle cells and glomeruloid microvascular proliferation. \\
\addlinespace[0.35em]
BRCA-PAM50 & BRCA non-basal & Unilateral & \textit{Defined by the opposition of the positive class; no ground-truth explanation.} \\
\addlinespace[0.35em]
CRC-MSI & CRC MSI-high & Unilateral & This class is characterized by a greater number of tumor-infiltrating lymphocytes, more peri-tumoral immune reaction, more mucin production, increased stromal plasma cells, and more of a medullary appearance \{solid pattern with amphophillic cytoplasm, increased mitosis, and increased lymphocytes\}. \\
\addlinespace[0.35em]
CRC-MSI & CRC MSS & Unilateral & \textit{Defined by the opposition of the positive class; no ground-truth explanation.} \\
\addlinespace[0.35em]
\end{longtable}

\endgroup

\section{Human evaluation details}
\label{app:human_eval}

The expert checklist subset comprises 47 explanations graded by the same pathologist. Thirty-six come from the tile-level pipelines across 8 datasets (12 \SIMILDense, 12 Cosine+, and 12 \SIMILSparse explanations); the remaining 11 come from \ourmethod on the seven TCGA tasks, graded on a single seed and paired with the matching seed judge scores. The full method-level means are shown in \cref{tab:appendix_human_means}, and human and LLM-judge correlations pooled over all 47 explanations are shown in \cref{tab:appendix_human_corr}.

\begin{table}[htbp]
\centering
\caption{\textbf{Expert checklist evaluation across all graded explanations.} The three tile-level pipelines and \ourmethod. Human precision, recall, and F1 are computed from pathologist annotations; LLM metrics are the corresponding automated judge scores on the same explanations.}
\label{tab:appendix_human_means}
\fitwidth{\begin{tabular}{lrrrrrrr}
\toprule
\textbf{Method} & \textbf{N} & \textbf{Human P} & \textbf{Human R} & \textbf{Human F1} & \textbf{Judge P} & \textbf{Judge R} & \textbf{Judge F1} \\
\midrule
\SIMILDense & 12 & 0.072 & 0.097 & 0.073 & 0.044 & 0.072 & 0.051 \\
Cosine+ & 12 & 0.089 & 0.186 & 0.110 & 0.103 & 0.155 & 0.121 \\
\SIMILSparse & 12 & 0.274 & 0.284 & 0.260 & 0.234 & 0.253 & 0.235 \\
\ourmethod & 11 & 0.384 & 0.541 & 0.437 & 0.474 & 0.614 & 0.516 \\
\bottomrule
\end{tabular}
}
\end{table}

\section{Judge and templating stochasticity}
\label{app:templating_judge_variance}

Each per-seed cell of \cref{tab:main-per-class} is one stochastic templating call followed by one stochastic judge call, so the across-seed std reported there folds LLM sampling noise into model variance.

To size that noise, we froze the \ourmethod model at an arbitrary seed and re-ran the template-then-judge pipeline \RubyTemplateJudgeNRepeats~times on the same profiles, which the \splice decomposition makes deterministic (\cref{tab:appendix_templating_judge_variance}). Templating and judging alone contribute a Judge-F1 std of $\sigma_{\mathrm{total}}=\RubyTemplateJudgeSigmaJfOne$, comparable to the \RubyTemplateJudgeSeedSigmaJfOne{} across-seed std of \cref{tab:main-per-class}: much of the nominal seed-to-seed variability is LLM sampling noise rather than model variance.

Freezing the templater as well and re-judging each class \RubyJudgeOnlyNRepeats~times isolates the judge, $\sigma_{\mathrm{judge}}=\RubyJudgeSigmaJfOne$; the templater follows by quadrature from the combined std, $\sigma_{\mathrm{templater}}=\sqrt{\sigma_{\mathrm{total}}^{2}-\sigma_{\mathrm{judge}}^{2}}=\RubyTemplaterSigmaJfOne$. The judge thus accounts for \RubyJudgeVarSharePct\% of the templating+judge variance and the templater \RubyTemplaterVarSharePct\%: both stages are modest and comparable, with phrasing the larger share.

\section{Qualitative explanation panels}
\label{app:explanation_panels}

To make the automated grading transparent, the panels in this appendix present each \ourmethod explanation shown to the pathologist during human evaluation. The colours encode the LLM judge's own decisions, parsed deterministically from the stored judge rationales behind the reported scores: each matched concept is given one pale tint that highlights the concept keyword on \emph{both} sides. Muted red text marks hallucinated claims (false positives, counted against precision) and grey text marks missed ground-truth concepts (false negatives, counted against recall). Each panel header repeats the judge precision, recall, and F1.

\definecolor{paleA}{HTML}{CBD5E8}
\definecolor{paleB}{HTML}{FDCDAC}
\definecolor{paleC}{HTML}{B3E2CD}
\definecolor{paleD}{HTML}{E7DBF0}
\definecolor{paleE}{HTML}{FFF2AE}
\definecolor{paleF}{HTML}{E6F5C9}
\definecolor{paleG}{HTML}{F1E2CC}
\definecolor{paleH}{HTML}{DDE3EF}
\definecolor{halluRed}{HTML}{A6485B}
\definecolor{missGrey}{HTML}{7A7A7A}
\providecommand{\hlc}[2]{{\sethlcolor{#1}\hl{#2}}}
\providecommand{\panellabel}[1]{\textcolor{black!80}{\scshape #1}}
\providecommand{\paneldivider}{\smallskip\noindent\textcolor{black!12}{\rule{\linewidth}{0.4pt}}\par\smallskip}
\tcbset{explpanel/.style={breakable, rounded corners, arc=2.5pt, boxrule=0.5pt, colframe=black!40, colback=white, fonttitle=\normalsize\mdseries, coltitle=black!85, colbacktitle=black!5, left=6pt, right=6pt, top=4pt, bottom=4pt, before skip=8pt, after skip=8pt}}
{\noindent\footnotesize\textbf{Legend.} \hlc{paleA}{matched concept}~(same tint links the model's words to the matching ground-truth phrase)\quad \textcolor{halluRed}{hallucinated}~(not in the ground truth)\quad \textcolor{missGrey}{missed}~(ground-truth phrase the model did not state)\par\medskip}
\begin{tcolorbox}[explpanel,title={\textbf{BRCA basal-like}\hfill \small\textcolor{black!55}{P~0.50~$\cdot$~R~0.55~$\cdot$~}\textcolor{black!70}{\textbf{F1~0.52}}}]
\footnotesize
\panellabel{Model explanation}\enspace Sections show prominent \hlc{paleC}{necrosis (comedo/dirty)} with \hlc{paleA}{solid sheets} of \hlc{paleB}{large pleomorphic atypical cells}, including \hlc{paleE}{keratinizing squamoid elements}, and an accompanying \hlc{paleD}{dense inflammatory infiltrate} with \textcolor{halluRed}{large atypical lymphoid cells (centroblasts/immunoblasts)} in a lichenoid/diffuse pattern. Additional features include \textcolor{halluRed}{epithelioid change} with \textcolor{halluRed}{occasional giant/``monster'' cells} and \textcolor{halluRed}{septation}, with \textcolor{halluRed}{foci suggesting germ cell/rhabdoid differentiation}.\par
\paneldivider
\panellabel{Ground truth}\enspace This class exhibits \hlc{paleB}{high histologic grade} with \hlc{paleA}{solid architecture}, \textcolor{missGrey}{absence of tubule formation}, \hlc{paleB}{marked nuclear pleomorphism} and \textcolor{missGrey}{high mitotic rate}, frequent geographic or \hlc{paleC}{comedo-type necrosis}, a \textcolor{missGrey}{continuous pushing margin of invasion}, \hlc{paleD}{prominent stromal lymphocytic infiltration}, and may show metaplastic elements such as \hlc{paleE}{squamous} or \textcolor{missGrey}{spindle cells} and \textcolor{missGrey}{glomeruloid microvascular proliferation}.
\end{tcolorbox}

\begin{tcolorbox}[explpanel,title={\textbf{BRCA lobular}\hfill \small\textcolor{black!55}{P~0.60~$\cdot$~R~0.67~$\cdot$~}\textcolor{black!70}{\textbf{F1~0.63}}}]
\footnotesize
\panellabel{Model explanation}\enspace A lobular-pattern lesion with prominent \hlc{paleA}{dense sclerotic/collagenous fibrosis} (thick collagen bundles/bands and fibrous septa) showing \hlc{paleB}{infiltrative plasmacytoid/signet-ring--like cells}, with only a \textcolor{halluRed}{mild interstitial lymphohistiocytic/plasma cell infiltrate} and \textcolor{halluRed}{occasional fat necrosis}.\par
\paneldivider
\panellabel{Ground truth}\enspace This class is characterized by \hlc{paleA}{small, non-cohesive cells individually dispersed throughout fibrous connective tissue} or arranged in \textcolor{missGrey}{single-file linear cords} that invade the stroma \textcolor{missGrey}{(Indian file pattern)}. The classical form is characterized by a \hlc{paleB}{proliferation of small cells that lack cohesion}.
\end{tcolorbox}

\begin{tcolorbox}[explpanel,title={\textbf{CRC MSI-high}\hfill \small\textcolor{black!55}{P~0.57~$\cdot$~R~0.50~$\cdot$~}\textcolor{black!70}{\textbf{F1~0.53}}}]
\footnotesize
\panellabel{Model explanation}\enspace This class is characterized by mucinous epithelial proliferation with prominent \hlc{paleC}{mucin production/deposition}, including \hlc{paleC}{abundant blue (acid) mucin} within a myxoid/mucinous stroma and \textcolor{halluRed}{glandular/intestinal-type architecture (mucous glands, intestinal glands)} with \textcolor{halluRed}{cohesive epithelial clusters} and \textcolor{halluRed}{occasional papillary growth}.  An associated inflammatory component with \hlc{paleA}{intraepithelial lymphocytes} and \hlc{paleB}{lymphoid aggregates} may be present.\par
\paneldivider
\panellabel{Ground truth}\enspace This class is characterized by a greater number of \hlc{paleA}{tumor-infiltrating lymphocytes}, more \hlc{paleB}{peri-tumoral immune reaction}, more \hlc{paleC}{mucin production}, increased \textcolor{missGrey}{stromal plasma cells}, and more of a \textcolor{missGrey}{medullary appearance} \{\textcolor{missGrey}{solid pattern} with \textcolor{missGrey}{amphophillic cytoplasm}, \textcolor{missGrey}{increased mitosis}, and \hlc{paleA}{increased lymphocytes}\}.
\end{tcolorbox}

\begin{tcolorbox}[explpanel,title={\textbf{LGG astrocytoma}\hfill \small\textcolor{black!55}{P~0.50~$\cdot$~R~0.29~$\cdot$~}\textcolor{black!70}{\textbf{F1~0.36}}}]
\footnotesize
\panellabel{Model explanation}\enspace A glial-rich lesion with prominent astrocytic and microglial components in a neuropil background, showing \textcolor{halluRed}{variable to increased cellularity} with scattered pleomorphic, \hlc{paleB}{hyperchromatic atypical cells} (including \textcolor{halluRed}{spindle forms and occasional multinucleated giant cells}). There are accompanying \hlc{paleA}{degenerative/myxoid changes} and \textcolor{halluRed}{rare atypical mitoses}.\par
\paneldivider
\panellabel{Ground truth}\enspace This class is characterized by \textcolor{missGrey}{well-differentiated fibrillary glial cells} often set in a \hlc{paleA}{loosely structured, microcystic matrix}, featuring \textcolor{missGrey}{enlarged}, \hlc{paleB}{hyperchromatic nuclei} with \textcolor{missGrey}{irregular contours} and \textcolor{missGrey}{uneven chromatin patterns} that display \textcolor{missGrey}{variable cellular processes}.
\end{tcolorbox}

\begin{tcolorbox}[explpanel,title={\textbf{LGG oligodendroglioma}\hfill \small\textcolor{black!55}{P~0.62~$\cdot$~R~0.83~$\cdot$~}\textcolor{black!70}{\textbf{F1~0.71}}}]
\footnotesize
\panellabel{Model explanation}\enspace Cellular neoplasm composed of \hlc{paleA}{uniform, monomorphic small round} ``blue'' cells in \textcolor{halluRed}{solid sheets} showing \textcolor{halluRed}{prominent perivascular pseudorosettes}, with \hlc{paleC}{perinuclear halos/clear cell change} and \hlc{paleB}{finely granular} (``salt-and-pepper'') chromatin.  \textcolor{halluRed}{Minimal inflammation} with associated \hlc{paleE}{microvascular/pericytic vasculature} and occasional \hlc{paleD}{calcifications/psammoma bodies} may be present.\par
\paneldivider
\panellabel{Ground truth}\enspace This class is characterized by \hlc{paleA}{uniformly round, monotonous nuclei} with \hlc{paleB}{delicate salt-and-pepper chromatin} and \textcolor{missGrey}{distinct nuclear membranes} that--along with \hlc{paleC}{clear cytoplasm}--create a pathognomonic "\hlc{paleC}{fried-egg}" or \hlc{paleC}{honeycomb appearance}, often accompanied by \hlc{paleD}{microcalcifications} and a distinct '\hlc{paleE}{chicken-wire}' \hlc{paleE}{capillary pattern}.
\end{tcolorbox}

\begin{tcolorbox}[explpanel,title={\textbf{Liver HCC}\hfill \small\textcolor{black!55}{P~0.57~$\cdot$~R~0.62~$\cdot$~}\textcolor{black!70}{\textbf{F1~0.60}}}]
\footnotesize
\panellabel{Model explanation}\enspace A \hlc{paleC}{trabecular}/\hlc{paleD}{solid} neoplasm of \hlc{paleB}{large polygonal cells} with abundant eosinophilic (occasionally clear/oncocytic) cytoplasm, set within a \hlc{paleA}{sinusoidal capillary network} with \textcolor{halluRed}{Kupffer cells} and \hlc{paleB}{bile production}, often with \textcolor{halluRed}{background steatosis}. \textcolor{halluRed}{Mitotic activity may be increased}.\par
\paneldivider
\panellabel{Ground truth}\enspace This class is characterized by a \hlc{paleA}{well-vascularized tumor} made up of \hlc{paleB}{malignant hepatocytes} with \textcolor{missGrey}{loss of normal hepatic architecture} \{\textcolor{missGrey}{portal tracts and reticulin}\}, \hlc{paleA}{increased arterialization}, and characteristic \hlc{paleC}{trabecular}, \hlc{paleD}{solid/compact}, \textcolor{missGrey}{pseudoglandular} or \textcolor{missGrey}{macrotrabecular} growth patterns.
\end{tcolorbox}

\begin{tcolorbox}[explpanel,title={\textbf{Liver cholangiocarcinoma}\hfill \small\textcolor{black!55}{P~0.43~$\cdot$~R~0.43~$\cdot$~}\textcolor{black!70}{\textbf{F1~0.43}}}]
\footnotesize
\panellabel{Model explanation}\enspace Class characterized by an infiltrative \hlc{paleA}{gland-forming epithelial neoplasm} with \hlc{paleA}{irregular/atypical ductal structures} and \textcolor{halluRed}{cribriform architecture}, set in a \hlc{paleB}{desmoplastic fibrohyalinized stroma}. Tumor cells show \textcolor{halluRed}{marked cytologic atypia with vesicular chromatin and prominent macronucleoli} (including \textcolor{halluRed}{bizarre/pleomorphic forms}), with possible \textcolor{halluRed}{comedo-type necrosis}.\par
\paneldivider
\panellabel{Ground truth}\enspace This class is characterized by \hlc{paleA}{biliary differentiation}, showing \hlc{paleA}{ductal/tubular} or \textcolor{missGrey}{cord-like} patterns \{sometimes \textcolor{missGrey}{micropapillary}\}, abundant \hlc{paleB}{desmoplastic fibrous stroma}, and \textcolor{missGrey}{small- to medium-sized cuboidal or columnar cells} with \textcolor{missGrey}{pale to eosinophilic or vacuolated cytoplasm}.
\end{tcolorbox}

\begin{tcolorbox}[explpanel,title={\textbf{NSCLC adenocarcinoma}\hfill \small\textcolor{black!55}{P~0.40~$\cdot$~R~0.80~$\cdot$~}\textcolor{black!70}{\textbf{F1~0.53}}}]
\footnotesize
\panellabel{Model explanation}\enspace Class shows an adenocarcinoma phenotype with prominent irregular \hlc{paleA}{glandular}/\hlc{paleB}{acinar} and \textcolor{halluRed}{alveolar growth}, often with \hlc{paleC}{papillary/micropapillary} architecture, lined by \textcolor{halluRed}{atypical pneumocyte‑like cuboidal to columnar cells} with \textcolor{halluRed}{ground‑glass/inclusion‑like cytoplasmic change} and \textcolor{halluRed}{macronucleoli}.  \textcolor{halluRed}{Variable mucin production} and occasional \hlc{paleD}{cribriforming} are noted, with \textcolor{halluRed}{scattered neuroendocrine‑type cells}.\par
\paneldivider
\panellabel{Ground truth}\enspace This class can be recognized by \hlc{paleA}{glandular formation} and the presence of \textcolor{missGrey}{lepidic}, \hlc{paleB}{acinar}, \hlc{paleC}{papillary} or \hlc{paleD}{complex gland patterns}.
\end{tcolorbox}

\begin{tcolorbox}[explpanel,title={\textbf{NSCLC squamous}\hfill \small\textcolor{black!55}{P~0.30~$\cdot$~R~0.67~$\cdot$~}\textcolor{black!70}{\textbf{F1~0.41}}}]
\footnotesize
\panellabel{Model explanation}\enspace Lesion shows predominantly squamous differentiation with \hlc{paleB}{intercellular bridges}/desmosomes and \hlc{paleA}{keratinization}, forming \textcolor{halluRed}{solid nests/sheets} with \textcolor{halluRed}{peripheral palisading} and a \textcolor{halluRed}{basaloid component}.  Foci of \textcolor{halluRed}{comedo-type necrosis} and \textcolor{halluRed}{increased/abnormal mitotic activity} may be present within a \textcolor{halluRed}{fibrous stroma} with \textcolor{halluRed}{variable lichenoid inflammatory infiltrate/ulceration}.\par
\paneldivider
\panellabel{Ground truth}\enspace This class can be recognized by evidence of \hlc{paleA}{keratinization}, \textcolor{missGrey}{squamous pearls}, and/or \hlc{paleB}{intercellular bridges}
\end{tcolorbox}

\begin{tcolorbox}[explpanel,title={\textbf{RCC papillary/chromophobe}\hfill \small\textcolor{black!55}{P~0.29~$\cdot$~R~0.40~$\cdot$~}\textcolor{black!70}{\textbf{F1~0.33}}}]
\footnotesize
\panellabel{Model explanation}\enspace A \hlc{paleA}{predominantly papillary neoplasm} with \textcolor{halluRed}{true fibrovascular cores} and \textcolor{halluRed}{associated tubular/glandular structures} \textcolor{halluRed}{lined by simple cuboidal to columnar epithelium}. The cells are often oncocytic with abundant \hlc{paleB}{eosinophilic granular cytoplasm} and \textcolor{halluRed}{relatively open/vesicular chromatin}, with occasional \textcolor{halluRed}{micropapillary/cribriform areas}.\par
\paneldivider
\panellabel{Ground truth}\enspace This class will lack the \textcolor{missGrey}{network of small blood vessels} and may have a \hlc{paleA}{papillary architecture}, cells with \hlc{paleB}{eosinophillic cytoplasm}, cell with a \textcolor{missGrey}{thick cytoplasmic membrane}, and/or \textcolor{missGrey}{foamy macrophages}.
\end{tcolorbox}

\begin{tcolorbox}[explpanel,title={\textbf{RCC clear-cell}\hfill \small\textcolor{black!55}{P~0.43~$\cdot$~R~1.00~$\cdot$~}\textcolor{black!70}{\textbf{F1~0.60}}}]
\footnotesize
\panellabel{Model explanation}\enspace Lesion composed predominantly of uniform clear cells with \hlc{paleB}{abundant clear/vacuolated cytoplasm} (clear cell change), often in \hlc{paleA}{nests/trabeculae} with thin septa and a \hlc{paleC}{delicate capillary network}.  \textcolor{halluRed}{Foci of sebaceous/foamy (xanthomatous) differentiation} may be present, sometimes within a \textcolor{halluRed}{partially circumscribed/capsulated} architecture with \textcolor{halluRed}{mild acanthosis} and \textcolor{halluRed}{scant perivascular lymphocytes}.\par
\paneldivider
\panellabel{Ground truth}\enspace This class is characterized by \hlc{paleA}{sheets and nests of tumor cells} with \hlc{paleB}{moderate to voluminous clear cytoplasm} with \hlc{paleC}{interspersed networks of small blood vessels}.
\end{tcolorbox}

\section{Qualitative TCGA-to-CPTAC explanation transfer}
\label{app:cptac_explanation_panels}

For each task, the panels below compare a representative paired run of the same TCGA-trained
model on its internal TCGA cohort and the external CPTAC cohort. Shared tints identify the same
ground-truth concept across cohorts, muted red marks explanation content not represented in the
written ground truth, and grey marks ground-truth content absent from the explanations. The panel
headers report the independently judged precision, recall, and F1 for each cohort.

\definecolor{paleA}{HTML}{CBD5E8}
\definecolor{paleB}{HTML}{FDCDAC}
\definecolor{paleC}{HTML}{B3E2CD}
\definecolor{paleD}{HTML}{E7DBF0}
\definecolor{paleE}{HTML}{FFF2AE}
\definecolor{paleF}{HTML}{E6F5C9}
\definecolor{paleG}{HTML}{F1E2CC}
\definecolor{paleH}{HTML}{DDE3EF}
\definecolor{halluRed}{HTML}{A6485B}
\definecolor{missGrey}{HTML}{7A7A7A}
\providecommand{\hlc}[2]{{\sethlcolor{#1}\hl{#2}}}
\providecommand{\panellabel}[1]{\textcolor{black!80}{\scshape #1}}
\providecommand{\paneldivider}{\smallskip\noindent\textcolor{black!12}{\rule{\linewidth}{0.4pt}}\par\smallskip}
\tcbset{explpanel/.style={breakable, rounded corners, arc=2.5pt, boxrule=0.5pt, colframe=black!40, colback=white, fonttitle=\normalsize\mdseries, coltitle=black!85, colbacktitle=black!5, left=6pt, right=6pt, top=4pt, bottom=4pt, before skip=8pt, after skip=8pt}}
{\noindent\footnotesize\textbf{Legend.} \hlc{paleA}{matched concept}~(the same tint links an explanation phrase to the corresponding ground-truth phrase)\quad \textcolor{halluRed}{unmatched}~(not represented in the written ground truth)\quad \textcolor{missGrey}{missed}~(ground-truth phrase absent from the explanation). Each panel compares one representative paired run of the same TCGA-trained model on TCGA and CPTAC.\par\medskip}

\begin{tcolorbox}[explpanel,title={\textbf{BRCA basal-like}\hfill \scriptsize\textcolor{black!55}{TCGA P~0.56~$\cdot$~R~0.60~$\cdot$~}\textcolor{black}{\textbf{F1~0.58}}\textcolor{black!55}{\; $\rightarrow$ \;CPTAC P~0.36~$\cdot$~R~0.30~$\cdot$~}\textcolor{black}{\textbf{F1~0.33}}}]
\footnotesize
\panellabel{TCGA explanation}\enspace This class is characterized by \hlc{paleA}{solid sheets} of large \hlc{paleB}{pleomorphic atypical cells} with prominent \hlc{paleC}{comedo/dirty necrosis} and \hlc{paleD}{keratinizing squamoid differentiation} (intercellular bridges) in an epithelial component, often with \textcolor{halluRed}{fibrous septation}. A conspicuous diffuse \hlc{paleE}{lymphoid infiltrate} rich in \textcolor{halluRed}{large transformed lymphoid cells (centroblasts/immunoblast-like)} and \textcolor{halluRed}{scattered giant ``monster'' cells} is noted, with \hlc{paleC}{patchy necrosis}/\textcolor{halluRed}{apoptosis} and occasional \textcolor{halluRed}{heterologous-appearing elements (rhabdomyoblast-/lipoblast-like cells)}.\par
\paneldivider
\panellabel{CPTAC explanation}\enspace This class is dominated by abundant \hlc{paleC}{necrotic/``dirty'' debris with comedo-type necrosis} and \textcolor{halluRed}{fat necrosis}, often with admixed \textcolor{halluRed}{bacteria and occasional fungal elements (pseudohyphae/yeast)}. The epithelium shows \hlc{paleD}{keratinizing}, \hlc{paleB}{pleomorphic}/\textcolor{halluRed}{dyskeratotic and acantholytic change (including orthokeratosis/koilocytosis)}, with additional \textcolor{halluRed}{stromal hyaline/elastic tissue and focal myxoid or septated adipocytic features}.\par
\paneldivider
\panellabel{Ground truth}\enspace This class exhibits \textcolor{missGrey}{high histologic grade} with \hlc{paleA}{solid architecture}, \textcolor{missGrey}{absence of tubule formation}, \hlc{paleB}{marked nuclear pleomorphism} and \textcolor{missGrey}{high mitotic rate}, frequent geographic or \hlc{paleC}{comedo-type necrosis}, a \textcolor{missGrey}{continuous pushing margin of invasion}, \hlc{paleE}{prominent stromal lymphocytic infiltration}, and may show metaplastic elements such as \hlc{paleD}{squamous} or \textcolor{missGrey}{spindle cells} and \textcolor{missGrey}{glomeruloid microvascular proliferation}.
\end{tcolorbox}

\begin{tcolorbox}[explpanel,title={\textbf{CRC MSI-high}\hfill \scriptsize\textcolor{black!55}{TCGA P~0.43~$\cdot$~R~0.38~$\cdot$~}\textcolor{black}{\textbf{F1~0.40}}\textcolor{black!55}{\; $\rightarrow$ \;CPTAC P~0.60~$\cdot$~R~0.14~$\cdot$~}\textcolor{black}{\textbf{F1~0.23}}}]
\footnotesize
\panellabel{TCGA explanation}\enspace This class is characterized by a \hlc{paleC}{mucin-rich, blue/myxoid mucinous stroma} with prominent mucinous epithelium showing \textcolor{halluRed}{intestinal-type gland formation} and abundant \hlc{paleC}{mucin production/deposition}. Minor associated findings include \textcolor{halluRed}{cohesive epithelial clusters with focal micropapillary architecture} and sparse inflammatory cells (including \hlc{paleA}{intraepithelial lymphocytes}).\par
\paneldivider
\panellabel{CPTAC explanation}\enspace Lesion characterized by prominent mucinous epithelium with \hlc{paleC}{abundant extracellular mucin/myxoid background} and foci of mucin-rich cells including occasional \textcolor{halluRed}{signet-ring change}. Architectural growth shows \textcolor{halluRed}{micropapillary/papillary fronds and cohesive clusters}, with minor admixed elements (\textcolor{halluRed}{sebaceous differentiation}) and a \textcolor{halluRed}{chronic inflammatory infiltrate}.\par
\paneldivider
\panellabel{Ground truth}\enspace This class is characterized by a greater number of \hlc{paleA}{tumor-infiltrating lymphocytes}, more \textcolor{missGrey}{peri-tumoral immune reaction}, more \hlc{paleC}{mucin production}, increased \textcolor{missGrey}{stromal plasma cells}, and more of a \textcolor{missGrey}{medullary appearance} \{\textcolor{missGrey}{solid pattern} with \textcolor{missGrey}{amphophillic cytoplasm}, \textcolor{missGrey}{increased mitosis}, and \hlc{paleA}{increased lymphocytes}\}.
\end{tcolorbox}

\begin{tcolorbox}[explpanel,title={\textbf{NSCLC adenocarcinoma}\hfill \scriptsize\textcolor{black!55}{TCGA P~0.43~$\cdot$~R~0.80~$\cdot$~}\textcolor{black}{\textbf{F1~0.56}}\textcolor{black!55}{\; $\rightarrow$ \;CPTAC P~0.44~$\cdot$~R~0.60~$\cdot$~}\textcolor{black}{\textbf{F1~0.51}}}]
\footnotesize
\panellabel{TCGA explanation}\enspace Neoplasm composed of irregular, atypical \hlc{paleA}{glandular}/\hlc{paleB}{acinar structures} in an \textcolor{halluRed}{alveolar (air-space) setting}, often with \hlc{paleC}{papillary/micropapillary architecture} and occasional \hlc{paleD}{cribriforming}. Tumor cells show \textcolor{halluRed}{pneumocytic features with ground-glass/inclusion-like cytoplasm, prominent macronucleoli and nuclear membrane irregularity}, with variable \textcolor{halluRed}{mucin production}.\par
\paneldivider
\panellabel{CPTAC explanation}\enspace This class is characterized by a predominantly \hlc{paleA}{gland-forming neoplasm with irregular/atypical glands} and an \textcolor{halluRed}{alveolar pattern}, with \hlc{paleD}{cribriform} and \hlc{paleC}{papillary/micropapillary architecture}. The tumor shows \textcolor{halluRed}{mucinous (mucous gland) differentiation and apocrine-type features (decapitation secretion)}, with cells displaying \textcolor{halluRed}{ground-glass cytoplasm/intracytoplasmic inclusions and cytologic atypia including macronucleoli and nuclear membrane irregularity}.\par
\paneldivider
\panellabel{Ground truth}\enspace This class can be recognized by \hlc{paleA}{glandular formation} and the presence of \textcolor{missGrey}{lepidic}, \hlc{paleB}{acinar}, \hlc{paleC}{papillary} or \hlc{paleD}{complex gland patterns}.
\end{tcolorbox}

\begin{tcolorbox}[explpanel,title={\textbf{NSCLC squamous}\hfill \scriptsize\textcolor{black!55}{TCGA P~0.36~$\cdot$~R~0.67~$\cdot$~}\textcolor{black}{\textbf{F1~0.47}}\textcolor{black!55}{\; $\rightarrow$ \;CPTAC P~0.40~$\cdot$~R~1.00~$\cdot$~}\textcolor{black}{\textbf{F1~0.57}}}]
\footnotesize
\panellabel{TCGA explanation}\enspace Lesion shows predominant squamous differentiation with \hlc{paleB}{intercellular bridges/desmosomes} and \hlc{paleA}{keratinization} (including \hlc{paleC}{squamous eddies/whorls} and hypergranulosis), in a \textcolor{halluRed}{solid-nested growth pattern with peripheral palisading and basaloid areas}. There are focal \textcolor{halluRed}{urothelial/transitional features (umbrella cells)}, with \textcolor{halluRed}{mitotic activity and occasional comedo-type necrosis and ulceration}.\par
\paneldivider
\panellabel{CPTAC explanation}\enspace Lesion composed predominantly of squamous epithelium with overt squamous differentiation--\hlc{paleB}{intercellular bridges/desmosomes}, \hlc{paleA}{keratinization} with \hlc{paleC}{keratin pearls/squamous eddies} and hypergranulosis--with associated \textcolor{halluRed}{ulceration/crust}. Tumor nests show a \textcolor{halluRed}{basaloid component with peripheral palisading and solid growth}, with \textcolor{halluRed}{focal comedo-type necrosis and increased mitotic activity}.\par
\paneldivider
\panellabel{Ground truth}\enspace This class can be recognized by evidence of \hlc{paleA}{keratinization}, \hlc{paleC}{squamous pearls}, and/or \hlc{paleB}{intercellular bridges}.
\end{tcolorbox}

\newpage

\section{Evaluation of concept scoring}
\label{app:concept_attr_eval}

\begin{figure}[ht!]
  \includegraphics[width=\linewidth]{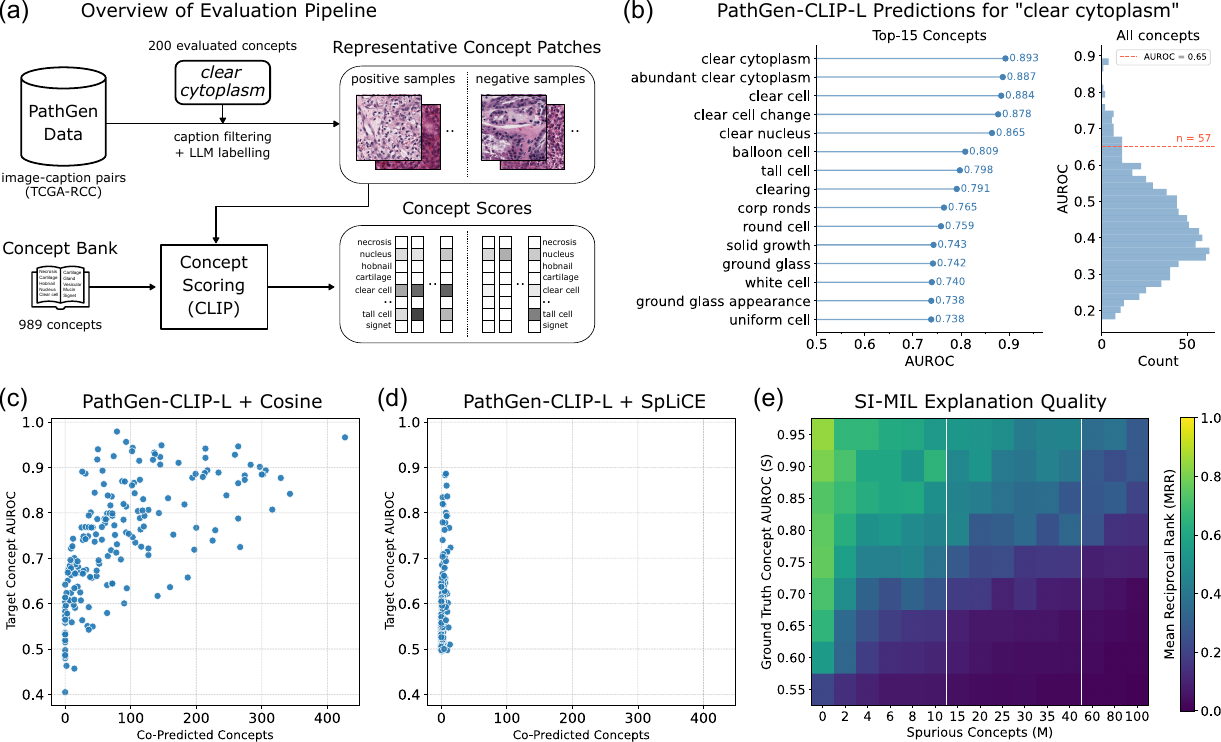}
  {\phantomsubcaption\label{fig:concept-eval-a}}%
  {\phantomsubcaption\label{fig:concept-eval-b}}%
  {\phantomsubcaption\label{fig:concept-eval-c}}%
  {\phantomsubcaption\label{fig:concept-eval-d}}%
  {\phantomsubcaption\label{fig:concept-eval-e}}%
  \caption{\textbf{Evaluation of concept scoring with representative patch data}. \textbf{(a)}~Overview of the concept scoring evaluation pipeline. We sampled representative patch data with negative control images for selected pathological concepts from the PathGen dataset \citep{pathgen2024}. \textbf{(b)}~Evaluation results for the PathGen-CLIP-L model with cosine similarity scoring for representative patches of the ``clear cytoplasm'' concept. The AUROC indicates to what degree the scores for the corresponding concept in the bank identify the ``clear cytoplasm'' images. \textbf{(c)/(d)}~Concept predictability and co-predictions for all $200$ evaluated concepts scored by PathGen-CLIP-L with cosine similarities / \splice decomposition. We define co-predicted concepts as the number of concepts in the bank that identify the images of a given concept with an AUROC of at least $0.65$. \textbf{(e)}~Explanation quality of an SI-MIL model for simulated concept scores with varying AUROC and co-prediction levels. The color gradient represents the mean reciprocal rank (MRR) of the global explanation score of a ground truth concept over $100$ repetitions (\textit{higher is better}).}
  \label{fig:concept_eval}
\end{figure}

\subsection{Measuring the quality of concept scoring}
\label{app:attribution_faithfulness}

\paragraph{Motivation}

A central element of our pipeline is the translation of dense histopathology image features into interpretable concept profiles, which is achieved through a pathology-specific CLIP model scoring the morphological concepts from our concept bank. To better understand how well this mechanism works, we collected patch data representative of selected concepts for a targeted assessment.

\paragraph{Experiment}

As data source, we used the kidney projects (TCGA-KIRC, TCGA-KIRP, TCGA-KICH) of the PathGen dataset \citep{pathgen2024}. We selected $200$ kidney-relevant morphological concepts from our concept bank. For each concept $C_{eval}$, we collected $200$ images on which the concept is present (positive samples) and $200$ control patches (negative samples) (\cref{fig:concept-eval-a}). Positive images were randomly sampled from those image-caption pairs that contain the concept in its caption, while negative samples from those that do not. To mitigate potential label noise, we prompted GPT-5.2 (Azure OpenAI, API version 2024-12-01-preview) to confirm the visual presence of a concept in each positive sample image. We then scored these representative samples for all morphological concepts in the bank with the PathGen-CLIP-L model. For each scored concept ($C_{pred}$), we then analyzed whether the concept scores were consistently higher on the positive samples than on the negative samples, measured via the AUROC. Intuitively, we expect a high AUROC for the concept that corresponds to the evaluated one ($C_{pred}=C_{eval}$) or any semantically / biologically closely related concept, but a low AUROC if $C_{pred}$ and $C_{eval}$ are not connected. We define two metrics:
\begin{itemize}
  \item
  \textbf{Target Concept AUROC}: We use GPT-5.2 to identify morphological concepts that are semantically equivalent to $C_{eval}$ (=target concepts). The target concept AUROC is the highest AUROC achieved by any of these concepts for $C_{eval}$. It represents how accurately the CLIP model identifies the visual presence $C_{eval}$.
  \item
  \textbf{Co-Predicted Concepts}: The co-predicted concepts are the amount of concepts $C_{pred}$ that achieve an AUROC of at least $0.65$ for $C_{eval}$. It measures how many concepts in the bank are (correctly or incorrectly) predictive for the concept $C_{eval}$.
\end{itemize}

\paragraph{Results}

\begin{sloppypar}
Concept prediction results for $C_{eval}=\text{``clear cytoplasm''}$ scored with cosine similarity are presented in \cref{fig:concept-eval-b}. The target concepts $C_{pred} \in \{ \text{``clear cytoplasm''}, \text{``abundant clear cytoplasm''} \}$ were the most predictive concepts, reaching AUROCs above $0.88$. This shows that the model can identify ``clear cytoplasm'' in kidney slides with high accuracy. At the same time, various other concepts in the bank were also highly predictive for the representative ``clear cytoplasm'' patches. Some of these have a close semantic connection to the correct concept (e.g., ``clear cell'', ``clear cell change''), while others were unexpected (e.g., ``clear nucleus'', ``baloon cell'', ``corp ronds''). This points to the existence of systematic model- or data-specific spurious concept activations.
\end{sloppypar}

To investigate these effects systematically, we compared the target concept AUROC to the number co-predictions across all $200$ evaluated concepts (\cref{fig:concept-eval-c}). Although the CLIP model predicted many concepts with high accuracy ($0.731$ mean AUROC), they frequently came with a high amount of co-predicted concepts ($83.0$ on average). In $63$ cases, more than $100$ concepts in the bank were predictive of $C_{eval}$ according to the CLIP model. Note that, given their quantity, most of these predictive concepts are likely erroneous. This co-prediction imposes significant noise on a downstream discovery model, which may not be able to identify the correct morphological concept among a large number of spurious concepts. \splice sparsification \citep{splice}, applied at the default $\lambda$ used throughout the paper (a budget of $25$ active concepts per decomposition), dramatically reduced the amount of co-predicted concepts to $1.9$ on average (\cref{fig:concept-eval-d}). Despite coming at the cost of lower target concept prediction performance ($0.582$ mean AUROC), the increased sparsity enabled stronger global explanation quality (\cref{fig:main-results}).

\subsection{Connecting concept scoring quality to explanation outcomes}

\paragraph{Motivation}

To quantify the effects of concept prediction accuracy and co-prediction noise on output explanation quality, we designed a simulation experiment using artificial data with controllable characteristics, allowing us to assess if correct concepts come through to the final explanation in an SI-MIL model \citep{simil}.

\paragraph{Experiment}

We generated artificial data that mimic a slide-level prediction task with tile-level concept score attributions. In particular, assume we have $2 \cdot N$ slides ($N$ slides representing class A / B) and $K$ patches per slide, resulting in $K \cdot N$ patches representative of class A / B). For each patch, we sampled a concept vector $\textbf{c} \in \mathbb{R}^d$ as follows:
\begin{itemize}
  \item Ground truth concept ($c_{1}$): sample i.i.d.\ from mean-shifted Gaussians, i.e., $c_{1} \sim \mathcal{N}(\delta, 1)$ for class A and $c_{1} \sim \mathcal{N}(0, 1)$ for class B patches. This way, the expected AUROC between patches of class A and class B is $S = \Phi\!\left(\frac{\delta}{\sqrt{2}}\right)$, where $\Phi$ is the CDF of the normal distribution. Hence, to model any desired concept scoring AUROC $S$, we choose $\delta=\sqrt{2}\Phi^{-1}(S)$.
  \item
  Spurious concepts ($c_2, \ldots, c_{M+1}$): sample i.i.d.\ from mean-shifted Gaussians, i.e., $c_{k} \sim \mathcal{N}(\delta_k, 1)$ for class A and $c_{k} \sim \mathcal{N}(0, 1)$ for class B patches. For each spurious concept, we sample an AUROC $S'_k \sim \mathcal{N}(\mu_{spurious}, \sigma_{spurious}^2)$ and choose $\delta_k=\sqrt{2}\Phi^{-1}(S'_k)$.
  \item
  Random concepts ($c_{M+2}, \ldots, c_{D}$): sample i.i.d.\ from a normal distribution, i.e., $c_{k} \sim \mathcal{N}(0, 1)$ for all patches.
\end{itemize}
We then trained a SI-MIL concept branch to predict whether a slide is class A or class B, and inspected the global explanation scores as in our SI-MIL pipelines (\cref{meth:simil}). Specifically, we measured the (reciprocal) rank of the explanation score of the ground truth concept in comparison to the other concepts. Intuitively, the rank score measures if the ground truth concept comes through to the final explanation in the SI-MIL model depending on the accuracy and spurious co-prediction noise of the concept attribution scores. Ideally, the ground truth concept should always receive the highest explanation score, i.e., rank 1.

We chose $N=300$ / $N=100$ for training / test, $K=20$, $D=989$, $\mu_{noise}=0.7$, $\sigma_{noise}=0.05$ and iterated over various levels of $S$ and $M$. The SI-MIL concept head was trained with an L1 coefficient of $0.05$, batch size $32$, and learning rate $0.0002$ for $30$ epochs. For stability, we repeated each experiment $100$ times with randomly resampled data and report the mean reciprocal rank (MRR) of the concept explanation score of the ground truth concept.

\paragraph{Results}

Both the ground truth concept AUROC $S$ of as well as the number of spurious concepts $M$ had a strong impact on the explanation quality of the SI-MIL model (\cref{fig:concept-eval-e}). As the number of spurious concepts $M$ increases, the mean reciprocal rank of the ground-truth concept in the final explanation decreases rapidly, particularly if the concept scores for this concept are imperfect ($S \ll 1.0$). For example, if the concept attribution model scores separate the ground truth concept with an AUROC of $0.75$ while co-predicting at least $15$ spurious concepts, the explanation will in most cases not recognize the ground truth concept as most predictive ($\text{MRR} \le 0.3$). Note that concepts with target concept AUROC of $0.75$ frequently come with more than $50$ co-predicted concepts when using PathGen-CLIP-L with cosine similarity concept scores (\cref{fig:concept-eval-c}).
This indicates that using a CLIP model without sparsification leads to too many spurious activations that prevent the identification of genuinely predictive morphological concepts. Next to accuracy, sparsity in the concept scoring --- as achieved through \splice --- is therefore crucial for effective discovery.

\clearpage
\begin{landscape}
\section{Supplementary atlas of generated morphological hypotheses}
\label{app:discovery_atlas}

\ourmethod was applied to \RubyDiscoveryAtlasCandidateTaskCount\ candidate TCGA molecular classification tasks. \RubyDiscoveryAtlasExcludedTaskCount\ tasks whose held-out classifier did not exceed test AUROC \RubyDiscoveryAtlasAurocThreshold\ (i.e.\ not reliably solved, so any generated hypothesis would not be grounded in a real learned decision rule) were excluded before hypothesis generation.

The atlas comprises \RubyDiscoveryAtlasHypothesisCount\ class-level morphological hypotheses across \RubyDiscoveryAtlasTaskCount\ tasks and \RubyDiscoveryAtlasOrganCount\ organs.
It includes all positive-class hypotheses and the available negative-class hypotheses.
These outputs are unvalidated hypotheses; test AUROC reports task-level classifier discrimination.

\begingroup
\footnotesize
\setlength{\tabcolsep}{2pt}
\setlength{\LTpre}{0.5em}
\setlength{\LTpost}{0pt}
\renewcommand{\arraystretch}{1.08}
\begin{longtable}{@{}P{0.10\linewidth}P{0.19\linewidth}P{0.15\linewidth}C{0.055\linewidth}P{0.44\linewidth}@{}}
\caption{\textbf{Generated morphological hypotheses across the 26-task atlas.} Test AUROC is task-level and is repeated when both class hypotheses are available.}\label{tab:discovery-atlas}\\
\toprule
\textbf{Organ} & \textbf{Task} & \textbf{Class} & \textbf{Test AUROC} & \textbf{Generated morphological hypothesis}\\
\midrule
\endfirsthead
\multicolumn{5}{c}{\tablename\ \thetable\ continued}\\
\toprule
\textbf{Organ} & \textbf{Task} & \textbf{Class} & \textbf{Test AUROC} & \textbf{Generated morphological hypothesis}\\
\midrule
\endhead
\midrule
\multicolumn{5}{r}{Continued on next page}\\
\endfoot
\bottomrule
\endlastfoot
Breast & \textbf{Estrogen receptor status}\newline{\scriptsize\texttt{tcga\_brca\_er\_status}} & ER-positive & 0.781 & Lesion composed of ductal/glandular structures with prominent myoepithelial cells, showing cribriform and lobular/tubular architecture with retraction artifact. The stroma is collagenous and often sclerotic/desmoplastic, with occasional apocrine/eccrine-type differentiation and scattered neuroendocrine-like cells.\\
\addlinespace[0.35em]
Breast & \textbf{Estrogen receptor status}\newline{\scriptsize\texttt{tcga\_brca\_er\_status}} & ER-negative & 0.781 & This class shows a high-grade neoplasm dominated by diffuse sheets of large atypical/pleomorphic cells with bizarre, eccentrically placed nuclei (``monster'' cells) and brisk mitotic activity with comedo-type necrosis. There are accompanying large atypical lymphoid elements (centroblasts/immunoblast-like cells) and areas suggesting squamous differentiation with keratinization/hypergranulosis and ulceration.\\
\addlinespace[0.35em]
Breast & \textbf{HER2 status}\newline{\scriptsize\texttt{tcga\_brca\_her2}} & HER2-positive & 0.770 & An infiltrative adnexal carcinoma with prominent apocrine/eccrine differentiation, composed of irregular malignant glands in a desmoplastic (scar-like) stroma. Tumor cells show marked cytologic atypia with pleomorphic/bizarre ``monster'' nuclei and macronucleoli, with focal clear/vacuolated cytoplasm and occasional mucin/signet-ring change; scattered foreign-body giant cell reaction and focal keratinization may be present.\\
\addlinespace[0.35em]
Breast & \textbf{Histological subtype}\newline{\scriptsize\texttt{tcga\_brca\_subtyping}} & Invasive lobular carcinoma & 0.947 & This class is characterized by discohesive signet-ring/plasmacytoid (``lobular'') infiltrates set within prominent dense sclerotic/collagenous stroma, sometimes with interstitial/diffuse spread. Scattered single melanocytes/melanin and mixed inflammatory elements (plasma cells/lymphohistiocytic cells) may be present, with occasional myxoid change.\\
\addlinespace[0.35em]
Breast & \textbf{Homologous recombination deficiency}\newline{\scriptsize\texttt{tcga\_brca\_hrd}} & HRD-positive & 0.829 & This class is characterized by solid nests/sheets of markedly pleomorphic large atypical cells with vesicular chromatin and prominent macronucleoli, including bizarre ``monster'' forms, often with comedo-type necrosis. Tumor cells may show apocrine/clear-cell change with occasional keratinization/intercellular bridges, and are accompanied by a lymphoid/lichenoid inflammatory infiltrate with large atypical lymphoid cells/centroblasts.\\
\addlinespace[0.35em]
Breast & \textbf{PIK3CA driver mutation}\newline{\scriptsize\texttt{tcga\_brca\_pik3ca\_driver}} & PIK3CA driver-mutant & 0.646 & This class is dominated by dense, thick, hyalinized/sclerotic collagen bundles forming scar-like fibrotic bands with a mild desmoplastic/myofibroblastic response and minimal inflammation. Associated findings include fat necrosis with occasional foreign-body type reaction/calcifications and entrapped adnexal/ductal (eccrine) structures.\\
\addlinespace[0.35em]
Breast & \textbf{Progesterone receptor status}\newline{\scriptsize\texttt{tcga\_brca\_pr\_status}} & PR-positive & 0.727 & Predominantly ductal adnexal differentiation with dual-layered ducts showing prominent myoepithelial cells, forming cribriform and small tubulo-lobular architectures. The lesion is set in dense sclerotic/hyalinized collagenous stroma with retraction spaces and occasional calcific/psammomatous bodies, with focal apocrine/eccrine features.\\
\addlinespace[0.35em]
Breast & \textbf{Progesterone receptor status}\newline{\scriptsize\texttt{tcga\_brca\_pr\_status}} & PR-negative & 0.727 & This class shows a high-grade malignant neoplasm composed of solid sheets/nests of markedly pleomorphic atypical large cells with vesicular chromatin, bizarre ``monster'' forms and brisk mitotic activity, often with comedo-type necrosis. There are focal squamous features (keratinization/intercellular bridges) and heterologous elements (lipoblasts, rhabdomyoblasts), within a prominent diffuse/lichenoid lymphohistiocytic infiltrate containing atypical large lymphoid cells (centroblast/immunoblast-like).\\
\addlinespace[0.35em]
Breast & \textbf{TP53 driver mutation}\newline{\scriptsize\texttt{tcga\_brca\_tp53\_driver}} & TP53 driver-mutant & 0.864 & This class is characterized by a high-grade pleomorphic malignant neoplasm composed of very large atypical cells with vesicular chromatin, prominent macronucleoli, bizarre ``monster'' nuclei and occasional rhabdomyoblastic differentiation, showing solid growth with comedo-type necrosis.  A prominent accompanying inflammatory component (lichenoid/diffuse lymphohistiocytic and granulomatous infiltrates with large atypical lymphoid cells/centroblasts) and focal desmoplastic response may be present, with occasional keratinization/squamoid features.\\
\addlinespace[0.35em]
Colorectal & \textbf{APC driver mutation}\newline{\scriptsize\texttt{tcga\_crc\_apc\_driver}} & APC driver-mutant & 0.797 & Sections show small intestinal-type mucosa with villi and crypts lined by tall columnar enterocytes with brush border, accompanied by Paneth cells. The supporting stroma and wall are rich in smooth muscle bundles with prominent collagen deposition/fibrous thickened collagen and admixed mature adipose tissue, with focal necrosis/debris and occasional bacilli.\\
\addlinespace[0.35em]
Colorectal & \textbf{BRAF driver mutation}\newline{\scriptsize\texttt{tcga\_crc\_braf\_driver}} & BRAF driver-mutant & 0.743 & Gland-forming intestinal-type mucinous epithelium composed of tall columnar cells with abundant basophilic/blue mucin and mucin deposition. Architecture is predominantly irregular tubular glands with prominent cribriforming and focal micropapillary change, accompanied by scattered intraepithelial lymphocytes and small lymphoid aggregates.\\
\addlinespace[0.35em]
Colorectal & \textbf{Consensus molecular subtype}\newline{\scriptsize\texttt{tcga\_crc\_cms4}} & CMS4 & 0.636 & This class is characterized by a prominent desmoplastic, collagen-rich stromal response with myofibroblastic proliferation in loose connective tissue. Associated features include nuclear molding and occasional adnexal (sebaceous) elements with foci of fat necrosis.\\
\addlinespace[0.35em]
Colorectal & \textbf{Consensus molecular subtype}\newline{\scriptsize\texttt{tcga\_crc\_cms4}} & CMS1, CMS2, or CMS3 & 0.636 & Intestinal-type mucosa with prominent gland/crypt architecture, including tubular gland formation lined by enterocytes with brush border, villi, goblet cells, and Paneth cells. There is epithelial proliferation with mitotic activity (including atypical forms) and areas of irregular/cribriform glandular architecture with mucinous epithelium.\\
\addlinespace[0.35em]
Colorectal & \textbf{Hypermutation status}\newline{\scriptsize\texttt{tcga\_crc\_hypermutated}} & Hypermutated & 0.800 & Mucin-rich epithelial neoplasm with abundant blue mucin/mucin deposition and mucinous epithelium, showing epithelial proliferation in cribriform and focal micropapillary/acinar patterns. Tumor cells exhibit vesicular chromatin with macronucleoli, with accompanying intraepithelial lymphocytes and scattered lymphoid aggregates/white blood cells.\\
\addlinespace[0.35em]
Colorectal & \textbf{Microsatellite instability}\newline{\scriptsize\texttt{tcga\_crc\_msi}} & MSI-high & 0.969 & Lesion characterized by an epithelial proliferation showing intestinal-type glandular differentiation and prominent mucinous epithelium with abundant extracellular mucin (often blue) producing a myxoid/mucinous stromal background. There may be cohesive clusters with focal micropapillary architecture and a mild inflammatory component including intraepithelial lymphocytes/white blood cells.\\
\addlinespace[0.35em]
Colorectal & \textbf{TP53 driver mutation}\newline{\scriptsize\texttt{tcga\_crc\_tp53\_driver}} & TP53 driver-mutant & 0.741 & This class is characterized by areas of mature adipose tissue with fat necrosis and necrotic/comedo-type debris, accompanied by a prominent desmoplastic stromal reaction with dense collagen deposition and fibroblasts. The lesional cells show high N:C ratio with nuclear molding and palisading (occasionally pseudostratified columnar-type architecture), set within fibrous and smooth muscle bundles.\\
\addlinespace[0.35em]
Head and neck & \textbf{Human papillomavirus status}\newline{\scriptsize\texttt{tcga\_hnsc\_hpv}} & HPV-positive & 0.835 & Lesion composed of basaloid ``blue'' epithelial cells with scant cytoplasm arranged in solid nests/sheets showing peripheral palisading and retraction clefting, consistent with a basal cell carcinoma--type pattern. A prominent lichenoid/dense lymphoid infiltrate with intraepithelial lymphocytes is present, with occasional surface scale crust/ulceration and focal necrosis.\\
\addlinespace[0.35em]
Head and neck & \textbf{Human papillomavirus status}\newline{\scriptsize\texttt{tcga\_hnsc\_hpv}} & HPV-negative & 0.835 & A keratinizing squamous epithelial lesion composed of atypical keratinocytes with abundant eosinophilic (``pink'') cytoplasm and prominent intercellular bridges/desmosomes, showing dyskeratosis/keratin pearl--type maturation with compact hyperkeratosis and hypergranulosis. Koilocytic change (HPV-type cytopathic effect) is also a notable feature.\\
\addlinespace[0.35em]
Kidney & \textbf{BAP1 mutation}\newline{\scriptsize\texttt{tcga\_kirc\_bap1\_driver}} & BAP1-mutated & 0.762 & Lesion dominated by cells with abundant clear/vacuolated cytoplasm, including sebaceous-type differentiation and xanthomatous/foamy histiocytes. Overall this suggests a clear cell--rich, lipid/glycogen-laden process with possible delicate capillary networking.\\
\addlinespace[0.35em]
Kidney & \textbf{BAP1 versus PBRM1 mutation}\newline{\scriptsize\texttt{tcga\_kirc\_bap1\_vs\_pbrm1}} & BAP1-mutant & 0.961 & Lesion characterized by prominent sebaceous differentiation with cells showing abundant clear/vacuolated cytoplasm, forming frond-like papillary projections on fibrovascular stalks with a papillary architecture and mild acanthosis.\\
\addlinespace[0.35em]
Kidney & \textbf{BAP1 versus PBRM1 mutation}\newline{\scriptsize\texttt{tcga\_kirc\_bap1\_vs\_pbrm1}} & PBRM1-mutant & 0.961 & The lesion is characterized by prominent tubular structures with tubular atrophy and focal cystic/microcystic dilatation lined by simple cuboidal epithelium, set within dense regular fibrous connective tissue with a thick-walled vascular component and a fibrous/capsular rim. There is conspicuous adipocytic differentiation (mature fat with lipoblast-like cells) and xanthomatous change with occasional cholesterol clefts.\\
\addlinespace[0.35em]
Kidney & \textbf{PBRM1 driver mutation}\newline{\scriptsize\texttt{tcga\_kirc\_pbrm1\_driver}} & PBRM1 driver-mutant & 0.649 & Lesion composed of sebaceous/clear cells with abundant vacuolated (clear) cytoplasm, including xanthomatous (``xanthoma'') cells, arranged in frond-like architecture with a stalk and prominent capillary network with red blood cells/hemorrhage. There may be associated acanthosis and occasional stromal encapsulation.\\
\addlinespace[0.35em]
Brain & \textbf{Lower-grade glioma subtype}\newline{\scriptsize\texttt{tcga\_lgg\_subtyping}} & Oligodendroglioma & 0.934 & A monomorphic small round/clear-cell neoplasm showing prominent perivascular pseudorosettes, with round nuclei featuring salt-and-pepper/granular chromatin and frequent perinuclear halos/clear spaces. The tumor grows in solid sheets with sharp cell borders, minimal inflammation, and may show microvascular proliferation with focal proteinaceous material and calcification.\\
\addlinespace[0.35em]
Brain & \textbf{Lower-grade glioma subtype}\newline{\scriptsize\texttt{tcga\_lgg\_subtyping}} & Astrocytoma & 0.934 & A glial neoplasm with prominent astrocytic differentiation in a neuropil background, showing variable to increased cellularity with microglial elements. Tumor cells exhibit scattered nuclear enlargement and hyperchromasia with mild pleomorphism/spindle--stellate morphologies, set in a focal myxoid/reticular degenerative matrix with occasional atypical mitotic activity.\\
\addlinespace[0.35em]
Liver & \textbf{CTNNB1 driver mutation}\newline{\scriptsize\texttt{tcga\_lihc\_ctnnb1\_driver}} & CTNNB1 driver-mutant & 0.731 & This class is characterized by hepatic architecture with prominent sinusoids/sinusoidal capillaries and bile, including Kupffer cells, with hepatocellular trabecular arrangement and associated steatosis/vacuolar change. Cells often show abundant eosinophilic (oncocytic/apocrine-like) cytoplasm with occasional macronucleoli and a lobular--trabecular growth pattern.\\
\addlinespace[0.35em]
Liver & \textbf{TP53 driver mutation}\newline{\scriptsize\texttt{tcga\_lihc\_tp53\_driver}} & TP53 driver-mutant & 0.638 & Lesion composed of epithelioid cells with conspicuous macronucleoli arranged around prominent sinusoidal capillary/sinusoid-type vasculature.  There is supportive fibrovascular stroma with scattered mononuclear inflammatory cells (including neutrophils) and occasional eosinophilic globules and osteoclast-type giant cells, with focal hobnail- or glomus/granulosa-lutein--like cellular morphology.\\
\addlinespace[0.35em]
Lung & \textbf{EGFR versus KRAS mutation}\newline{\scriptsize\texttt{tcga\_luad\_egfr\_vs\_kras}} & EGFR-mutant & 0.838 & Papillary to micropapillary epithelial neoplasm with fibrovascular cores, composed of cuboidal to tall columnar (hobnail) cells showing prominent nucleoli and a honeycomb/ductal--glandular architecture.  Notable acantholysis with acantholytic dyskeratosis and focal apocrine-type decapitation secretion is present, with associated interstitial fibrosis and fibroblastic stroma and occasional neuroendocrine-type cells.\\
\addlinespace[0.35em]
Lung & \textbf{EGFR versus KRAS mutation}\newline{\scriptsize\texttt{tcga\_luad\_egfr\_vs\_kras}} & KRAS-mutant & 0.838 & Pulmonary parenchyma with an alveolar pattern and prominent pneumocytes lining enlarged air spaces, with scattered alveolar macrophages and interstitial inflammation producing a ground-glass type appearance. Occasional viral-type inclusions with necrotic/nuclear debris may be present, with minor associated bronchiolar/mucous gland--mucinous epithelium change and focal fibroblastic/fibrous tissue reaction.\\
\addlinespace[0.35em]
Lung & \textbf{TP53 driver mutation}\newline{\scriptsize\texttt{tcga\_luad\_tp53\_driver}} & TP53 driver-mutant & 0.811 & This class is characterized by a high-grade malignant infiltrate composed of pleomorphic epithelioid/large atypical cells with vesicular chromatin and prominent macronucleoli, including bizarre ``monster'' forms, with foci of solid nesting and comedo-type necrosis. Associated features include neuroendocrine-type nuclear molding, keratinization with acantholysis/acantholytic dyskeratosis and ulceration, and frequent pigment-laden histiocytes/macrophages with occasional osteoclast-like giant cell reaction and mixed lichenoid/granulomatous inflammation.\\
\addlinespace[0.35em]
Stomach & \textbf{ARID1A driver mutation}\newline{\scriptsize\texttt{tcga\_stad\_arid1a\_driver}} & ARID1A driver-mutant & 0.611 & A predominantly lymphoid/lymphoplasmacytic inflammatory process with a dense, lichenoid band-like infiltrate and intraepithelial lymphocytes, accompanied by ulceration and a superimposed neutrophilic component. Scattered immunoblasts with vesicular chromatin are noted within the inflammatory background.\\
\addlinespace[0.35em]
Stomach & \textbf{Hypermutation status}\newline{\scriptsize\texttt{tcga\_stad\_hypermutated}} & Hypermutated & 0.841 & An intestinal-type gland-forming epithelial neoplasm with irregular/cribriform glands and occasional micropapillary/solid areas, composed of tall columnar mucin-producing cells showing vesicular nuclei with macronucleoli, high N:C ratio, pleomorphism and budding (with focal comedo-type necrosis). The background shows prominent lymphoid/lymphoplasmacytic inflammation with intraepithelial lymphocytes, sometimes with ulceration and neutrophilic infiltrates.\\
\addlinespace[0.35em]
Stomach & \textbf{Lauren histological subtype}\newline{\scriptsize\texttt{tcga\_stad\_lauren\_diffuse}} & Diffuse type & 0.768 & Class characterized by signet(-ring) cells with abundant clear/vacuolated cytoplasm, set within prominent fibromuscular/smooth muscle stroma with myofibroblastic spindle cells and collagenous to myxoid change. There is a dense lymphoid/lymphoplasmacytic (lichenoid) inflammatory infiltrate, with associated fibrosis/edema.\\
\addlinespace[0.35em]
Stomach & \textbf{Lauren histological subtype}\newline{\scriptsize\texttt{tcga\_stad\_lauren\_diffuse}} & Intestinal type & 0.768 & Neoplasm with prominent gland formation composed of irregular, atypical intestinal-type glands/tubules lined by tall columnar mucin-producing epithelium (enterocytic differentiation with lumina and occasional villous features). Architectural complexity includes cribriforming and budding with comedo-type necrosis, accompanied by high-grade cytologic atypia (high N:C ratio, vesicular chromatin, macronucleoli, nuclear pleomorphism/irregular membranes).\\
\addlinespace[0.35em]
Stomach & \textbf{Microsatellite instability}\newline{\scriptsize\texttt{tcga\_stad\_msi}} & MSI-high & 0.857 & This class is characterized by an intestinal-type gland-forming epithelial neoplasm with irregular/tubular glands, budding, and occasional cribriform or micropapillary architecture, lined by tall columnar cells showing vesicular nuclei with prominent nucleoli and increased N:C ratio/pleomorphism. It is accompanied by a prominent lymphoid/lymphoplasmacytic infiltrate with intraepithelial lymphocytes and may show ulceration, neutrophilic inflammation, and focal comedo-type necrosis.\\
\addlinespace[0.35em]
\end{longtable}

\endgroup
\end{landscape}

\end{document}